\documentclass[12pt]{iopart}
\usepackage{iopams}
\usepackage{setstack}
\usepackage{graphicx}
\usepackage[latin1]{inputenc}
\usepackage{harvard}

\usepackage{bm}

\newcommand{\bfr}{\bi{r}}
\newcommand{\bp}{\bi{p}}
\newcommand{\bk}{\bi{k}}

\newcommand{\fr}{\frac}

\newcommand{\df}{\stackrel{\rm def}{=}}

\newcommand{\fraczero}[2]{{#1} \atop {#2}}

\newcommand{\FTF}{{\cal F}_{\rm TF}}
\newcommand{\TTF}{{\cal T}_{\rm TF}}
\newcommand{\tTF}{{t}_{\rm TF}}
\newcommand{\Eext}{{\cal E}_{\rm ext}}
\newcommand{\Vext}{{U}_{\rm ext}}
\newcommand{\Ecoul}{{\cal E}_{\rm coul}}
\newcommand{\DEF}{\equiv}

\newcommand{\smeq}{\! = \!}

\newcommand{\smpl}{\! + \!}
\newcommand{\smmi}{\! - \!}

\newcommand{\kt}{k_{{\rm B}}T}

\newcommand{\Hint}{{\hat H}_{{\rm int}}}

\newcommand{\area}{\mathrm{a}}
\newcommand{\Area}{\mathrm{A}}
\newcommand{\ssA}{\mathcal{A}}
\newcommand{\betaRMT}{{\beta_{\rm RMT}}}
\newcommand{\energy}{\epsilon}
\newcommand{\G}{{\cal G}}   
\newcommand{\orbittime}{t}
\newcommand{\Volume}{{\cal A}}

\newcommand{\cl}{\chi_{\scriptscriptstyle L}}
\newcommand{\kf}{k_{\scriptscriptstyle F}}
\newcommand{\vf}{v_{\scriptscriptstyle F}}
\newcommand{\pf}{p_{\scriptscriptstyle F}}
\newcommand{\Ef}{\epsilon_{\scriptscriptstyle F}}
\newcommand{\lambdaf}{\lambda_{\scriptscriptstyle F}}
\newcommand{\bq}{\bi{q}}

\newcommand{\kb}{k_{\scriptscriptstyle B}}

\newcommand{\dO}{{\Delta \Omega}}  
\newcommand{\Ha}{{\cal H}}     
\newcommand{\Fo}{{\cal F}}     

\newcommand{\Hop}{H_{\rm 1p}}
\newcommand{\kF}{k_{\scriptscriptstyle F}}
\newcommand{\lF}{\lambda_{{\rm F}}}
\newcommand{\pmc}{\mu_{{\rm mc}}}
\newcommand{\Eth}{E_{{\rm Th}}}
\newcommand{\maslov}{\zeta}
\newcommand{\Vsc}{V_{\rm sc}}
\newcommand{\hV}{{\hat V}_{\rm sc}}
\newcommand{\Vcoul}{V_{\rm coul}}
\newcommand{\Umf}{U_{\rm mf}}
\newcommand{\Vsr}{V_{\rm short range}}

\newcommand{\nubeta}{\nu_{\beta}}

\eqnobysec

\begin{document}
\bibliographystyle{jphysicsB}

\review{Many-Body Physics and Quantum Chaos}

\author{Denis Ullmo}
\address{ Universit\'e Paris-Sud, LPTMS, UMR8626, 91405 Orsay Cedex,
  {\em and}}
\address{ CNRS, LPTMS, UMR8626, 91405 Orsay Cedex}
\ead{denis.ullmo@u-psud.fr}

\begin{abstract}
  Experimental progresses in the miniaturisation of electronic devices
  have made routinely available in the laboratory small electronic
  systems, on the micron or sub-micron scale, which at low temperature
  are sufficiently well isolated from their environment to be
  considered as fully coherent. Some of their most important
  properties are dominated by the interaction between
  electrons. Understanding their behaviour therefore requires a description of
  the interplay between interference effects and interactions.

 The goal of this review is to address this relatively broad issue,
 and more specifically to address it from the perspective of the
 quantum chaos community.  I will therefore present some of the
 concepts developed in the field of quantum chaos which have some
 application to study many-body effects in mesoscopic and nanoscopic
 systems.  Their implementation is illustrated on a few examples of
 experimental relevance such as persistent currents, mesoscopic
 fluctuations of Kondo properties or Coulomb blockade.  I will
 furthermore try to bring out, from the various physical
 illustrations, some of the specific advantages on more general
 grounds of the quantum chaos based approach. 

\end{abstract}

\pacs{05.45.Mt,71.10.-w,73.23.Hk,73.23.Ra,72.10.Fk}


\submitto{\RPP}

\maketitle

\tableofcontents

\section{Introduction}

The title of this review may seem self contradictory in two respects.
To begin with, it associates chaos, a purely classical notion, with
quantum physics. Furthermore it implies that this association, which
as we will see refers traditionally to the study of low-dimensional
non-interacting quantum systems, will be considered in the context of
many-body physics.

The first of these contradictions is however mainly a question of
semantics.  Indeed, if in an early period of development of the field
of quantum chaos, some of the issues addressed had to do with the
possible existence of true chaotic dynamics for quantum systems, this
was relatively quickly answered, essentially in the negative.  Quantum
chaos now mainly refers to the study of the consequences, for a
quantum system, of the more or less chaotic nature of the dynamics of
its classical analogue.  It has followed two main avenues. The first
one is based on semiclassical techniques -- specifically the use of
semiclassical Green's function in the spirit of Gutzwiller's trace
formulae
\cite{Gutzwiller71,Balian72,Berry77b,Gutzwiller91,GutzwillerBook} --
which provides a link between a quantum system and its $\hbar \to 0$
limit. The second is associated with the Bohigas Giannoni Schmit
conjecture \cite{Bohigas84,Bohigas84jpl,Bohigas91}, or related
approaches \cite{Peres84}, which states that the spectral
fluctuations of classically chaotic systems can be described using the
proper ensembles of random matrices.

Some of the beauty of quantum chaos is that it has developed a set of
tools which have found applications in a large variety of different
physical contexts, ranging from molecular and atomic physics
\cite{Wintgen86,Delande86,Wintgen87,Delande91}, to acoustics
\cite{Derode95,Fink97,Fink00}, nuclear physics
\cite{Bohigas02,Olofsson06}, cold atoms
\cite{Mouchet01,Hensinger01,Steck01,Steck02}, optical \cite{Nockel97,Gmachl98}
 or microwave \cite{Stockmann90,Sridhar91,Kudrolli95,Alt95,Pradhan00}
 resonators, and of course mesoscopic physics
\cite{RichterPhysRep96,Richter00Book,Alhassid00RMP}.  With few exceptions (see
nevertheless \cite{Bohigas02,Olofsson06}) most of these physical
systems share the property of being correctly described by
non-interacting, low-dimensional, models.

This is true in spite of the fact that random matrix ensembles were
introduced in the early fifties by Wigner \citeaffixed{PorterBook}{see
the series of  articles reprinted in} to
explain the statistics of slow neutron resonances, and were therefore
applied in the context of strongly interacting systems.  In that case
however, it was less the notion of chaos than the one of complexity
(large number of degrees of freedom, strong and complicated interactions) which
was proposed by Wigner to justify this approach.

At any rate, the scope of this review will be concerned with a very
different type of interacting system, namely the Landau-Fermi liquid,
for which the system can be explain as a set of quasi-particles
interacting weakly through a (renormalized) interaction amenable to
perturbative treatment.  More specifically, what we have in mind are
various realizations of fully coherent, confined electron gasses, with
a density high enough that a Landau-Fermi-liquid type description
applies. These are typically semiconductor quantum dots or small
metallic nano-particles, within which a few tens to a few thousands of
electrons interact through a screened Coulomb interaction.

Although this screened interactions between electrons is weak, and is
therefore well described by a standard perturbative approach (first
order perturbation in the simplest cases, or eventually with some
re-summation of higher order terms in other situations), some
important physical processes are actually largely dominated by them.
Moreover, the systems considered are only weakly affected by their
environment and can therefore be assumed fully coherent.  Because of
the confinement, translational invariance is then broken, and some new
and interesting physics is brought in by the fact that, in the
non-interacting limit from which the perturbation scheme is developed,
eigenstates are not just plane waves. The mesoscopic fluctuations
associated with confinement and interference need to be taken into
account for the eigenstates and one-particle energies, either at a
statistical level or in a detailed way associated with a given
geometry.

How to describe these mesoscopic fluctuations, and implement their
consequences for many-body effects can be done in a variety of ways.
For diffusive systems, techniques based on diagrammatic perturbation
expansion in the disorder potential can be used
\cite{AltshulerAronov85,Aleiner02PhysRep,AkkermansMontambauxBook}, as
well as approaches based on the supersymmetric $\sigma$-model
\cite{EfetovBook,Mirlin00PhysRep}, which is also appropriate for the
description of ballistic chaotic systems \cite{Blanter01b} (see also
\cite{Andreev96,Leyvraz97,Agam97} in this context).  In this review, I
shall however limit myself to the methods coming from the quantum
chaos community.  One reason for this limitation of scope is that
there already are very good and complete reviews which give a
excellent account of the other approaches.  Another is that the quantum
chaos perspective is in many useful cases more intuitive, and somewhat
simpler to apply from a technical point of view.  As a consequence,
this will make it possible to present in an essentially self-contained
way the technical tools required to understand a large class of
many-body effects relevant for these quantum dots or nano-particles,
{\em as long as they are in the Landau-Fermi-liquid regime}.  The goal
is that it should be possible to follow almost all of the review with
a graduate level in quantum mechanics and, in some cases, basic
notions of many-body theory such as what can be found in classic
textbooks such as Fetter and Walecka \cite{Fetter&Walecka}.  This, I
hope, will make it particularly convenient for experimentalists or
theoreticians who wish to enter into this field.

Another advantage of the quantum-chaos based approach is that it is by
nature more flexible, and is therefore not limited to chaotic or
diffusive dynamics.  How much physics is missed by other points of
view because of this limitation can be debated, and I will return to
this discussion at the end of this review.  However, if for metallic
nano-particles the choice of a description in term of disordered
(diffusive) system is dictated by the actual physics of these
materials, it is clear that for semiconductor quantum dots, one reason
for why so much focus has been put on chaotic dynamics is that it is
the only one that could be addressed by the more traditional
techniques of solid state physics.  Having a tool which makes it
possible to consider other kinds of dynamics at least gives the
possibility of asking the question of whether anything new, or
interesting, can be found in these other regimes.

The structure of this review is therefore the following. A first
section will be devoted to the description of the basic tools
necessary to study the physical problems we want to consider.  As we
want to address different energy scales, or from an experimental point
of view different temperature ranges, it will be necessary to introduce
a few complementary point of views.  Semiclassical techniques, and in
particular the use of semiclassical Green's function, will be well
adapted to temperature ranges significantly larger than the mean level
spacing $\Delta$.  They will be the subject of
section~\ref{sec:semiclassics}. The low ($ T < \Delta$) temperature
regime however requires a modeling of individual energies and
wave-functions, and are therefore better described, in the hard chaos
regime, by statistical approaches such as random matrix theory and the
random plane wave model.  These latter will be introduced in
section~\ref{sec:RMT}.  Finally section~\ref{sec:screening} will
provide a discussion of the screening
of the Coulomb interaction.

I will then turn to the description of a few examples of physical
systems where the physics is dominated by the interplay of interaction
effects and mesoscopic fluctuations.  The choice of these examples is
of course quite arbitrary, and the criterion I have used to select
them is essentially my familiarity with the issue.  Therefore, there
will be a strong bias toward questions I have actually worked on,
which should not be interpreted as a statement about their relative
importance.  I will start with a discussion of the orbital magnetic
response, with some general considerations in
section~\ref{sec:OrbMag1} followed by a few specific examples of
diffusive and ballistic systems in section~\ref{sec:OrbMag2}.  One
important difficulty to be addressed here is the renormalization of
the interaction due to higher order terms in the Cooper channel, and
this issue will be discussed in detailed in both the diffusive and
ballistic regimes.  In section~\ref{sec:Kondo}, I will wander briefly
away from Fermi liquid systems and address the mesoscopic fluctuations
associated with the physics of a Kondo impurity \cite{Kondo64,HewsonBook}
placed in a finite-size system.  The last physical example will be, in
Section~\ref{sec:CB}, the role of interactions in the fluctuation of
peak spacing in Coulomb blockade
\cite{Beenakker91,Weinmann96,Grabert&DevoretBook,Kouwenhoven97Book}
experiments.  After a general introduction of the universal
Hamiltonian picture, I will cover the various physical effects which
needs to be further considered if one expect to understand
experimental peak spacing and ground-state spin distributions for
these systems.

Finally, the concluding section will contain some general discussion,
and in particular will come back to the issue of non-chaotic dynamics.

\setcounter{footnote}{1}

\section{Basic tools}
\label{sec:basic}

\subsection{Semiclassical formalism}
\label{sec:semiclassics}

Consider a system of indistinguishable Fermions governed by the
one-particle Hamiltonian
\begin{equation} \label{eq:H0}
\Hop = -\frac{\hbar^2}{2m_e} \Delta + U(\bfr) \; ,
\end{equation}
and interacting weakly through the two body potential $V(\bfr-\bfr')$.
A systematic perturbative expansion can be constructed to arbitrary
order (if necessary) in terms of the unperturbed Green's function
\footnote{All the Green's functions used in this review will be
  unperturbed Green's functions.  I shall therefore not use any
  subscript to distinguish them from the interacting ones.}
\begin{equation}
G(\bfr,\bfr';\energy) \df \langle \bfr | \frac{1}{\energy - \Hop } | \bfr' \rangle = 
\sum_\kappa \frac{\varphi_\kappa(\bfr)
  \varphi_\kappa^*(\bfr')}{\energy - \epsilon_\kappa } 
\; ,
\end{equation}
where in the last expression $\epsilon_\kappa$ and $\varphi_\kappa$
are respectively the energies and eigenstates of $\Hop$.  In a clean
bulk system $U(\bfr) \equiv 0$ so that the eigenstates are just plane
waves, and the expression of the Green's function becomes trivial in
the momentum representation.  For confined (coherent) systems however,
translational invariance is lost and there is in general no simple
expression for the exact  eigenstates and eigenfunctions.  It
therefore becomes necessary to find some approximation scheme for the
unperturbed Green's function itself before considering a
perturbation expansion in the interaction.

\subsubsection{Semiclassical Green's function}

In many regimes of interest a semiclassical approach can be used to
fulfil this role. This includes the case where the confining
potential $U(\bfr)$ is a smooth function on the scale of the Fermi
wavelength $\lF$, but also for instance when it contains a weak,
eventually short range, disorder, as long as $\lF$ is much
smaller than the elastic mean free path $\ell$.  Under these
conditions, the retarded Green's function $G^R(\energy) \df
\lim_{\eta \to 0} G(\energy+i\eta)$ can be written as a sum over all {\em
  classical} trajectories $j$ joining $\bfr'$ to $\bfr$ at energy $\energy$
\cite{GutzwillerBook,Gutzwiller91}
\begin{eqnarray}
G^R(\bfr,\bfr';\energy) & \simeq &  \sum_{j:\bfr'\to\bfr} G^R_j(\bfr,\bfr';\energy) 
\nonumber \\
 G^R_j(\bfr,\bfr';\energy) & \df &
\frac{2\pi}{(2i\pi\hbar)^{(d+1)/2}}
   D_j(\energy) \exp \left( iS_j(\energy)/\hbar - i
  \maslov_j \pi/2 \right ) \; , \label{eq:1/2classGreen}
\end{eqnarray}
with $d$ the number of degrees of freedom,
\begin{equation} \label{eq:Sj}
 S_j(\energy) = \int_{\bfr'}^\bfr \bp \cdot d\bfr
\end{equation} 
the classical action along the trajectory $j$, and 
\begin{equation} \label{eq:Dj}
D_j(\energy) = \left| \frac{1}{\dot r_1 \dot r'_1} {\rm det}' 
\left[ -\fr{\partial^2 S}{\partial \bfr \partial \bfr'} \right]
\right|^{1/2} 
\end{equation}
a determinant describing the stability of trajectories near $j$.  In
(\ref{eq:Dj}) $\bfr = (r_1, \cdots , r_d)$, and the prime on the
determinant indicates that the first component (i.e.\, first row
$\partial^2 S/\partial r_1 \partial r'_i$ and first column $\partial^2
S/\partial r_i \partial r'_1$, $i=1,\cdots,d$) is omitted.  Finally,
the Maslov index $\maslov_j$ essentially counts the number of caustics
(i.e. places where the determinant $D_j$ is zero) on the trajectory
$j$ between $\bfr'$ and $\bfr$ \footnote{For the kind of kinetic plus
  potential Hamiltonian we consider here, the Maslov index increase by
  one at each crossing of a caustic.  Note however that for more
  general Hamiltonian it may however also decrease.}.  For two
dimensional systems ($d=2$), the determinant takes the particularly
simple form
\begin{equation} \label{eq:Dj2d}
D_j(\energy) = \left| \frac{1}{\dot r_\| \dot r'_\|} \frac{1}{\left( \partial
  r_\bot/ \partial p'_\bot \right)_{r_\bot} } \right|^{1/2} \; ,
\end{equation}
where $r_\|$ and $r_\bot$ are the $r$-components  respectively parallel
and orthogonal to the trajectory.

\subsubsection{Simple properties of the classical action}

Many important characteristic features of the semiclassical Green's
function, and therefore of the fermion gas, can be directly deduced
from basic properties of the classical action
\cite{ArnoldBook,GoldsteinBook}. In particular: \\ {\bf i) the
  variation with respect to the energy }
\begin{equation} \label{eq:dS/dE}
\frac{\partial S_j}{\partial \energy} = \orbittime_j \; ,
\end{equation}
where $\orbittime_j$ is the time elapsed  to go from
$\bfr'$ to $\bfr$ along trajectory $j$ at energy $\energy$;
\\ {\bf ii) the variation with respect to the  position}
\begin{equation} \label{eq:dS/dr}
\frac{\partial S_j}{\partial \bfr'} = -\bp'_j \qquad \frac{\partial
  S_j}{\partial \bfr} =  \bp_j \; ;
\end{equation}
and finally
\\ {\bf iii) the effect of a perturbation.} \\
Indeed, let us assume that the one particle Hamiltonian can be written as the
sum of a main term $H_0$ and a small perturbation $H_1$
\begin{equation}
H_{\rm 1p} = H_0 + H_1 \; ,
\end{equation}
and let us denote by $S^0_j$ the action calculated for the trajectory
$j$, i.e. $(\bfr^0_j(t),\bp^0(t))_j$, $t \in [0,t^0]$,
joining $\bfr'$ to $\bfr$ under the Hamiltonian $H_0$.  We then have
\begin{equation} \label{eq:class_pert}
\delta S_j \df S_j - S_j^0 \simeq - \int_0^{t_0} dt \,
H_1(\bfr^0_j(t),\bp^0_j(t))  \; .
\end{equation}
Note there is no need for $H_1$ to be small on the quantum scale, and
therefore (\ref{eq:class_pert}) remains applicable much beyond the
limit of quantum perturbation theory.\footnote{It should be stressed
  that the exponential sensitivity to perturbations of chaotic
  trajectories does not prevent finding a perturbed trajectory
  following closely the unperturbed one and joining the same endpoints
  in configuration space (but with a slightly different momenta).
  See for instance \citeasnoun{Cerruti02} for a recent discussion of this
  (old) question.}

To illustrate how the above properties can be used in our context,
let us consider for instance the (unperturbed) local density of states,
\begin{equation}
  \nu_{\rm loc}(\bfr;\energy) \df \sum_\kappa |\varphi_\kappa(\bfr)|^2
  \delta(\energy - \epsilon_\kappa) = - \frac{1}{\pi} {\rm Im}
  G^R(\bfr,\bfr;\energy) \; .
\end{equation} \label{eq:nuloc}
Using (\ref{eq:1/2classGreen}), $\nu_{\rm loc}$ can be expressed
as a sum over all the closed orbits starting and
ending at the point $\bfr$.  In this process, the ``direct'' orbit $j_0$,
whose length goes to zero as $\bfr \to \bfr'$, needs however to be treated
separately, as the determinant $D_{j_0}$ diverges.  On the other hand,
the contribution of this orbit to the Green's function can be
identified to the free Green's function for a constant potential. It
therefore just gives rise to the smooth (bulk-like) contribution 
\begin{equation} \label{eq:nu-0}
\nu^{(d)}_0(\bfr;\energy) =  \int \frac{d\bp}{(2\pi\hbar)^d} \delta(\energy -
H(\bp,\bfr)) =  \frac{m_e k(\bfr)^{d-2}}{(2\pi)^n \hbar^2} 
 \frac{d \pi^{d/2} }{\Gamma(d/2+1)}  \; . 
\end{equation}
($\nu^{(2)}_0 = m_e/2\pi\hbar^2$, $\nu^{(3)}_0 = m_e k/2\pi^2\hbar^2$),
where $d$ is the dimensionality and the last equality holds for the
usual kinetic plus potential Hamiltonian, with $k(\bfr) = \sqrt{2m_e
  (\energy-U(\bfr))}/\hbar$.
I shall in the following  use the notation $\rho(\energy) = \int d\bfr
\nu_{\rm loc}(\bfr;\energy)$ for the total density of states, and 
\begin{equation} \label{eq:rho-0}
\rho_0(\energy) =  \int \frac{d\bp \, d\bfr}{(2\pi\hbar)^d} \delta(\energy - H(\bp,\bfr))
\end{equation}
for its Weyl (smooth) part.

The local density of states can therefore be separated into a smooth
and an oscillating part
\begin{equation}
   \nu_{\rm loc}(\bfr;\energy) = \nu_0 (\bfr;\energy) + \nu_{\rm osc}(\bfr;\energy)
\end{equation}
where $\nu_{\rm osc}(\bfr;\energy)$ is expressed semiclassically as a sum
over all finite length closed orbits
\begin{eqnarray} \label{eq:nu-osc}
\nu_{\rm osc}(\bfr;\energy) & = & \sum_{j \neq j_0:\bfr\to\bfr}
\nu_{j}(\bfr;\energy) + {\rm c.c.}\\
\nu_{j}(\bfr;\energy) & = & 
\frac{-i}{(2i\pi\hbar)^{(d+1)/2}}
   D_j(\energy) \exp \left( i S_j(\energy)/\hbar - 
  i \maslov_j \pi/2 \right ) \; .
\end{eqnarray}

For energies $\energy$ close, on the classical scale, to some reference
energy $\bar \energy$, one can therefore use (\ref{eq:dS/dE}) to write
\begin{equation} \label{eq:rho(E+dE)}
\nu_{j}(\bfr;\bar \energy + \delta \energy) = \nu_{j}(\bfr;\bar \energy) 
\exp(i \, \delta \energy \, \orbittime_j / \hbar) \; .
\end{equation}
Thus, the local density of states appears as the bulk contribution plus
some oscillating terms which, with  (\ref{eq:dS/dE}), have
a period in energy $2\pi \hbar / \orbittime_j$ determined by the {\em travel
  time} of the closed orbits.

In the same way, Friedel oscillations near the boundary of the system
or near an impurity can be understood as a direct consequence of
(\ref{eq:dS/dr}), applied to the trajectory bouncing on the
obstacle and coming back directly to its starting point.  Quite
generally one can write for the contribution of the orbit $j$ to the
local density of states
\begin{equation}
\nu_{j}(\bfr + \delta\bfr;\energy ) = \nu_{j}(\bfr;\energy) 
\exp(i (\bp'_j - \bp_j) \delta\bfr/ \hbar) \; ,
\end{equation}
so that locally $\nu_{\rm loc}(\bfr;\energy)$ appears as a sum of plane waves
the wave vectors of which are determined by the difference between the
final and initial momentum of the corresponding returning orbits.
Periodic orbits, which are such that $\bp'_j = \bp_j$, have no
variation locally and therefore will give rise to the dominant
contribution to the total density of states $\rho(\energy)$
\cite{Gutzwiller70,Gutzwiller71}.  Friedel oscillations on the other
hand correspond to trajectories which, after bouncing off the boundary
of the system or some impurity, travel back directly to the initial
point, so that $\bp'_j = - \bp_j$.  In this case the corresponding
plane wave contribution has a wave vector with modulus twice the
wave vector $k(\bfr) = \sqrt{(\energy - U(\bfr))/2m_e} /\hbar$.  In the particular
case of a two dimensional system with a straight hard wall (with
Dirichlet boundary condition) at $x=0$, direct application of
(\ref{eq:Dj2d}) and (\ref{eq:nu-osc}) gives (for $k x \geq 1$)
\begin{equation}
\nu_{\rm osc}(\bfr\!=\!(x,y); \energy)  = - {\sqrt{2\pi} \nu_0^{(2)}}
\frac{\sin(2 k x + \pi/4) }{\sqrt{2 k x}} \; ,
\end{equation}
from which the Friedel oscillations in the density of particles
$n(\bfr)$ is derived by integration over the energy.
 More general cases (e.g.\ curved boundary) are easily obtained by
calculating the corresponding value of ${\partial p_\bot}/{\partial
  r'_\bot }$ (and eventually Maslov indices).

Finally as an illustration of the third property (\ref{eq:class_pert}),
let us compute the variation of the density of states under the
modification of an external magnetic field $\bi{B} = {\bnabla}
\times \bi{A}$.  When the magnetic field is changed from $\bi{B} \to
\bi{B} + \delta \bi{B}$ (with the corresponding change of the vector
potential $\bi{A} \to \bi{A} + \delta \bi{A}$), one has
\begin{equation}
 H_0 = \frac{1}{2m_e} (\bp - e \bi{A})^2 + U(\bfr) \; ,
\end{equation}
and in first order in $\delta \bi{A}$
\begin{equation}
 H_1 = \frac{e}{m_e} (\bp - e \bi{A}) \cdot \delta \bi{A} = 
e \bi{v} \cdot \delta \bi{A} \; ,
\end{equation}
with $\bi{v} = \dot \bfr = (\partial H/\partial \bp)$ the velocity.
The variation of the action along a closed trajectory $j : \bfr \to \bfr$ is
therefore given by
\begin{equation}  \label{eq:delta_S(B)}
 \frac{1}{\hbar} \delta S_j = \frac{e}{\hbar}\oint_{j : \bfr \to \bfr}
 dt \,
 (\delta \bi{A} \cdot \bi{v})  = \frac{e}{ \hbar}\oint_{j : \bfr \to \bfr}
 d\bi{l} \cdot \delta \bi{A}  = \frac{2\pi \delta \phi_j}{\phi_0}  \; ,
\end{equation}
where, $\delta \phi_j$ is the flux of $\delta B$ across the trajectory
$j$ and $\phi_0 \df 2 \pi \hbar /e$ is the quantum flux.  The
variation of the contribution of the orbit $j$ to the local density of
states can therefore be written as
\begin{equation}\label{eq:delta_rho_j(B)}
\nu_{j}(\bfr; \energy ; \bi{B} + \delta \bi{B}) = \nu_{j}(\bfr;\energy; \bi{B}) 
\exp(i \delta \phi_j / \phi_0) \; .
\end{equation}

\subsubsection{Sum rule for the determinant $D_j$}

The computation of the contribution of some orbit $j$ to the
semiclassical Green's function for a given geometry implies in
practice the determination of the action of the orbit, which is
usually not too difficult, but also of the stability determinant $D_j$
and the Maslov index $\maslov_j$ which for generic systems may involve
some technicalities (see e.g.\ \citeasnoun{Bogomolny88} for an illustration on
the examples of the stadium and elliptic billiards, and \citeasnoun{Creagh90} for a
detailed discussion about the Maslov indices).  It turns out that in
practice a large number of results can be obtained without an explicit
calculation of these quantities, but can be derived from a
sum rule (M-formula) for the determinants $D_j$, analog in spirit to
the Hannay - Ozorio de Almeida sum rule \cite{Hannay84,OzorioBook},  and which
in a similar way expresses the conservation of Liouville measure by the
classical flow.  The M-formula can be expressed as
\cite{Argaman96}
\begin{equation} \label{eq:Mformula}
\sum_{j:\bfr' \to \bfr}\frac{ |D_j(\energy)|^2}{(2\pi\hbar)^d} \delta(t
- \orbittime_j) 
= \nu_0^{(d)}(\bfr') P^\energy_{\rm cl}(\bfr,\bfr',t) \; ,
\end{equation}
where $\nu_0^{(d)}(\bfr')$ is the bulk density of states per unit area
(and spin) (see (\ref{eq:nu-0})) for the local value of $\kF$ 
and $P^\energy_{\rm cl}(\bfr,\bfr',t)$ is the classical (density of)
probability that a trajectory launched in $\bfr'$ is in the neighbourhood
of $\bfr$ at time $t$.

The sum rule (\ref{eq:Mformula}) is particularly useful for
diffusive systems, for which $P^\energy_{\rm cl}(\bfr,\bfr',t)$ is solution of
a diffusion equation (with diffusion coefficient $D$)
\begin{equation} \label{eq:diffusion}
(\partial_t + D \Delta_\bfr ) P^\energy_{\rm cl}(\bfr,\bfr',t) = \delta(\bfr-\bfr')
  \delta(t) 
\end{equation}
with boundary conditions $\partial_{\vec n} P_{\rm cl} = 0$ at the
boundary of the system (if any).  We shall see in the next subsection
that it can be also applied usefully for ballistic chaotic systems.

\subsubsection{Thouless Energy}

When considering a confined system of (for now) non-interacting
particles, one might first, before any actual calculation, try to
understand what are the energy scales
affected by the confinement.  On the low-energy side, this is clearly
bounded by the mean level spacing, the finiteness of which is the most
obvious consequence of the fact that the system is bounded.  On the
high energy end, a direct implication of (\ref{eq:rho(E+dE)}) is
that if the Green's function is smoothed on an energy scale $\energy$, only
the contributions of  trajectories with a time of travel $\orbittime < \hbar/
\energy$ survive the averaging process.   Therefore the minimal time
$\orbittime_{\rm min}$ such that a classical particle feels the presence of
the boundary determines the maximal energy scale $\Eth$ such that the
quantum system is affected by this latter.  This energy scale, $\Eth$, is
referred to as the Thouless energy.  

For ballistic systems, $\orbittime_{\rm min}$ is essentially the time of
flight $\orbittime_{\rm fl}$ across the system, which is also the time scale
of the shortest closed orbit for a generic point $\bfr$ inside the
system.  As a consequence $\Eth$ is also in this case the energy scale
beyond which no fluctuations exist.

For diffusive systems $\orbittime_{\rm min}$ is the time necessary to
diffuse to the boundary of the system, time at which the solution of
the diffusion equation (\ref{eq:diffusion}) starts to differ from a
free space diffusion.  In that case however, this scale is different
(and usually significantly larger) than the time associated with the
shortest closed orbit, which is rather of the order of the momentum
randomisation time $\orbittime_{\rm tr}$.

As I shall illustrate below, a significant number of results can be
derived for singly connected chaotic or diffusive quantum dots using
the sum rule (\ref{eq:Mformula}) in conjunction with a simple
approximation for the probability $P^\energy_{\rm cl}(\bfr,\bfr',t)$. Indeed,
for time shorter than $\orbittime_{\rm min}$ the presence of the boundaries
can be ignored and the 
free flight (free diffusion) expression can be used for the
ballistic (diffusive) case.  On the other hand the classical probability 
$P^\energy_{\rm cl}(\bfr,\bfr',t)$ can usually be taken as being independent
of the initial condition for time larger than $\orbittime_{\rm min}$ and
simply proportional to the phase space volume $\int d \bp \,
\delta \left( \energy-H(\bp,\bfr) \right)$.  For a billiard system of volume
$\Volume$ this gives for instance for the return probability
\begin{eqnarray} 
P^\energy_{\rm cl}(\bfr,\bfr,t) \, = \, 0  \; \;  & \qquad & t < \orbittime_{\rm
  min} 
\label{eq:PclChaosa}\\
P^\energy_{\rm cl}(\bfr,\bfr,t) \, = \, 1/\Volume & \qquad & t > \orbittime_{\rm min} 
\label{eq:PclChaosb}
\end{eqnarray}
for chaotic system, and in the diffusive case
\begin{eqnarray}
P^\energy_{\rm cl}(\bfr,\bfr,t) \, = \, (4\pi D t)^{-d/2}  & \qquad & t < \orbittime_{\rm min} \\
P^\energy_{\rm cl}(\bfr,\bfr,t) \, = \, 1/\Volume \qquad & \qquad & t > \orbittime_{\rm min} \; .
\end{eqnarray}

\subsection{Random matrix and random-plane-wave models in the hard-chaos regime}
\label{sec:RMT}

The semiclassical approach introduced in the previous subsection is
the natural tool to describe energy scales significantly larger than
the mean level spacing $\Delta$.  It is however not convenient, would
it be only because it is usually not convergent, when the properties
of a single wave-function are considered, and more generally when
quantum properties are investigated on the scale not larger than a few
mean level spacings (see however \citeasnoun**{Tomsovic08} in this
respect). 

In the hard-chaos regime (and quite often in the diffusive regime), it
is however possible to use an alternative route, based on
the statistical description of the eigenstates and eigenfunctions
fluctuations.

\subsubsection{Random matrices}

The most basic model is to assume that the fluctuations of physical
quantities for the quantum system under consideration are well
described by ensembles of random matrices, such as the Gaussian
orthogonal, unitary or symplectic ensembles.  These ensemble have been
introduced by Wigner in the context of nuclear physics to account for
the complexity (i.e.: large number or degrees of freedom, large and
complicated interactions) characteristic of the nuclei.  Studies of
billiard systems have however led Bohigas, Giannoni and Schmit
\cite{Bohigas84,Bohigas84jpl} to conjecture that even ``simple''
(i.e.: low dimensional, with innocent looking Hamiltonians) systems
would display the spectral fluctuations of these Gaussian ensembles as
long as they have a chaotic dynamics.  This conjecture, although still
not formally proven, is supported by recent semiclassical calculations
showing that the form factor (the Fourier transform of the two point
correlation function) predicted by the random matrix ensembles can be
recovered in all order of a perturbation expansion within a periodic
orbit theory \cite{Berry85,Richter02prl,Muller04,Heusler04,Muller05}.
I has furthermore been verified numerically on a large variety of
systems, giving a kind of experimental demonstration that classical
chaos, rather than complexity, is the origin of these characteristic
fluctuations properties.

Wigner Gaussian ensembles are constructed by first considering the set
of $N \times N$ (in the limit $N \to \infty$) Hamiltonian matrices
corresponding to the symmetries with respect to time reversal of the
physical systems.  Those are Hermitian matrices $H = H^{(0)} + iH^{(i)}
$ for time-reversal non-invariant systems (Gaussian unitary ensemble
(GUE)), real symmetric matrices $H = H^{(0)}$ for spinless time-reversal invariant systems (Gaussian orthogonal ensemble, (GOE)) , and
quaternion real matrices $H = H^{(0)} \otimes {\bf 1} + H^{(1)}
\otimes \sigma_x + H^{(2)} \otimes \sigma_y + H^{(3)} \otimes \sigma_z
$ for spin 1/2 time-reversal invariant but non-rotationally invariant
systems (Gaussian symplectic ensemble, (GSE)).  ($H^{(0)}$ is a real
symmetric matrix, and the $H^{(\alpha)}, \alpha > 0$ are real
antisymmetric matrices.)  The associated probability is then
constructed i)~assuming that each matrix element $h_{ij}^{(\alpha)}$
($i \leq j$) is independent; ii)~in such a way that the probability is
invariant under the group transformation corresponding to a change of
basis (unitary transformations for GUE, orthogonal transformations for
GOE, and symplectic transformations for GSE).  This leads to
\cite{MehtaBook}
\begin{equation} \label{eq:P(H)}
P_\betaRMT(H) dH = K_{N\betaRMT} \exp \left[-{\rm Tr}(H^2)/4v^2 \right] dH
\end{equation}
where  $K_{N\betaRMT}$ is a normalization constant, $\betaRMT$ 
indexes the symmetry class ($\betaRMT=1,2,4$ corresponding respectively
to GOE, GUE and GSE), $v$ is an energy scale determined in practice by the
physical mean level spacing, and $dH$ is the natural measure
\begin{eqnarray}
dH = \prod_{i \leq j} \prod_\alpha d h_{ij}^{(\alpha)} \; .
\end{eqnarray}
From the probability distribution (\ref{eq:P(H)}) various spectral
correlation functions can be derived \cite{MehtaBook}.  For instance, the
distribution of the (scaled) nearest neighbor $s= (\epsilon_{n+1} -
\epsilon_{n})/\Delta$ can be shown to be well approximated by the
Wigner-surmise distribution  
\begin{equation} \label{eq:Wigner_surmise}
P_{\rm nns}(s) = a_\betaRMT s^\betaRMT \exp(-c_\betaRMT s^2) \; ,
\end{equation}
with the numbers $(a_\betaRMT,c_\betaRMT)$ fixed by the constraints on
the normalization and on the mean \footnote{The parameters
  $(a_\betaRMT,c_\betaRMT)$ are equal to $(\pi/2,\pi/4)$ for
  $\betaRMT=1$, $(32/\pi^2,4/\pi)$ for $\betaRMT=2$, and
  $(2^{18}/3^6\pi^3, 64/9\pi)$ for $\betaRMT=4$ (see e.g.~ Appendix A
  of \citeasnoun{Bohigas91}).}.

The main content of (\ref{eq:P(H)}) is its universal character.
Indeed beyond the scale $v$ and the symmetries of the system, the
resulting distributions are completely independent of the particular
feature of the physical problem under consideration, as long as the
corresponding classical dynamics is chaotic.  This makes it possible
to obtain quantitative predictions for various physical configuration
without a precise knowledge of the system specific details. This is
presumably one of the reasons why so much focus has been put, both
theoretically and experimentally, on the hard chaotic regime.

\subsubsection{Random-plane-wave model}

As long as spectral statistics are concerned, and that energy scales much shorter
than the Thouless energy are considered, the random matrix models
have been shown to be extremely reliable.  The situation is 
however more ambiguous when one considers  wave-functions statistics.
On one hand, some properties, like the Porter-Thomas character 
\cite{Brody81}
\begin{eqnarray} 
  P(u = |\varphi_n(\bfr)|^2/\langle|\varphi|^2\rangle) & = & \qquad \quad
  \exp(-u) 
  \qquad \mbox{GOE} \label{eq:Porter_Thomas_GOE}
  \\
  & = & \frac{1}{\sqrt{2\pi u}} \exp(-u/2)
  \qquad \mbox{GUE} \label{eq:Porter_Thomas_GUE}
\end{eqnarray}
of the fluctuations around the mean value $\langle|\varphi|^2\rangle$ of
the eigenstates probability at a given position, are well observed in
numerical calculations, and can be derived straightforwardly from a
random matrix description.  On the other hand some other wave-function
statistics clearly cannot be addressed within a simple random matrix
model. Consider for instance a billiard-like quantum dot, which is
therefore characterized by a constant (in space) Fermi wave vector
$\kF$.  Wave-functions correlations are then characterized by a length
scale $\lF = 2\pi/\kF$ which is clearly absent in the random matrix
description.

To introduce this scale in the wave-function statistics let us
consider, for an arbitrary wave function $\varphi(\bfr)$, its Wigner
transform defined as
\begin{equation} \label{eq:wigner}
[\varphi]_{\rm W}(\bfr,\bp) \df \int d \bi{x} \exp(-i \bp \bi{x} /\hbar)
\varphi^*(\bfr + \bi{x}/2) \varphi(\bfr - \bi{x}/2) \; .
\end{equation}
The normalization of the wave function implies $(2\pi\hbar)^{-d}\int
d\bfr d\bp [\varphi_n]_{\rm W} = 1$.  If $\varphi$ is an eigenstate
with energy $\energy$, it can be shown \cite{Berry77a,Voros79} that
for chaotic systems $[\varphi]_{\rm W}(\bfr,\bp)$ converges in
probability in the semiclassical limit toward the micro-canonical
distribution
\begin{equation} \label{eq:pmc}
\pmc(\bfr,\bp) = \rho_0^{-1} \delta(\energy_n - H(\bfr,\bp))
\end{equation}
($\rho_0$ is the Weyl density of states (\ref{eq:rho-0}).)
the definition (\ref{eq:wigner}) can be inverted into
\begin{equation} \label{eq:wigner_inverse}
\varphi^*(\bfr ) \varphi(\bfr') = \frac{1}{(2\pi\hbar)^d} \int d\bp \exp(i
\bp\cdot(\bfr-\bfr')/\hbar) [\varphi]_{\rm W}(\bfr,\bp) \; .
\end{equation}
Replacing on average $[\varphi]_{\rm W}$ by $\pmc$ one immediately obtains an
explicit expression for the two-point correlation function $\langle
\varphi^*(\bfr ) \varphi(\bfr') \rangle $.  For instance for the usual
kinetic plus potential Hamiltonian $H = \bp^2/2m + U(\bfr)$, and for
distance short enough that the variation of the potential (and
therefore of $k(\bfr)$) can be neglected
\numparts
\begin{eqnarray} 
 \langle \varphi^*(\bfr ) \varphi(\bfr') \rangle
& = &  \frac{\nu_0(\bfr)}{\rho_0}\int_0^{2\pi} \frac{d\theta}{2\pi} \exp(i
k |\bfr \! - \! \bfr'| \cos(\theta)) \nonumber\\
& = & \frac{\nu_0(\bfr)}{\rho_0}J_0(k
|\bfr \! - \! \bfr'|) 
\qquad \qquad  \qquad\mbox{($d$ = 2)} \label{eq:correlationsa} \\
\langle \varphi^*(\bfr ) \varphi(\bfr') \rangle & = &\frac{\nu_0(\bfr)}{\rho_0} 
         \int_{0}^{+\pi} \frac{\sin\theta d\theta}{2} 
          \exp(i k |\bfr \! - \! \bfr'| \cos\theta) \nonumber \\
& =  & 
\frac{\nu_0(\bfr)}{\rho_0} 
\frac{\sin(k |\bfr \! - \! \bfr'|)}{(k |\bfr \! - \! \bfr'|)} 
\qquad \qquad \qquad \mbox{($d$ = 3)}
\; . \label{eq:correlationsb}
\end{eqnarray}
\endnumparts
This equation is obviously valid only for $|\bfr \!  - \! 
\bfr'| \ll L$, with $L$ the typical size of the system.  

More generally, (\ref{eq:wigner_inverse}) with the notion that on
average $[\varphi]_{\rm W}(\bfr,\bp) = \pmc(\bfr,\bp)$ makes it natural to
model the statistical properties of an eigenstate $\varphi_i$ close (on
the scale of $L$) to some point $\bfr$, by a superposition of a large
number $M$ of plane waves
\begin{equation} \label{eq:RPW}
\varphi_i(\bfr') = \sum_{\mu=1,M} a_{i\mu} \exp(i \bp_\mu \cdot (\bfr'-
\bfr)/\hbar)
\end{equation}
 where the $\bp_\mu$ are uniformly
distributed with the probability $\delta (\energy_n - H(\bfr,\bp))$ and, for
time-reversal non-invariant systems, the $a_i$ are complex random
numbers such that
\begin{equation} \label{eq:aimu}
\langle a_{i\mu} a^*_{i'\mu'} \rangle =\frac{1}{ M}
\frac{\nu_0(\bfr)}{\rho_0} \delta_{ii'} \delta_{\mu\mu'} \; .
\end{equation}
Time-reversal invariance can be taken into account by having $2M$
plane waves ($\mu=\pm 1,\ldots,\pm M$) with $\bp_{-\mu} = - \bp_\mu$,
$a_{-\mu} = a^*_\mu$, and $\langle a_{i\mu} a^*_{i'\mu'} \rangle =
(\nu_0(\bfr)/2\rho_0 M) \delta_{ii'} \delta_{|\mu||\mu'|}$.

I shall for instance use this approach in section \ref{sec:CB} to
include the residual interactions into the fluctuations of
Coulomb-blockade peak-spacings. In that case, the leading-order terms
in an expansion in $1/g$, where $g = \Eth / \Delta$ the dimensionless
conductance, can be derived straightforwardly from the model
(\ref{eq:RPW})-(\ref{eq:aimu}), supplemented only, following a kind of
``minimum information hypothesis'', by the assumption that the
$a_{i\mu}$ form a Gaussian vector.

Sub-leading terms are however
``aware'' of the finiteness of the system.  One then needs to further
modify the description of the wave-function fluctuations so that it
also includes the typical scale $L$ of the system under consideration.
A simple way to do this is to use only plane waves fulfilling the
quantization condition 
\begin{equation} \label{eq:quantcond}
\bp_\mu = 2 \pi \hbar \bi{n}_\mu / L
\end{equation}
with the set of integers $\bi{n}_\mu = (n_1, \cdots, n_d)$. While
doing so, one needs to give a width $\simeq \hbar / L$ to the Fermi
surface, and to include in the modelling of the eigenstates all the
plane wave with a kinetic energy in a band $\delta \energy = \alpha
\Eth$ around the Fermi energy, with $\alpha$ a constant of order one.
For the $M = \alpha g$ basis vector in this shell, one can then use a
random matrix description.  To leading order in $1/M$, the eigenstates
$\phi_i$ then fulfil (\ref{eq:RPW})-(\ref{eq:aimu}), as well as the
Gaussian character of the coefficient $a_{i\mu}$, but some
correlations of order $1/M$ are induced by the orthonormalisation
constraints \cite{Brody81}.  More importantly however, the width
$\delta p \sim \hbar /L$ of the Fermi surface modifies the
wave-function correlations at distances of order $L$ (for instance the
two point correlation function decreases more rapidly than what is
expressed by (\ref{eq:correlationsa})-(\ref{eq:correlationsb})).
Time-reversal symmetry can here be taken into account using a real
basis ($(\cos (\bp_\mu \cdot \bfr), \sin (\bp_\mu \cdot \bfr))$
instead of $(\exp (\pm i \bp_\mu \cdot \bfr)$), and a GOE matrix.

This model, which is defined both by the choice of the basis vectors
and by the use of random matrices is what I will refer to as the
random-plane-wave (RPW) model.  It has to be borne in mind that this
is a model, justified by physical considerations, but that should be
in principle validated by comparing the fluctuations derived within
this model to those obtained for the eigenfunctions of actual chaotic
systems such as quantum billiards
\citeaffixed{McDonald88,Kudrolli95,Pradhan02,Urbina03,Miller05}{see
  for instance in this respect}.  In particular,  relatively delicate
questions concerning the normalization of the wave-functions may be
important for some statistical quantities, which then requires
further modifications of the random-plane-wave model described above
\cite{Urbina04,Urbina07,Tomsovic08}.

\subsection{Screening of the Coulomb interaction in quantum dots}
\label{sec:screening}

For a degenerate electron gas, the  ``strength'' of the  Coulomb
interaction 
\begin{equation} \label{eq:bare}
\Vcoul(\bfr-\bfr') = \frac{e^2}{|\bfr - \bfr'|}
\end{equation}
between particle is usually expressed in terms of the gas parameter
$r_s = r^{(d)}_0/a_0$, where $a_0 \df \hbar^2/m_e e^2$ is the (3$d$)
Bohr radius and $r^{(d)}_0$ is the radius of a $d$-dimensional sphere
containing on average one particle. Expressing, for instance for $d \!
= \! 2$ or $3$, the density of particle $n^{(d)}_0 \equiv
(2\pi\hbar)^{-d} \mathrm{g_s} \int d\bp \, \Theta(\energy -\bp^2/2m)$ as
($\mathrm{g_s} = 2$ is the spin degeneracy, and $\Theta$ the Heaviside
function)
\begin{eqnarray}
n^{(2)}_0 \df \frac{1}{\pi r_0^{(2)}} & = & \frac{\mathrm{g_s}
  \kf^2}{4\pi} \; , \\
n^{(3)}_0 \df \frac{1}{\frac{4}{3} \pi r_0^{(3)}} & = & \frac{\mathrm{g_s}
  \kf^3}{6\pi^2} \; ,
\end{eqnarray}
we see that, up to a constant of order one, $r^{(d)}_0$ is essentially
the inverse of the Fermi wave vector $\kf$.  As a consequence, $r_s$
is, again up to a constant of order one, proportional to the ratio
$(e^2/r^{(d)}_0) / (\hbar^2 \kf^2 / 2m_e)$ of the Coulomb energy
between two electrons at typical inter-particle distance and the
kinetic energy. The parameter $r_s$ therefore actually measures the
relative strength of the Coulomb interaction.

Even for small $r_s$, $\Vcoul(\bfr-\bfr')$ is long range, and for this
reason large, in the sense that it cannot be taken into account by a
low-order expansion.  When physical properties are considered at an
energy scale much smaller than the Fermi energy, it is however known
(and well understood) that, for bulk systems, this interaction is
renormalised because of screening into a much weaker effective
interaction $\Vsc(\bfr-\bfr')$.  Approximations for $\Vsc(\bfr-\bfr')$ can
be obtained using for instance the random phase approximation (RPA)
\cite{Fetter&Walecka} giving in the zero-frequency low-momentum limit
\begin{equation} \label{eq:Vsc}
  \Vsc(\bfr) =  \int \frac{d\bq}{(2\pi)^2} \ \hV (\bq) \ 
     \exp[i\bq\!\cdot\!\bfr] \ ,
\end{equation}
\begin{eqnarray}
    \hV (\bq)   & = & \frac {2 \pi e^2 }  {|\bq| +
      \kappa_{(2)}}  
    \qquad (d=2) \label{eq:Vq2d} \; , \\
    \hV (\bq)   & = & \frac {4\pi e^2 }  {\bq^2 +
      \kappa_{(3)}^2}
    \qquad (d=3)  \; , \label{eq:Vq3d} 
\end{eqnarray}
with $\kappa_{(2)} = (2\pi e^2) (\mathrm{g_s} \nu_0^{(2)})$ and $\kappa^2_{(3)} =
(4\pi e^2) (\mathrm{g_s} \nu_0^{(3)})$ the screening wave vectors
\footnote{The equations (\ref{eq:Vq2d}) and (\ref{eq:Vq3d}) are actually the
  expression of the screened interaction in the Thomas-Fermi
  approximation.  To avoid confusion with the Thomas-Fermi
  approximation for the mean field potential, I will nevertheless
  refer to them, although slightly improperly, as the RPA-screened
  interaction.}.  One way to understand the screening mechanism is to
view it in the spirit of the renormalization-group approach, where the
effective Coulomb interaction that should be used for low-energy
processes is produced by the  integration of the ``fast modes'' (high
energy degrees of freedom) of the electron gas \cite{ShankarRMP94}.

For finite systems, the situation is slightly more complicated
because the renormalization process which transforms the bare
interaction (\ref{eq:bare}) into the screened one (\ref{eq:Vq2d}) or
(\ref{eq:Vq3d}) also produces a mean field potential $\Umf(\bfr)$ which
modifies the one particle part of the electrons Hamiltonian.
Since both processes (screening and creation of the mean field
potential) take place at the same time, their interplay  is a priori
not completely obvious.

In the semiclassical limit, and more precisely whenever the screening
length $\kappa^{-1}$ is much smaller than the typical size $L$ of the
system, the common wisdom -- that I shall follow here whenever
necessary -- is however simply to state that since the characteristic
scales of variation of $\Vsc$ and of $\Umf$ are parametrically
different (the former $\kappa^{-1}$ is a quantum scale, when the
latter $L$ is classical), one could nevertheless use the same screened
interaction as for the bulk, and furthermore assume that $\Umf(\bfr)$
is correctly approximated by a Thomas-Fermi approximation.  This
latter amounts to minimise, with a fixed number of particles, the
density functional
\begin{equation} \label{eq:FTF}
 \FTF[n] = \TTF[n] + \Eext[n] + \Ecoul[n] \; , 
\end{equation}
where
\begin{eqnarray*}
	\Ecoul[n] & = & 
	\frac{e^2}{2}\int d\bfr d\bfr' \frac{n(\bfr)n(\bfr')}{|\bfr-\bfr'|}
        \\
        \Eext[n]  & = & \int d\bfr \, \Vext(\bfr) n(\bfr) 
\end{eqnarray*}
($\Vext(\bfr)$ is the external confining potential), and the kinetic
energy term, originating from the Pauli exclusion principle, is given
by
\begin{eqnarray}
  \TTF[n] & \DEF & \int d\bfr \, \tTF(n(\bfr)) \nonumber \\
  \tTF(n) & \DEF & \int_0^n  dn' \, \rme(n')   \label{eq:TTF} \; ,
\end{eqnarray}
with $\rme(n)$  the inverse of the function $n(\rme)$ defined as
\begin{equation} \label{eq:nu}
   n(\rme) \DEF \mathrm{g_s} \int \frac{d\bp}{(2\pi \hbar)^d} \,
   \Theta(\rme-\bp^2/2m_e) \; .
\end{equation}
 In particular $\rme(n)= (\hbar^2/2m_e) (4\pi n/\mathrm{g_s})^2$
for $d=2$, et $\rme(n)= (\hbar^2/2m_e) (6\pi^2 n/\mathrm{g_s})^{2/3}$
for $d=3$.

The self-consistent equations obtained by minimising the Thomas-Fermi
functional then read
\begin{eqnarray} 
  \Umf(\bfr) &  = & \Vext(\bfr) + \int d\bfr' n(\bfr') \Vcoul(\bfr,\bfr') 
  \label{eq:TFSC1}\\
  n(\bfr) & = &\int \frac{d\bp}{(2\pi\hbar)^d}
  \Theta(\mu-\Umf(\bfr)-\bp^2/2m_e) \; . \label{eq:TFSC2}
\end{eqnarray}

Note however, currently there is no general microscopic derivation of
the above picture.  More precisely, our confidence in having the
Thomas-Fermi approximation as a correct starting point for the
computation of $\Umf$ is due to the fact that this approximation can
be derived in a quite general framework starting from a density functional
description (in e.g. the local density approximation) and neglecting
the effect of interferences \cite{Ullmo01prba,Ullmo04prb}.  The
``common wisdom'' prescription given above therefore essentially
amounts to trusting the density functional approach on the classical
scale $L$ (although it might be less reliable on the quantum scale
$\lF$ ; cf.\ for instance the discussion in \citeasnoun{Ullmo04prb}),
keeping the usual (bulk) form of the screened interaction on the
quantum scale, and assuming that the two scales are not going to
interfere in any significant way.  That there is no microscopic
derivation of this ``common wisdom'' prescription is presumably not
too much of an issue as far as qualitative or statistical
descriptions are concerned, but might become a limitation when accurate
simulation tools are required to describe quantitatively the
particular behaviour of a specific mesoscopic system.

In this respect, one should note that there is a class of
systems (namely billiards with weak disorder) for which it is possible
to perform a renormalization procedure
\cite{Blanter97,Aleiner02PhysRep} where the fast modes are integrated
out so that only the interesting low-energy physics remains.  It is
then possible to see how both the mean field and the screened
interaction emerge from this procedure.  The generalisation to a more
general case, for which $\Umf(\bfr)$ is not well approximated by a
constant, is however not completely straightforward, and is still an
open problem.  A relatively extensive discussion of this question is
given in \ref{sec:appendix}.

\setcounter{footnote}{1}


\section{Orbital magnetism: general formalism}
\label{sec:OrbMag1}

As a first illustration of physical systems where the interrelation
between interference effects and interactions plays a fundamental
role, we shall consider in this section the orbital part -- by
opposition to the Zeeman part, associated with the coupling of the
magnetic field to the spin degrees of freedom -- of the magnetic
response at finite temperature $\kt = \beta^{-1}$ of mesoscopic
objects in the ballistic or diffusive regime.

Since the Bohr-van Leeuwen theorem \cite{Leeuwen21}, it is known that
the magnetic response of a system of classical charged particles is
exactly zero. This is a simple consequence of the fact that when
writing the classical partition function $ Z_{\rm cl} = \int d\bp
d\bfr \exp\left( -\beta H_{\rm cl} (\bfr,\bp)\right) $ with $H_{\rm
  cl} (\bfr,\bp) = (\bp -e \bi{A}(\bfr))^2/2m + U(\bfr)$, the vector
potential $\bi{A}(\bfr)$, and thus any dependence in the magnetic
field, can be eliminated by a change of the origin of $\bp$ in the
integral over momentum. The same holds true for the Weyl density of
states since, up to an irrelevant multiplicative constant, it is
derived from $Z_{\rm cl}$ by an inverse Laplace transform.  As a
consequence, whatever is measured has to be related to quantum effect
\footnote{For instance, as discussed in \citeasnoun{RichterPhysRep96},
  the Landau diamagnetism can be understood as originating from
  quantum corrections to the Weyl term in the smooth density of
  states.}, and, in the case of bounded fully coherent systems as is
disordered systems, more specifically interference effects.

In the early nineties, progresses in the design and probing of micron
scale electronic systems, such as small metallic grains or quantum dots
patterned in semiconductor hetero-structures, made is technically
feasible to measure the magnetic response of coherent electronic
structures.  As the orbital magnetism is a particularly well adapted
probe of interference effects in these coherent structures, this
motivated a series of experimental work for (disordered/diffusive)
metallic grain \cite{Levy90,Chandrasekhar91} as well as, slightly later,
ballistic quantum dots \cite{Levy93,Mailly93}.

All of these experiments were able to give convincing evidences -- in
particular the magnetic field scale -- that the measured magnetic
response was indeed related to quantum interferences.  Moreover it was
relatively soon realized that although the magnetic response of a
single ring or dot can be dominated by terms for which the
interactions are irrelevant, the dominant non-interacting contribution
to the magnetic response was varying very rapidly with the size or
chemical potential of the system.  As a consequence, after averaging,
the mean response of an ensemble of micro-structures is, most probably,
driven by the contribution of the interactions.  In other words, the
magnetic response of ensembles of coherent electronic micro-structures
is due to the interplay between interference effects and interactions.

I stress however that ``most probably'' is the best that could be said
here.  Indeed, after the first series of experiments
\cite{Levy90,Chandrasekhar91,Levy93,Mailly93}, which has sparked a
host of theoretical works \footnote{See for instance
  \cite{Bouchiat89,Ambegaokar90,Ambegaokar90epl,VonOppen91,Eckern91,Schmid91,Oh91,Argaman93,VonOppen93,Gefen94,VonOppen94,Ullmo95,Montambaux96,RichterPhysRep96,RichterPRB96,Ullmo97,Ullmo98,VonOppen00}.},
enough puzzles remain to indicate that a full understanding of the
experimental data is lacking.  I shall come back in more detail to
this point at the end of section~\ref{sec:OrbMag2}.  It should be born
in mind however that what follows should not be understood as the
final ``theory'' of orbital magnetism in mesoscopic systems, but
merely as the predictions that can be obtained for the equilibrium
properties within a perturbative / Fermi-liquid description.

In practice, I shall therefore consider a model of particles confined
by some potential $U(\bfr)$, which is already assumed to contain the
smooth part of the Coulomb interaction within some self-consistent
scheme, interacting through the (residual) screened interaction
(\ref{eq:Vsc}), and subject either to a uniform magnetic field
$\bi{B}$, or, in the case of a ring, to a flux line $\Phi$.  Spin will be
included only as a degeneracy factor (i.e. Zeeman coupling will not be
considered).  I shall furthermore limit the discussion to
two-dimensional systems, but no drastically new effect is expected for
$d=3$.

Within the grand canonical formalism, our goal will be to compute
perturbatively in the interactions the field dependent part of the
grand potential 
\begin{equation}
\Omega  =  - \frac{1}{\beta}\ln Z_{\rm G.C} \label{eq:Omega}
\end{equation}
with $Z_{\rm G.C} = {\rm Tr} \exp (-\beta (\hat H - \mu \hat N))$ the
grand canonical partition function, and from there, for instance in
the uniform field case, the magnetisation
\begin{equation} \label{eq:Mz}
 \langle \hat M_z \rangle = - \frac{\partial \Omega}{\partial B_z} 
\end{equation}
(for completeness, equation (\ref{eq:Mz}), as well as its analog
(\ref{eq:Iorb}) for
persistent current, is re-derived briefly in
\ref{sec:AppB}) or the susceptibility
\begin{equation} \label{eq:susc}
        \chi = \frac{1}{\Volume}
        \left(\frac{\partial \langle \hat M_z \rangle}{\partial B}
        \right)_{T,\mu} 
        \; .
\end{equation}
($\Volume$ is the area of the micro-structure).

In the bulk, and more precisely when the cyclotron radius is larger
than the coherence length $L_\phi$ or the thermal length $L_T$ (to
be defined more precisely below), the magnetic response is given by
the (diamagnetic) Landau susceptibility $\cl = - \mathrm{g_s} e^2/24 \pi m_e$.
This latter originates from higher order in $\hbar$ corrections to the
Weyl (smooth) density of states \cite{Kubo64,Prado94} (see
also the discussion in section 3 of \citeasnoun{RichterPhysRep96}),
and I will use it below as the reference scale for the susceptibility.

\subsection{First order perturbations}
\label{subsec:first_order}

As mentioned in section \ref{sec:screening}, the effective strength of
the screened interaction is related to the parameter $r_s$
characterising the density of the electron gas.  In most
experimentally relevant cases, $r_s$ is of order one and the
high-density expansion is just a convenient way to order the various
contributions, but some re-summation of series of higher order diagrams
is necessary in order to get an accurate result. On the other hand,
it is interesting and pedagogical to start with the genuine
high-density asymptotics of small $r_s$. Then, provided the
momenta involved are of the order of the Fermi momentum $p_F$ (which
will be the case, except for the notable exception of {\em periodic}
orbits, see discussion below) $\hV(q)$ will be of order $r_s / \nu_0$
and the diagrammatic development of the thermodynamic potential is
indeed a development in $r_s$. In that case we are interested in the
first order (or Hartree-Fock) correction in the screened interaction,
which can be evaluated without drawing any Feynman diagram.

\subsubsection{First order perturbations}

Working in the grand-canonical ensemble at temperature $\kb T =
\beta^{-1}$, one can express the first correction to the thermodynamic
potential as a direct (Hartree) plus an exchange (Fock) contribution in
terms of the eigenfunctions $\varphi_u$ and eigenenergies $\energy_u$ of
the non-interacting problem \cite{Fetter&Walecka},
\begin{eqnarray} 
    \dO^{(1)} & = & \mathrm{g_s}^2 \Ha - \mathrm{g_s} \Fo \nonumber \\
    & = &  \frac{1}{2} \sum_{u,v} f_u f_v 
    [\mathrm{g_s}^2 \langle \varphi_u \varphi_v | V | \varphi_u
    \varphi_v \rangle -  
    \mathrm{g_s} \langle \varphi_u \varphi_v | V | \varphi_v  \varphi_u \rangle ]
    \; , \label{eq:HFgen1}
\end{eqnarray}
with $f_v = f(\energy_v\!-\!\mu) = [1 +\exp[\beta(\energy_v\!-\!\mu)]]^{-1}$ the
Fermi occupation factor. For sake of clarity, the spin degeneracy
factor $\mathrm{g_s}=2$ is made explicit.

Introducing
\begin{eqnarray} 
    n(\bfr,\bfr') & \equiv & \sum_v f_v \langle \bfr'|\varphi_v \rangle
	                          \langle \varphi_v | \bfr \rangle
                                  \nonumber \\
    & = & - \frac {1}{2 i \pi} \int d\energy f(\energy-\mu)
    [G^R(\bfr,\bfr';\energy) - G^A(\bfr,\bfr';\energy)] \; , 
\label{eq:nrr}
\end{eqnarray}
($n(\bfr) \equiv   n(\bfr,\bfr)$  is the local electron density  in the
non-interacting problem), one can re-express the direct and indirect
contributions as
\numparts
\begin{eqnarray} \label{eq:HFgen1a}
    \Ha & = &  \frac{1}{2} \int d\bfr\, d\bfr'\, n(\bfr) V(\bfr -
    \bfr') n(\bfr')  
		\label{eq:HFgen2} 
\end{eqnarray}
\begin{eqnarray} \label{eq:HFgen1b}
    \Fo & = &  \frac{1}{2} \int d\bfr\, d\bfr'\, n(\bfr,\bfr') V(\bfr -
    \bfr') n(\bfr',\bfr) 
\end{eqnarray} 
\endnumparts
As discussed in Sec.~\ref{sec:semiclassics}, $G^R(\bfr,\bfr';\energy)$ can be
semiclassically approximated as a sum over classical trajectories
travelling from $\bfr'$ to $\bfr$ at energy $\energy$.  The advanced Green's
function can be written in terms of  the retarded one as
\begin{equation} \label{eq:Ga}
    G^A(\bfr,\bfr';\energy) = [G^R(\bfr',\bfr;\energy^*)]^*
\end{equation}
and can therefore be interpreted as a sum running over all the
trajectories going backward in time from $\bfr$ to $\bfr'$.
	
Using the equation (\ref{eq:dS/dE}) which relates the variation in
energy of the action with the time of travel $\orbittime_j$ of the
trajectory $j$, we understand the integral in (\ref{eq:nrr}) as the
convolution between a function oscillating with a period
$2\pi\hbar/\orbittime_j$ and the Fermi function which varies from one to
zero on a scale $\beta^{-1} = \kt$.  Introducing the characteristic
 time (or length)  associated with temperature
\begin{equation} \label{eq:tT}
t_T= \frac{L_T}{\vf} = \frac{\hbar\beta}{\pi} \ , 
\end{equation}
($\vf$ is the Fermi velocity) we see that the contribution of a
trajectory $j$ will be exponentially damped as soon as $t_j \gg t_T$.
More precisely (see for instance the appendix~A of
\citeasnoun{RichterPhysRep96}), if $G^R_j$ is the contribution of the
trajectory $j$ to the sum (\ref{eq:1/2classGreen}), one has
\begin{equation} \label{eq:Tsmoothing}
    \int d\energy \, f(\energy-\mu) G^R_j(\bfr,\bfr')
    = \left( -\frac{i \hbar}{t_j} R(t_j/t_T) G^R_j(\bfr,\bfr') \right)
    \; ,  
\end{equation}
with $R(x) \df x/\sinh(x)$.  To proceed in the evaluation of the
equations (\ref{eq:HFgen1a})-(\ref{eq:HFgen1b}), let us introduce the
coordinates $\bar 
\bfr = (\bfr\!+\!\bfr')/2$ and $\delta \bfr = (\bfr\!-\!\bfr')$.
Since the interaction between electrons is taken to be short ranged,
one can assume the relevant $\delta\bfr$ to be small and, using
$(\partial S_j /\partial\bfr')= \bp_j^f$, $(\partial
S_j/\partial\bfr)= -\bp_j^i$, with $\bp_j^i, \bp_j^f$ the initial and
final momentum of the trajectory $j$, one can approximate
\begin{equation}
    G^R_j(\bar \bfr \pm \delta \bfr/2,\bar \bfr \pm \delta \bfr'/2) =
    G^R_j(\bar \bfr ,\bar \bfr ) \exp \left[\frac{i}{2\hbar}
     \left(\pm \bp_j^f\!\cdot\!\delta \bfr'  \mp
       \bp_j^i\!\cdot\!\delta \bfr
    \right) \right] \; .
\end{equation}
The integral over $\delta \bfr$ therefore yields the Fourier transform
of $\Vsc(\bfr-\bfr')$ and, neglecting the terms $G^AG^A$ and $G^RG^R$
which will eventually average to zero, one can semiclassically express
the first order correction to the thermodynamic potential a sum over
all pairs $(k,l)$ of {\em closed} orbits
\numparts 
    \begin{eqnarray}  \fl
    \Ha & = & \frac{1}{(2\pi)^3 \hbar} \int d \bfr \sum_{kl} Q_{k} Q_l
    \cos \left( \psi_{k} - \psi_{l} \right)
    \hV \left( (\bp_k^f\!-\!\bp_k^i +
      \bp_l^f\!-\!\bp_l^i) /{2\hbar}\right) \; , \label{eq:HFgen3a} \\
\fl    \Fo & = & \frac{1}{(2\pi)^3 \hbar} \int d \bfr \sum_{kl}Q_{k} Q_l
    \cos \left( \psi_{k} - \psi_{l} \right)
     \hV \left( (\bp_k^f\!-\!\bp_k^i +
      \bp_l^f\!-\!\bp_l^i) /{2\hbar}\right) \; ,  \label{eq:HFgen3b} 	
  \end{eqnarray}
\endnumparts
where $ Q_{j} = {R(t_j/t_T) D_j}/{t_j}$ and $\psi_{j} = (S_k
/\hbar \!  - \! \maslov_j {\pi}/{2}) $.   The field dependence of
the above expression can then be obtained from (cf.\ \ref{eq:delta_S(B)})
\begin{equation} \label{eq:dS/dB}
   \frac{\partial S_j }{\partial B} = 2\pi \area_j/\phi_0  \; ,
\end{equation}
with $\area_j$ the area enclosed by the orbit $j$. 

For a generic pair of trajectories $(k,l)$ the term $\cos[(S_k \pm
S_l)/\hbar]$ will be a highly oscillating function of the coordinate
$\bfr$. Performing the integration over position, the stationary phase
condition reads $(\bp_k^f\!-\!\bp_k^i) \pm (\bp_l^f\!-\!\bp_l^i) = 0$,
and unless $k$ and $l$ are related by a symmetry, this will correspond
to isolated points, each of which yields a contribution $\hbar^{1/2}$
smaller than the original prefactor.  On the other hand, special
pairings where $S_k=S_l$ will kill the oscillating phase. Such a
condition is trivially satisfied when $k=l$ but this also kills any
field dependence.  A second possibility is to pair a given trajectory
with its time reversed.  This is a nontrivial pairing since the
resulting term is field dependent.  Keeping only these contributions
the direct and exchange terms can be written as
\numparts 
  \begin{eqnarray} \label{eq:HFgen4a} \fl
  \Ha_{(D)} & = &  \frac{1}{(2 \pi)^3 \hbar} \int d\bfr 
   \sum_j \left( \frac{\hbar R(t_j/t_T) }{t_j} \right)^2
   |D_j|^2 \cos \left( \frac{4\pi \area_j B}{\phi_0} \right)
   \hV \left( \frac{\bp'_j - \bp_j}{\hbar} \right) \ , 
   \end{eqnarray}
   \begin{eqnarray} \label{eq:HFgen4b} \fl
  \Fo_{(D)} & = &  \frac{1}{(2 \pi)^3 \hbar} \int d\bfr 
   \sum_j \left( \frac{\hbar R(t_j/t_T) }{t_j} \right)^2
   |D_j|^2 \cos \left( \frac{4\pi \area_j B}{\phi_0} \right)
   \hV \left( \frac{\bp'_j + \bp_j}{\hbar} \right ) \; ,
   \end{eqnarray}
\endnumparts
where the sub-index $(D)$ indicates the {\em diagonal} approximation,
with sums running over {\em individual} trajectories $j$ (and not
pairs as in (\ref{eq:HFgen3a})- (\ref{eq:HFgen3b})). A third possible
pairing appears when we can match the actions of $k$ and $l$, even if
the trajectories are not the same or time-reversed of each other. Such
situation arises in integrable systems with families of trajectories
degenerate in action, and is discussed in details in
\citeasnoun{Ullmo98}.

\subsubsection{The high density limit}

The sums in equations (\ref{eq:HFgen4a})-(\ref{eq:HFgen4b}) run over
all {\em closed} 
(not necessarily periodic) trajectories [more precisely, we have twice
the sum over all time-reversed pairs].  Therefore a priori, and
in contrast to the non-interacting theory, periodic orbits do not
play any particular role.  It is interesting to note however that if
the high density limit is to be taken seriously (i.e.~$r_s \to 0$),
then again periodic orbits are singled out.  Indeed, in this case
${\hV}\left((\bp'_j\!-\!\bp_j) / \hbar \right)$ is of order
$r_s\nu_0^{-1}$ except when $\bp'_j - \bp_j = 0$, that is when the
orbit is periodic, in which case $\hV (0) = \nu_0^{-1}$. Note that
$\bp'_j+\bp_j=0$ implies that the trajectory $j$ is self retracing,
and thus has a zero enclosed area. As a consequence for the exchange
term all contributions to the magnetic response are of order $r_s$.

Therefore in the high density limit, the integrand in
(\ref{eq:HFgen4a}) is significantly larger in the neighbourhood of
periodic orbits.  For chaotic systems this will be compensated by the
fact that these orbits are isolated.  As a consequence the relative
weight of their neighbourhood may depend on the particular system under
consideration.  For integrable systems however, for which periodic
orbits come in families whose projection on the configuration space
has a non-zero measure, the magnetic response induced by
electron-electron interactions will be dominated by periodic orbits
when $r_s \ll 1$, and will reach a finite limit as $r_s \to 0$.

\subsection{Correlations effects}
\label{subsec:correlations}

As discussed at the beginning of this chapter, realistic values of
the parameter $r_s$ (appropriate for metals or GaAs/AlGaAs
hetero-structures) force us to consider high-order effects in the
diagrammatic expansion of the thermodynamic potential. Therefore,
correlation effects are important, and need to be taken into account.

A value of $r_s \simeq 1$ means that the range of the screened
potential is of the order of the Fermi wavelength.  In other words,
the screened interaction has a local character, and can be written as
\begin{equation} \label{eq:localV} 
   V(\bfr - \bfr') = \frac{\lambda_0 }{ \mathrm{g_s} \nu_0 } \delta (\bfr -\bfr') \, 
\end{equation}
where $ \mathrm{g_s} \nu_0$  is the total density  of states  (i.e. including
the  spin degeneracy factor $\mathrm{g_s}$) and $\lambda_0$ is a constant of order one
that will also serve   to label the order of perturbation.  

\begin{figure} 
\begin{center}
\includegraphics[totalheight=4cm]{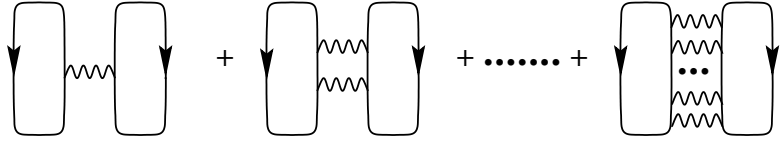}
\end{center}
\caption{Direct terms of the Cooper series for the perturbative
  expansion of the thermodynamic potential $\Omega$.}
\label{fig:cooper}
\end{figure}

The perturbative expansion of the thermodynamic potential can be
represented in the usual way (cf. for instance section 15 of
\citeasnoun{AGD}) by Feynman diagrams, with straight lines standing for the
finite temperature (Matsubara) Green's functions and the wavy lines
for the interaction $V$.  In the same way as in the theory of
superconductivity, it turns out that it is essential to consider the
full Cooper series which associated diagrams are represented in
figure~\ref{fig:cooper} \cite{AGD,Aslamazov75}.  One way to see this
is to realize that, since $r_s$ is actually not a small parameter,
what is done here is more a semiclassical expansion (i.e.\ in powers
of $\hbar$) than one in the strength of the interaction. Therefore one
should  perform a simple counting of the powers of $\hbar$ of such
contributions.  The Cooper diagram of order $k$ implies $k$
interaction lines (each of which yields a factor $\nu_0^{-1}$), $n$
pairs of Green's functions, and $(k+1)$ summations of Matsubara
frequencies (each of them associated with a factor $\beta^{-1}$).  As
far as powers of $\hbar$ are concerned, $|G|^2 \sim \nu_0/\hbar$
(whatever the dimension).  Therefore the only delicate point here is
to realize that each temperature factor $\beta^{-1}$ should be
accounted for as an $\hbar$, since in the mesoscopic regime considered
here, the time $t_T = \hbar\beta / \pi$ introduced above should be of
the order of some characteristic time $t_c$ of the system (say the
time of flight), and thus $\beta^{-1} \sim \hbar / t_c \propto \hbar$
(again as far as powers of $\hbar$ are concerned).  The RPA series can
be seen, in the same way, to be of the same order in $\hbar$, but the
corresponding terms turn out to have negligible magnetic field
dependence, and can therefore be omitted from the calculation of the
magnetic response.  Moreover one can convince oneself that all other
diagrams would, at some given order $k$, have either a smaller number
of Green's function or a larger number of frequency summations, and
therefore are of higher order in $\hbar$.  

Noting that, because the interaction (\ref{eq:localV}) is local, the
direct and exchange Cooper diagrams differ only by their sign and by a
spin degeneracy factor, the magnetic response can be derived from the
Cooper series contribution to the thermodynamic potential
\footnote{Note that the diagrammatic rules for $\Omega$ differ
  slightly from the ones for correlation functions.  There is in
  particular a factor $1/k$ associated with each term of order $k$,
  thus the $\log$.}
\begin{eqnarray} \label{eq:cooperon_series}
     \Omega^C & = &\frac{\mathrm{g_s}^2 - \mathrm{g_s} }{ 2 \beta}
     \sum_{k=1}^\infty 
                \frac{\lambda_0^k}{k} \sum_{\omega_m < \Ef}
		\int d\bfr_1 \ldots d\bfr_k \Sigma(\bfr_1, \bfr_k; \omega_m)
		\ldots \Sigma(\bfr_2, \bfr_1 ; \omega_m) 
		\nonumber \\
     & = & \frac{\mathrm{g_s}^2 - \mathrm{g_s} }{2 \beta}
     \sum_{\omega_m < \Ef} {\rm Tr} \left\{  
      \ln [1+ \lambda_0 \Sigma(\bfr, \bfr' ; \omega_m) ] \right\} \;,
      \label{eq:omega^C} 
\end{eqnarray}
where $\omega_m = 2\pi m/ \beta$ are [bosonic] Matsubara frequencies,
\begin{equation}  \label{eq:sigma}
     \Sigma(\bfr, \bfr' ; \omega_m)=\frac{1}{ \beta
     \mathrm{g_s} \nu_0 } \sum_{\epsilon_n < \Ef} 
     {\G}(\bfr,\bfr';\epsilon_n) {\G}(\bfr,\bfr';\omega_m-\epsilon_n)  
\end{equation} 
is the [free] particle-particle propagator, and the finite
range $\sim \lF$ of the interaction introduces a cutoff on the
summation over Matsubara frequencies at the corresponding energy scale
$\Ef$.

The trace over the space coordinates is a short way of expressing the
expansion in all orders in $\lambda_0 \Sigma$. The concept of
particle-particle propagator, as well as the Cooper series contribution,
comes from the Cooper pairs in the theory of superconductivity. The
main difference in our case is that now the interaction is repulsive
(thus the plus sign in the trace) and that we have lost translational
invariance (therefore we cannot trade the operators for ordinary
functions by going to the momentum representation).

\subsubsection{Semiclassical evaluation of the particle-particle propagator}

To proceed further with our semiclassical formalism, it is useful to write 
the finite-temperature Green's function between points $\bi{r}$ and $\bi{r'}$
for a fermionic Matsubara frequency (or rather, energy)
$\epsilon_n\!=\!(2n\!+\!1)\pi/\beta$  
 in terms of the retarded and advanced Green's functions as
\begin{equation} \label{eq:ftGreen}
   \G(\bfr, \bfr' ;\epsilon_n) \!= 
   \Theta (\epsilon_n) G^R(\bfr,\bfr';\Ef \!+\!i\epsilon_n) 
 + \Theta(-\epsilon_n) G^A(\bfr,\bfr';\Ef \!+\!i\epsilon_n) \; ,
\end{equation}
with $\Ef$ the Fermi energy. The retarded and advanced Green's functions
are related through (\ref{eq:Ga}) and expressed, in a
semiclassical approach, as expansions over all trajectories $j$
joining $\bi{r'}$ and ${\bf r}$ at energy $\energy$
(see (\ref{eq:1/2classGreen})). The complex energy-arguments of
(\ref{eq:ftGreen}) force us to perform some analytic continuation.
However, if the Matsubara energies are much smaller than $\Ef$, one
can expand the classical action and use (\ref{eq:dS/dE}) to obtain
\begin{equation} \label{eq:Green_ie}
   G^{R}_{j}(\bfr,\bfr';\Ef\!+\!i\epsilon_n) = 
   G^{R}_{j}(\bfr, \bfr';\Ef) \times\exp\left[ -\frac{\epsilon_n
       t_j}{\hbar}\right]  \; .
\end{equation}
Note that, as in (\ref{eq:tT}) temperature introduces the time
scale $ t_T\!=\!\hbar\beta / \pi$ which exponentially suppresses the
contributions of long paths through the term $\epsilon_n t_j /
\hbar\!=\!(2n\!+\!1) t_j / t_T$.  Therefore, only small Matsubara
frequencies need to be considered, and the assumption used for the
perturbative treatment of the energy is consistent.

To compute the magnetic susceptibility at $B\!=\!0$, the field
dependent part of the semiclassical Green's function can also be treated
perturbatively, and using (\ref{eq:dS/dB}) one can write
\begin{equation} \label{eq:Green_ieB}
  \fl  G^{R}_{j}(\bfr,\bfr';\Ef\!+\!i\epsilon_n;B) = 
 G^{R}_{j}(\bfr,\bfr';\Ef;B\!=\!0)
   \times\exp\left[ -\frac{\epsilon_n t_j}{\hbar}\right] \times \exp
   \left[ i 2 \pi \frac{B \area_j}{\phi_0} \right] \; ,
\end{equation}
where $\area_j$ is the effective area enclosed by the orbit (circulation
of the vector potential between $\bi{r}$ and $\bi{r'}$) and $\phi_0$ the 
flux quantum. The weak-field semiclassical approximation to 
(\ref{eq:ftGreen}) is then given by
\begin{eqnarray} \label{eq:ftsgf}
 \fl  \G(\bfr, \bfr';\epsilon_n,B) \!= \theta (\epsilon_n)
\sum_{j : \bi{r} \to \bi{r'}} \frac{D_j}{\sqrt{-2i\pi\hbar^3}} \,
e^{iS_j/\hbar - i\pi\maslov_j/2} 
   \times\exp\left[ -\frac{\epsilon_n t_j}{\hbar}\right] \times \exp
   \left[ i 2 \pi \frac{B \area_j}{\phi_0} \right] + \nonumber \\
 \fl + \theta(-\epsilon_n) 
\sum_{j' : \bi{r'} \to \bi{r}} \frac{D_{j'}}{\sqrt{-2i\pi\hbar^3}} \,
e^{-iS_{j'}/\hbar + i\pi\maslov_{j'}/2} 
   \times\exp\left[ \frac{\epsilon_n t_{j'}}{\hbar}\right] \times \exp
   \left[ - i 2 \pi \frac{B \area_{j'}}{\phi_0} \right] \; ,
\end{eqnarray}
where trajectories $j$ and $j'$ travel from ${\bfr}$ to ${\bfr'}$ in
opposite directions, at energy $\Ef$, and in the absence of
magnetic field.

Note the usefulness of (\ref{eq:ftsgf}) goes beyond the problem of
orbital magnetism  discussed here, as it provides a calculational
approach to any perturbative problem where the single-particle
classical dynamics is known.

\begin{figure} 
\begin{center}
\includegraphics[totalheight=6cm]{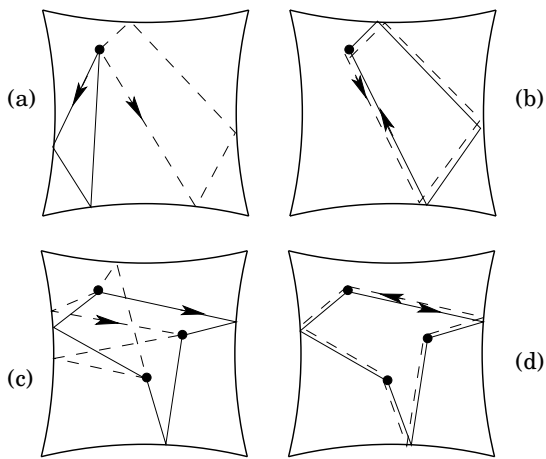}
\end{center}
\caption{ Pairs of orbits contributing to
  $\Omega^C$ (see (\ref{eq:cooperon_series}) for a (non
  integrable)  billiard.  Top row : first order contributions.  Bottom
  row : third order contributions (there are therefore three pairs of
  orbits connected at interaction points in both (c) and (d)).  Left
  column : generic case.  Right column : pairing of time reversed
  trajectories (diagonal contribution), for which the dynamical phases
  cancel. (Courtesy of Harold Baranger.)
}
\label{fig:harold}
\end{figure}

The particle-particle propagator $\Sigma(\bfr,\bfr';\omega_m)$ can now
be evaluated semiclassically from  (\ref{eq:ftsgf}) and
(\ref{eq:sigma}).  In general this involves a sum over all pairs
of classical trajectories joining $\bfr'$ to $\bfr$.  An illustration is
shown in figure~\ref{fig:harold}.  As in
section~\ref{subsec:first_order} however, most of these pairs 
yield highly oscillating contributions which  average to zero when
integrated over position, and one should only consider the
non-oscillating terms which maintain a field dependence.  One way to
do this is, again, to pair time reversed trajectories, which implies
that in the sum over the fermionic Matsubara frequencies in
(\ref{eq:sigma}), only the $\epsilon_n$ such that $\epsilon_n
(\omega_m- \epsilon_n) < 0$ should be kept. This diagonal part
$\Sigma^{(D)}$ of the particle-particle propagator can then be written
as
\begin{eqnarray}    
    \Sigma^{(D)}(\bfr, \bfr';\omega_m) & \simeq & \frac{\kt}{
    \mathrm{g_s} \hbar} \sum_{ j : \bi{r} \to \bi{r'}}
   \frac{|D_j |^2}{m_e} \, \exp \left[ i4\pi \frac{B
       \area_j}{\phi_0}\right] 
   \nonumber \\
   & \times & \sum_{\epsilon_n (\omega_m- \epsilon_n) < 0}^{\epsilon_n<\Ef}
   \exp\left[-\frac{(|\epsilon_n|\!+\!|\omega_m\!-\!\epsilon_n|)
       t_j}{\hbar}\right] \; .   \label{eq:sigma_Dbis}
\end{eqnarray}
Summing over $\epsilon_n$ in the contribution of trajectory $j$, one
gets
\begin{equation}
 \fl  \sum_{\epsilon_n (\omega_m- \epsilon_n) < 0}^{\epsilon_n<\Ef}
   \exp\left[-\frac{(|\epsilon_n|\!+\!|\omega_m\!-\!\epsilon_n|) t_j}{
       \hbar}\right] 
   = \exp\left[-\frac{\omega_m t_j}{\hbar}\right] \frac {R(2 t_j/
     t_T)}{2 t_j/ t_T} 
   \left(1 -
     \exp\left[-\frac{(\Ef\!-\!\omega_m)t}{\hbar}\right]\right) \; , 
\end{equation}
where the function $R$ and the temperature time $t_T$ were introduced
in the discussion of (\ref{eq:Tsmoothing}). The last factor
$(1-\exp[-(\Ef\!-\!\omega_m)t/\hbar])$ originates from the upper bound
$\Ef$ of the Matsubara sum.  If one assumes $\omega_m \! \ll \!\Ef$,
this factor  removes from $\Sigma^{(D)}(\bfr, \bfr')$ all the
contributions of trajectories of length smaller than $\Lambda_0 =
\lambdaf / \pi$, thus preventing the particle-particle propagator from
diverging as $\bfr \to \bfr'$.  Replacing it by a hard
cutoff at $\Lambda_0$ one obtains
\begin{equation}  \label{eq:sigma_D}
 \fl  \Sigma^{(D)}(\bfr, \bfr';\omega_m)\simeq \frac{\kt}{
    \mathrm{g_s} \hbar } \sum_{\fraczero{ j : \bi{r} \to \bi{r'}
    }{L_j > \Lambda_0}} 
   \frac{|D_j |^2}{m_e}\, \frac{R(2 t_j/ t_T) }{2 t_j / t_T} 
   \times\exp \left[ i4\pi \frac{ B \area_j}{\phi_0}\right] 
   \exp[-\frac{\omega_m t_j}{\hbar}]\; .  
\end{equation}

The semiclassical form for $\Sigma^{(D)}(\bfr, \bfr';\omega_m)$ shares
with $\Ha_D$ and $\Fo_D$ (equations (\ref{eq:HFgen4a})-(\ref{eq:HFgen4b})) the
property of being a semiclassical expansion which does not
oscillate rapidly (on the scale of $\lambdaf$) as a function of the
coordinates, as would be the case for the Green's functions
(\ref{eq:ftsgf}).

\setcounter{footnote}{1}

\section{Orbital magnetism: Diffusive and ballistic systems}
\label{sec:OrbMag2}

\subsection{Diffusive Systems}
\label{subsec:diffusive}

The semiclassical approach described above does not rely on any
assumption concerning the character of the underlying classical
dynamics. It is therefore applicable to (integrable or chaotic)
ballistic structures \cite{Ullmo98,VonOppen00} as well as to diffusive systems
\cite{Montambaux96,Ullmo97}.
Because diffusive motion is in some sense relatively
simple, it is natural to consider first the orbital magnetism of
interacting systems whose non--interacting classical dynamics is
diffusive.  More specifically, I will discuss the interaction
contributions to the persistent current of metal rings and to the
susceptibility of singly--connected two--dimensional diffusive
systems. We shall see that in this case, the semiclassical approach
recovers, in a very transparent and intuitive way, results
previously obtained by quantum diagrammatic calculations
\cite{Ambegaokar90,Eckern91,Aslamazov75,Altshuler83,Oh91}.  Applied to
diffusive dynamics, the semiclassical approach is indeed on the same
level of approximation. Moreover, by making the connection with the
classical dynamics, it provides a physically intuitive picture of the
interplay between disorder and interactions.

We assume here that the Fermi wavelength $\lambda_F$ is the shortest
length scale, in particular $\lambda_F < \ell$ with $\ell$ the elastic mean
free path, and that the magnetic field is classically weak, i.e., that
the cyclotron radius at the Fermi energy is such that $R_c\gg \ell$. Then
the paths entering into (\ref{eq:sigma_D}) can be approximated by
those of the system at zero field, but include the presence of the
disorder potential.

For diffusive systems it proves convenient to relate $\Sigma^{(D)}$ to
the (classical) conditional probability $P^\energy_{\rm
  cl}({\bfr,\bfr'};t|A)$ to propagate from $\bfr'$ to $\bfr$ in a time
$t$ and enclosing an area $A$ since this probability satisfies a simple
diffusion equation. For that purpose let us introduce an additional
time and area integration $1 = \int dt \delta(t-t_j)\int d\Area
\delta(\Area-\area_j)$ in (\ref{eq:sigma_D}) in order to make use of
the sum rule (\ref{eq:Mformula}), which, for a two dimensional kinetic
plus potential Hamiltonian, and including the constraint on the area,
reads
\begin{equation} \label{eq:Mformula2d}
\sum_{j:\bfr' \to \bfr}\frac{ |D_j(\energy)|^2}{m_e} \delta(t -
\orbittime_j)\delta(\Area - \area_j) 
= 2\pi P^\energy_{\rm cl}(\bfr,\bfr',t | A) \; .
\end{equation}
One therefore obtains
\begin{equation}  \label{eq:sigma_Ddiff}
\fl   \Sigma^{(D)}(\bfr, \bfr';\omega_m)\simeq
\frac{2\pi}{\mathrm{g_s}} \frac{\kt}{ 
    \hbar } \int d A\int_{t>\Lambda_0/\vf} dt \, P^\energy_{\rm
    cl}(\bfr,\bfr',t|A)  
    \frac{R(2 t/ t_T) }{2 t / t_T} 
   \times\exp \left[ i4\pi \frac{ B A}{\phi_0}\right] 
   \exp[-\frac{\omega_m t}{\hbar}]\; .  
\end{equation}

In the same way the $n$-th order
(diagonal) contribution to the thermodynamic potential in
(\ref{eq:omega^C}) can then be expressed through the joint return
probability $P(\bfr_1,\bfr_n,\ldots,\bfr_1;t_n,\ldots,t_1|A)$ to visit
the $n$ points $\bfr_i$ under the conditions that $t_i$ is the time
elapsed during propagation from $\bfr_i$ to $\bfr_{i+1}$, and that the
total enclosed area is $A$. For diffusive motion the probability is
multiplicative, namely
\begin{equation} \label{Pmulti}
  \int d \bfr_1 \ldots
   d \bfr_n P(\bfr_1,\bfr_n,\ldots,\bfr_1;t_n, \ldots, t_1|A) = \int d \bfr
  P(\bfr,\bfr; t_{\rm tot}|A)
\end{equation}
with $t_{\rm tot} = \sum t_i$. Upon inserting the sum rule
(\ref{eq:Mformula2d}) and the relation (\ref{Pmulti}) into
(\ref{eq:sigma_D}), the contribution from the diagonal terms
$\Sigma^{(D)}$ to $\Omega$ (see (\ref{eq:omega^C}))  yields
\begin{eqnarray} \label{omega_sc}
 \Omega^{(D)} & =&  \sum_n \Omega^{(D)}_n \nonumber \\
             & = & 
     \frac{1}{\beta} \int d \bfr \int dt \coth\left(\frac{
     t}{t_T}\right) \, K(t) \, {\ssA}(\bfr,t;B) \; .
\end{eqnarray}
The factor $\coth(t/t_T)$ (with $t_T$ defined by (\ref{eq:tT})) arises
from the $\omega$-sum in (\ref{eq:omega^C}) which here can be performed
explicitly. In (\ref{omega_sc}) the functions $K$ and ${\ssA}$ are
defined as
\begin{eqnarray}
\fl K(t) & \equiv & \sum_n K_n(t) \quad ; \quad
 K_n(t) \equiv \frac{(-\lambda_0)^n}{n} \,
   \left\{ \int  \prod_{i=1}^n \left[ \frac{dt_i R(2 t_i/ t_T) }{\mathrm{g_s} t_i}\right]
   \, \delta(t-t_{tot}) \right\} \; , 
   \label{KAa} \\
\fl {\ssA}(\bfr,t;B) & \equiv & \int d A \cos\left(\frac{4\pi B A}{\phi_0}\right)
   P(\bfr,\bfr;t|A)  \, .
   \label{KAb}
\end{eqnarray}
$K(t)$ accounts for temperature effects while ${\ssA}(\bfr,t;B)$ 
contains the field dependence and the classical return probability.

\subsubsection{Renormalization scheme for Diffusive Systems}

\begin{figure} 
\begin{center}
\includegraphics[totalheight=15cm]{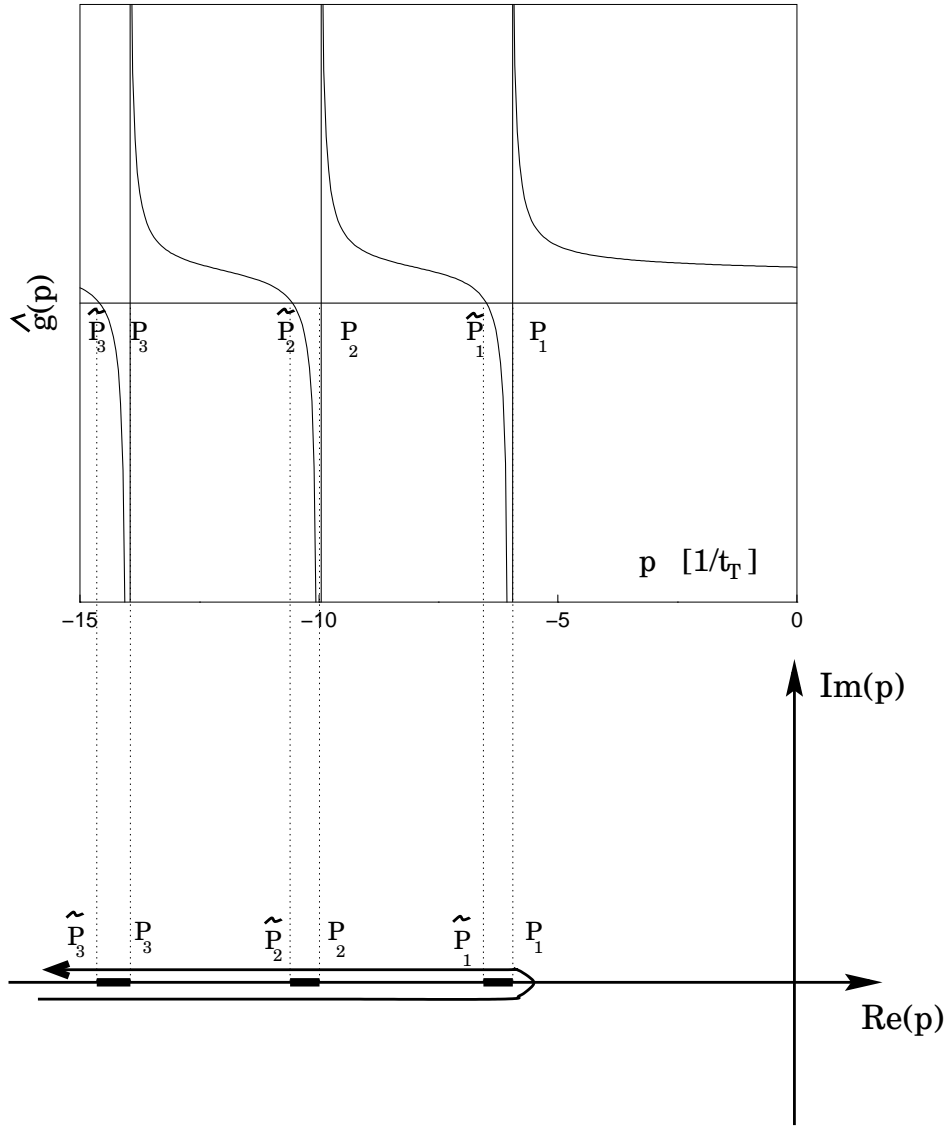}
\end{center}
\caption{
Top: graph of the function $\hat g(p)$  (for $n_f=200$).
Bottom: integration path for the inverse Laplace transform
(\ref{eq:K_t_bis}) in the complex $p$ plan.}
\label{fig:contour}
\end{figure}

This semiclassical approach allows us further to obtain the
renormalization of the coupling
constant~\cite{Altshuler83,AltshulerAronov85,Eckern91,Ullmo97} for
diffusive systems by resuming the higher-order diagrams of the Cooper
series. To this end let us first introduce the Laplace transform of
$K_1(t)$,
\begin{equation} \label{hatf} 
  \hat f(p)   = \frac{4 \lambda_0}{\mathrm{g_s}} \sum_{n=0}^{n_F} 
  \frac{1}{p t_T +2(2n+1)} \; , 
\end{equation}
where 
\begin{equation}
n_F = \frac{\beta \Ef}{ 2 \pi } = \frac{\kf L_T}{4} \; .
\end{equation}
The full kernel $K(t)$ is given by the inverse Laplace transform
\begin{equation} \label{renorm}
   K(t) = \frac{1}{2\pi i} \
  \int_{-i \infty}^{+i \infty} \!\! dp \, e^{+pt} \, \ln [1 + \hat f (p)] \; .
\end{equation}
To evaluate the above integral, let us  define
\begin{equation}
  \hat g (p)  \equiv  1 + \hat f (p)  \; ,
\end{equation}
and furthermore denote the singularities of $\hat g (p)$ by
\begin{equation}
  p_n = -  \frac{2(2n+1)}{t_T}
\end{equation}
with $n = 0,\ldots ,n_F$. Let $\tilde p_n$ be the corresponding zeros
($\tilde p_n$ is assumed to lie between $p_n$ and $p_{n+1}$).  On the
real axis, $\hat g$ is a real function which is negative within each
interval $[\tilde p_n, p_n]$ (with the notation $\tilde p_{n_F} = -
\infty$) and positive elsewhere.  As a consequence, $\ln \hat g (p)$
is analytic everywhere in the complex plane except for the branch cut
$[\tilde p_n, p_n]$.  The phase jump across the branch cuts is $2\pi$,
since the imaginary part of $\hat g(p)$ is positive above and negative
below the real axis.  Deforming the contour of integration as sketched
in figure~\ref{fig:contour} one therefore finds
	\begin{eqnarray}
	K(t) & = & \lim_{\epsilon \to 0}  \int_{-\infty}^0 \frac {dp}{2i\pi} 
            \left( 
	    \ln [ \hat g (p-i \epsilon)] - \ln [ \hat g (p+i \epsilon) ]
	    \right)  e^{pt}\\
	     & = & \sum_{n=0}^{n_F} \int_{\tilde p_n}^{p_n} dp e^{pt}
             \label{eq:K_t_bis} \\
	     & = &  \frac{1}{t} \sum_{n=0}^{n_F} 
	          \left [  e^{p_nt} - e^{\tilde p_n t} \right] \; .
		  \label{eq:app:sum}
	 \end{eqnarray}
For $n \ll n_F$ one has $\delta_n \equiv t_T (p_n - \tilde p_n) \ll 1$ and 
thus to first order in $\delta_n$:
	\begin{equation}
	 1 + \frac{\lambda_0}{\mathrm{g_s}} \sum_{n' \neq n}^{n_F} \frac{1}{n'-n}
	 - \frac{ 4 \lambda_0}{\mathrm{g_s}} \frac{1}{\delta_n} = 0 \; .
	 \end{equation}
The above condition gives
	\begin{equation} \label{eq:delta_n}
	\delta_n = \frac {4}{\mathrm{g_s}/\lambda_0 + \Psi(n_F+1) - \Psi(2n+1)}
	\end{equation}
with $\Psi$ the digamma function.  

In the {\em high temperature regime} $t_T \le t$, all the $n$'s
actually contributing to the sum (\ref{eq:app:sum}) are such that the
denominator in (\ref{eq:delta_n}) is dominated by
$\Psi(n_F+1)\simeq\ln(n_F)$.  One obtains in this case 
	\begin{eqnarray}
	K(t) & = &   \frac{1}{t} \sum_{n=0}^{n_F} 
	           e^{p_n t} \left [ 1- e^{-\delta_n t / t_T} \right] \\
             & \simeq &
		   \frac{1}{t_T}  \frac{4}{\ln(\kf L_T / 4)}
		   \sum_{n=0}^{n_F}  e^{p_n t} \; .
	\end{eqnarray}

In the {\em low temperature regime} $t_T/t \gg 1$, the typical $n$
contributing to (\ref{eq:app:sum}) is $n_0\equiv{t_T/4t}$ (that is
still assumed much smaller than $n_F$).  Because of the slow variation
of the logarithm, one can in this case replace $n$ by $n_0$ in
(\ref{eq:delta_n}). [A formal derivation would consist in replacing
the sum (\ref{eq:app:sum}) by an integral, change to the variable
$\ln(n)$, and use the stationary phase approximation.] This gives
\begin{equation}
  \delta_n \simeq \frac{4}{\ln (n_f) - \ln (2 n_0)} =
  \frac{4}{\ln(2 \kf \vf t)} \; ,
\end{equation}
and in the same way as above
\begin{equation}
  K(t) \simeq \frac{1}{t_T}  \frac{4}{\ln(2 \kf \vf t)}
		   \sum_{n=0}^{n_F}  e^{p_n t} \; .
\end{equation}

Noting that
\begin{equation}
  K_1(t) =  \frac{4 \lambda_0}{\mathrm{g_s} t_T}    \sum_{n=0}^{n_F}  e^{p_n t} 
\end{equation}
one identifies 
\begin{equation} \label{eq:Knorm}
  K(t) \simeq \frac{\mathrm{g_s}}{\lambda_0 \ln (\kf L^*)} K_1(t)
\end{equation}
with 
\begin{equation}
\label{eq:L^ast}
  L^* = \min(2 \vf t, L_T /4) \; .
\end{equation}
This relation is valid when $\ln \kf L^* \gg 1$. It is certainly
satisfied when $\ln \kf l \gg 1$ which may be regarded as a semiclassical
approximation in the diffusive regime.
The equation (\ref{eq:Knorm}) shows that the higher-order terms in $K(t)$
merely lead to a renormalization of the first-order contribution
$K_1(t)$: The coupling constant $\lambda_0=1$, entering into $K_1(t)$,
is renormalised to $\mathrm{g_s}/ \ln (\kf L^*)$.

In first order one has from (\ref{KAa})
        \begin{equation}
    \label{eq:K1}
        K_1(t)=\lambda_0 \frac{R(2t/t_T)}{2t} \, 
        \end{equation}
so that $K$ reduces to
    \begin{equation}
\label{eq:Knorm2}
K(t) \simeq \frac{R(2t/t_T)}{t \ln (\kf L^*)} \; .
    \end{equation}
    Equations (\ref{omega_sc})--(\ref{KAb}) together with
    (\ref{eq:Knorm2}) may serve as a general and convenient starting
    point to compute the orbital response of disordered mesoscopic
    systems. The specific character of the geometry enters into
    $\Omega^{(D)}$ solely through the Fourier transform (\ref{KAb}) of
    the return probability $P(\bfr, \bfr, t|A)$.  In the following we
    shall apply this approach to compute the magnetic response of two
    important types of mesoscopic structures.

\subsubsection{Persistent current of disordered rings}

While the magnetic response of a singly--connected system is usually 
described in terms of its susceptibility,  the response  of a 
ring-type structure threaded by a flux $\phi$ is  related to the
persistent current  
\begin{equation}
\label{eq:Iorb}
I \equiv -\frac{\partial \Omega}{\partial \phi}\; 
\end{equation}
(see Appendix~\ref{sec:AppB} for a derivation of this equation).  To
make contact with previous approaches \cite{Ambegaokar90,Montambaux96}
let us start with the computation of the contribution of the
first-order interaction term, $\Omega^{(D)}_1$, to $I$ which will be
denoted by
$I_1$.

Consider a (thin) disordered ring of width $b$, cross section $\sigma$
and circumference $L$. For $L \gg l,b$ the motion of
particles around the ring effectively follows a law for
one-dimensional diffusion.  The total area enclosed by a path is given
in terms of the number $m$ of windings around the ring. One
thus has
	\begin{equation}
	P(\bfr,\bfr;t|A) = \sum_{m=-\infty}^{+\infty}
	\frac{1 }{\sigma}
	\frac{1}{\sqrt{4\pi Dt}} \exp\left(-\frac{m^2L^2}{4Dt}\right) \,
	\delta \left(A - \frac{m L^2}{4\pi}\right) \; ,
	\end{equation}
where $D=\vf l/d$ is the diffusion constant in $d$ dimensions
and $\vf$ the Fermi velocity. Note that due  
to the disorder average, the classical return probability
does not depend on $\bfr$.  

Combining the expression (\ref{eq:K1}) for $K_1(t)$ with the 
$\coth$ function in (\ref{omega_sc}) one obtains
	\begin{equation} \label{omega1_ring}
 	\Omega^{(D)}_1 =  \frac{\lambda_0}{\mathrm{g_s}} \frac{2 L \hbar }{2\pi} 
 	\sum_{m=-\infty}^{+\infty} 
	\cos\left(\frac{4\pi m \phi}{\phi_0} \right)  g_m(T) \,
	\end{equation}
with
	\begin{equation}
	g_m(T) = \int_0^\infty  dt \frac{R^2(t/t_T)}{t^2}
	\frac{\exp\left[{-(mL)^2}/{(4Dt)}\right]}{\sqrt{4\pi Dt}} \, .
	\end{equation}
After taking the flux derivative according to (\ref{eq:Iorb}),
one recovers the first-order interaction contribution to the persistent
current \cite{Ullmo97}, 
	\begin{equation} \label{I1_ring}
 	I_1 = \frac{\lambda_0}{\mathrm{g_s}}
	\frac{2 L e }{\pi } \sum_{m=-\infty}^{+\infty}
   	m \sin \left( \frac{4\pi m \phi}{\phi_0} \right) g_m(T) \; .
	\end{equation}
This first-order result was first obtained by purely diagrammatic 
techniques by \citeasnoun{Ambegaokar90}  and 
semiclassically by \citeasnoun{Montambaux96}.

However, higher-order terms are essential for an appropriate
computation of the interaction contribution.
As shown in the preceding subsection,
these higher-order terms merely lead to a renormalization of
the coupling constant according to (\ref{eq:Knorm}). Thus
the persistent current from the entire interaction contribution
is reduced to \cite{Ullmo97}
\begin{equation} \label{I_ring}
 	I = \frac{2 L e }{\pi\ln(\kf L^\ast)} \sum_{m=-\infty}^{+\infty}
   	m \sin \left( \frac{4\pi m \phi}{\phi_0} \right) g_m(T) \; .
\end{equation}
For diffusive rings the length scale $\vf t$, entering in
(\ref{eq:L^ast}) for $L^*$, is given by $L_m \smeq \vf (m L)^2/4
D$, the average length of a trajectory diffusing $m$ times around the
ring. Hence one gets at low temperature ($L_T \gg L_m$) a
(renormalised) prefactor $\sim 2 / \ln(\kf L_m) $ for $I$.  At high
temperature, $L_T \ll L_m$, the prefactor includes ${2}/{ \ln(\kf L_T
  / 4)}$.  These two limits agree with quantum results obtained
diagrammatically by \citeasnoun{Eckern91}.  The functional form of
the temperature dependence (exponential $T$--damping
\cite{Ambegaokar90}) is in line with experiments
\cite{Levy90,Chandrasekhar91,Mohanty96}. However, the amplitude of the
full persistent current with renormalised coupling constant is a
factor $\sim 3-5$ to small compared to experiments.

\subsubsection{Susceptibility of two--dimensional diffusive systems}

In ring geometries the exponential temperature dependence of $I$ is
related to the existence of a minimal length, the circumference of the
ring, for the shortest flux-enclosing paths.  In singly connected
systems the geometry does not constrain returning paths to have a
minimal length, and therefore one expects a different temperature
dependence of the magnetic response.

Consider a two-dimensional singly connected quantum dot with diffusive
dynamics.  If one makes use of the general renormalization property of
diffusive systems, expressed by (\ref{eq:Knorm2}), the diagonal part
of the thermodynamic potential (see (\ref{omega_sc})), including the
entire Cooper series, reads
\begin{equation} \label{gen:omegaD}
  \Omega^{(D)} = \frac{1}{\beta} \,
  \int d {\bfr} \
	   \int_{\tau_{el}}^\infty dt \
  \frac{1}{\ln(\kf \, \vf t)} \
  \frac {t_T}{t^2} \  R^2\left(\frac{t}{t_T}\right)
   \ {\ssA}({\bfr},t;B) \, .
\end{equation}
The parameter $L^*$ appearing in (\ref{eq:Knorm}) has been here
replaced by $\vf t$ since the factor $R^2$ ensures that the main
contribution to the integral comes from $t < t_T$. In the above time
integral the elastic scattering time $\tau_{el} = l/\vf$ enters as a
lower bound. This cutoff must be introduced since for backscattered
paths with times shorter than $\tau_{el}$ the diffusion approximation
(\ref{TFreturn}) no longer holds. Short paths with $t < \tau_{el}$,
which may arise from higher-order interaction events, contribute to
the clean bulk magnetic response \cite{Aslamazov75,VonOppen00}.  This
latter term is, however, negligible compared to the disorder induced
interaction contributions considered here.

To evaluate the integral for $\ssA$,
the conditional return probability in two dimensions 
is conveniently represented in terms of the Fourier transform
\cite{Argaman93}
\begin{equation} \label{TFreturn}
      P(\bfr,\bfr,t|A) = \frac{1}{8 \pi^2} \ \int dk \ |k| \ 
     \frac{e^{ikA} }{\sinh(|k|Dt)} \; .
\end{equation}
Introducing the magnetic time
\begin{equation}
\label{tB}
t_B = \frac{\phi_0}{4 \pi B D} = \frac{L_B^2}{4\pi D} \; ,
\end{equation}
($t_B$ is related to the square of the magnetic length $L_B^2$ which
can be regarded as the area enclosing one flux quantum 
(assuming diffusive dynamics)), one obtains
\begin{equation}
\label{A2dim}
	{\ssA}({\bfr},t;B) = \frac{1}{4 \pi D} \frac{R(t/t_B)}{t} \; .
\end{equation}
The function $R$ occurring in (\ref{A2dim})
has a different origin than in (\ref{gen:omegaD}).

The magnetic susceptibility (\ref{eq:susc}) is obtained after
including the expression (\ref{A2dim}) in (\ref{gen:omegaD}) and
taking the second derivative with respect to the magnetic field. One
finds, after normalization to the Landau susceptibility of
non-interacting particles in a clean system,
\begin{equation}
\label{bulk:chi1}
 \frac{ \chi^{(D)} }{|\chi_{\rm L}|}
     = -\frac{12}{\pi} (\kf l)
	   \int_{\tau_{el}}^\infty
  \frac{d t}{t \ln(\kf\,\vf t)} R^2\left(\frac{t}{t_T}\right)
  R''\left(\frac{t}{t_B}\right) \; .
\end{equation}
Here,  $D=\vf l/2$, and $R''$ denotes the second derivative of $R$.

The above equation (\ref{bulk:chi1}) holds true only as long as the effective 
upper cutoff time $t^* \! \equiv \! \min(t_T,t_B)$,
introduced through the $R$-functions, remains smaller than
the Thouless time $t_c \smeq L^2/D$ with $L$ being the system size.
For times larger than $t_c$  the
two-dimensional diffusion approximation is no longer valid since 
the dynamics begins to behave ergodically.

Assuming $\tau_{el} \ll t^* < t_c$, the integral in
(\ref{bulk:chi1}) can be approximately evaluated by
replacing the upper bound by $t^*$, and by replacing $R(t/t_T)$ and
$R''(t/t_B)$ by $R(0)=1$ and $R''(0) = -1/3$, respectively. The
averaged magnetic susceptibility of a diffusive two--dimensional
quantum system reads under these approximations \cite{Ullmo97}
\begin{equation}
\label{chifinal}
 \frac{ \chi^{(D)} }{|\chi_{\rm L}|}
\simeq \frac{4}{\pi} (\kf l)
\ln \left\{  \frac {\ln [\kf \, \vf \min(t_T,t_B) ]} {\ln(\kf l)} \right\} \; .
\end{equation}
The magnetic response of diffusive systems is
paramagnetic and enhanced by a factor $\kf l$ compared to the clean Landau
susceptibility $\chi_{\rm L}$.

Contrary to the exponential temperature dependence of the
ring geometry discussed in the previous section 
one  finds a log-log temperature dependence
for $t_T < t_B$ as well as a log-log $B$ dependence  for $t_T > t_B$.
The log-log form of the result is produced by the $1/t \ln t$ dependence 
of the integral in (\ref{bulk:chi1}). It results from
the wide distribution of path-lengths in the system --- there are 
flux-enclosing paths with lengths ranging from about $\vf \tau_{el}$ up 
to $\vf t^*$.

The expression (\ref{chifinal}) agrees with results from Aslamazov and Larkin
\citeyear{Aslamazov75}, Altshuler, Aronov and Zyuzin
\citeyear{Altshuler83}, and Oh, Zyuzin and Serota
\citeyear{Oh91} which are obtained with quantum diagrammatic
perturbation theory. The equivalence between the semiclassical and
diagrammatic approaches, demonstrated here as well as the preceding
section for diffusive rings may be traced back to the fact that the
``quantum'' diagrammatic perturbation theory relies on the use of the
small parameter $1/\kf \ell$, and can therefore also be viewed as a
semiclassical approximation.

\subsection{Ballistic Systems}
\label{subsec:ballistic}

We turn now to the magnetic response of ballistic mesoscopic objects.
This problem has received considerably less attention than the
diffusive regime, partly because of the experimental difficulties
involved, but also presumably because from the theoretical point of
view, ballistic systems cannot be addressed with the more traditional
approaches of solid state physics.  Indeed, for diffusive systems, the
main virtue of the ``quantum-chaos'' based approach presented in this
review was to provide an alternative, maybe more intuitive, way to get
known results.  In contrast, the ballistic regime represents one
example for which it is actually the only way to get 
understanding of the physics involved.

The ballistic character of the underlying classical dynamics brings
about some important differences, especially since one no longer talks
about probabilistic concepts (like the return probability), but one
needs to input the information about specific trajectories.  Depending
on the geometry of the confining potential, one can have chaotic or
integrable motion, and the structure of the semiclassical expansions
differs in some respects in these two cases.  In particular, the existence
of families of trajectories for integrable dynamics will translate into
larger magnetic response.

\subsubsection{Renormalization scheme for ballistic systems}

As for diffusive systems, (\ref{eq:omega^C}) and (\ref{eq:sigma_D})
form the basic expressions from which the magnetic response can be
computed.  We just saw that in this latter case, they allow quite
straightforwardly to express the magnetic susceptibility of dots or
persistent current of rings in closed form.  For ballistic systems on
the other hand, even when the classical dynamics is simple enough to
yield $\Sigma^{(D)}(\bfr,\bfr';\omega_m)$ explicitly, taking the
logarithm of this operator as implied by (\ref{eq:omega^C}) will
require proceeding numerically.  It should be born in mind however
that this numerical calculation is significantly simpler than the one
we would have faced had we decided to start directly with a numerical
approach.  This can be understood for instance by considering that the
original operator $\Sigma(\bfr,\bfr';\omega_m)$, because of its
quantal nature, has a scale of variation determined by $\lambdaf$.
Discretizing this operator on a grid would therefore require using a
mesh containing at least a few points per $\lambdaf$, which would make
any computation rapidly intractable as the size of the problem
increases.  On the other hand, within the semiclassical approach, the
operator one has to deal with is $\Sigma^{(D)}(\bfr,\bfr')$ which, up
to one exception to which I shall return below, varies only
on a {\em classical} scale.  $\Sigma^{(D)}(\bfr,\bfr')$ is therefore
what one may call a ``classical operator''.  It permits the use of a
grid mesh whose scale is fixed by the classical dynamics within the
system, and therefore much larger than $\lambdaf$.

As mentioned above $\Sigma^{(D)}(\bfr,\bfr')$ is not yet completely
classical.  Indeed,  there still is, in (\ref{eq:sigma_D}), 
the scale $\Lambda_0 = \lambdaf/\pi$ which specifies that
trajectories shorter than $\Lambda_0$ should  be excluded from the sum
over trajectories joining $\bfr'$ to $\bfr$.  This last 
quantum scale can be  removed using a  simple renormalization scheme where
integration over short trajectories yields a decreased effective
coupling constant. To that end consider a new cutoff $\Lambda$ 
larger than $\Lambda_0$ but much smaller than any other 
characteristic lengths ($a$, $L_T$, or $\sqrt{\phi_0 / B}$). For  each path
$j$ joining $\bfr'$ to $\bfr$ with $L_j > \Lambda$, let 
$\Sigma_j(\bfr, \bfr')$ denotes its contribution to 
$\Sigma^{(D)}(\bfr, \bfr')$ and define 
\begin{eqnarray}
   \label{tilde} 
   \tilde\Sigma_j(\bfr,\bfr')
   \equiv&&\Sigma_j(\bfr,\bfr') - \lambda_0
   \int d \bfr_1  \ \Sigma_j(\bfr,\bfr_1)
   \hat \Sigma(\bfr_1,\bfr') \nonumber\\
   &&+\lambda_0^2\int  d \bfr_1 \
    d {\bfr}_2 \  \Sigma_j(\bfr, \bfr_2) \hat  \Sigma(\bfr_2,\bfr_1)
    \hat \Sigma(\bfr_1, \bfr') +\cdots \; .  
\end{eqnarray}
where the      $\bfr_i$       integration                  is            over
$\Lambda_0{\!}<{\!}|\bfr_{i-1}-\bfr_i|{\!}<\!\Lambda$                     (with
${\bfr}_0\!\equiv\!{\bfr'}$).  $\hat\Sigma(\bfr_1,\bfr')$ is defined by
(\ref{eq:sigma_D}) but with  the sum restricted to ``short'' trajectories
with lengths in the range  $[\Lambda_0,\Lambda]$;  $\Sigma_j(\bfr,\bfr_1)$
is  obtained from  $\Sigma_j(\bfr,\bfr')$  by continuously  deforming
trajectory  $j$. To avoid the awkward $\ln$ in (\ref{eq:omega^C}),
let us introduce 
\begin{equation} 
  \Gamma =  \frac{1}{\beta} \sum_{\omega_m} 
  {\rm    Tr} \left[
   \frac{1}{1+\lambda_0\Sigma^{(D)}(\bfr, \bfr';\omega_m)}
   \right] \; ,
\end{equation}
from which $\Omega^{(D)}$ can be derived through
\begin{equation}
   \label{integral}
   \Omega^{(D)}(\lambda_0) = \int_0^{\lambda_0}    \frac{d
   \lambda'_0}{\lambda'_0}\,   \Gamma(\lambda'_0)  \;  .  
\end{equation}
The replacement $\Sigma$ by $\tilde \Sigma$ in $\Gamma$ amounts to a
reordering of the perturbation expansion of $\Gamma$ in which short
paths are gathered into lower-order terms. Moreover, if $L_j \gg
\Lambda$, small variations in the spatial arguments do not modify
noticeably the characteristics of $\Sigma_j$. Approximating
$\Sigma_j(\bfr,\bfr_1)$ by $\Sigma_j(\bfr,\bfr')$ in (\ref{tilde}) and
using ${\hat\Sigma(\bfr_1,\bfr')}\simeq{1/4}\pi|\bfr_1-\bfr'|^2$ valid
for short paths, one obtains
\begin{equation}
	\lambda_0 \, \tilde \Sigma_j(\bfr, \bfr') \simeq
	\frac{ \lambda_0 \Sigma_j(\bfr, \bfr') }{1 +
	\lambda_0
	\int d{\bfr_1}
	\hat \Sigma(\bfr_1, \bfr')}
	\simeq
        \lambda(\Lambda)\,\Sigma_j(\bfr, \bfr')
\end{equation}
where the running coupling constant is defined by 
\begin{equation} \label{eq:lambda(Lambda)}
       \lambda(\Lambda) = \frac{\lambda_0}{1 + 
        (\lambda_0 / \mathrm{g_s}) \ln(\Lambda/\Lambda_0)} \; .
\end{equation}       
Therefore, these successive steps amount to a change of both the
coupling constant and the cutoff (since now trajectories shorter than
$\Lambda$ must be excluded) without changing $\Gamma$; that is,
\begin{equation}
   \label{RGgamma}
   \Gamma (\Lambda_0, \lambda_0) = \Gamma \left(\Lambda,
   \lambda(\Lambda)\right) \; .  
\end{equation} 
Through (\ref{integral}), this renormalization scheme can be applied 
to $\Omega^{(D)}$, and so to the average susceptibility.

In this way, we have eliminated the last ``quantum scale'' $\Lambda_0$ from
the definition of $\Sigma^{(D)}$: $\Lambda$ can be made much larger than
$\lambdaf$ while remaining smaller than all classical lengths. This
will serve two purposes.  From a quantitative point of view, it means
that discretizing the operator $\Sigma^{(D)}$ can be done on a
relatively coarse grid, and that therefore operations such as taking
the trace logarithm of $1 + \Sigma^{(D)}$ that may be necessary to get
the magnetic response can be done numerically much more easily.

Furthermore, it is qualitatively reasonable that the perturbation
series of $\Omega^{(D)}$ becomes convergent when $\Lambda$ is of order
of the typical size $L$ of the system, since by this point the spread
in length scales causing the divergence has been eliminated. Of course
the renormalization scheme above assumes, $\Lambda$ much smaller than
all the classical scales of the system, and using $\Lambda \simeq L$
is beyond the range for which a reliable quantitative answer can be
obtained.  At a qualitative level however, this implies that
higher-order terms in the diagonal contribution mainly renormalise the
coupling constant $\lambda_0$ into
\begin{equation} \label{eq:renormalized_int}
  \lambda(L) = \frac{\lambda_0}{1 + (\lambda_0/\mathrm{g_s}) \ln(\kf L)} \; .
\end{equation}
In the deep semiclassical limit $\ln(\kf L) \gg 1$, the original
coupling constant $\lambda_0$ drops from the final results and is
replaced by $\mathrm{g_s}/\ln(\kf L)$.

For integrable systems in which non-diagonal channels exist, simple
inspection shows that these latter are not renormalised.  Indeed the
corresponding higher-order terms in the perturbation expansion are
highly oscillatory.  The non-diagonal contributions are therefore of
order $\lambda_0$ rather than $\mathrm{g_s}/\ln(\kf L)$, and will as a
consequence dominate the magnetic response in the deep semiclassical
regime.

\subsubsection{Ballistic squares}

\begin{figure} 
\begin{center}
\includegraphics[totalheight=6cm]{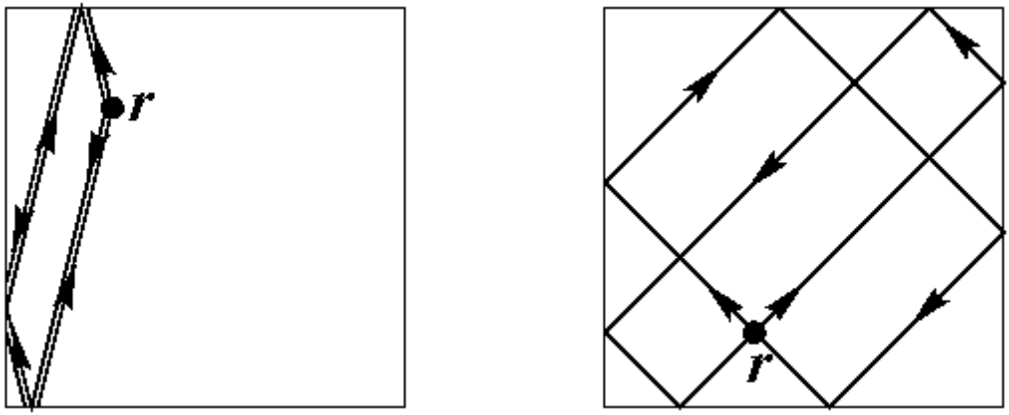}
\end{center}
\caption{ Typical pairs of real-space trajectories that contribute to
  the average susceptibility to first order in the interaction in the
  diagonal channel (left) and the non-diagonal channel (right).  This
  figure is taken from \citeasnoun{Ullmo98}.}
\label{fig:square} 
\end{figure}

As an illustration, let us consider square quantum dots. The choice of
squares is motivated first because this is the geometry used
experimentally \cite{Levy93}, and secondly because squares are
particularly amenable to a semiclassical treatment, since it is very
easy to enumerate all the classical trajectories \footnote{ As
a consequence of the two previous facts, orbital magnetism in
ballistic squares has been thoroughly studied within non-interacting
models
\cite{VonOppen94,Ullmo95,RichterPhysRep96,Gefen94,RichterPRB96,RichterJMathPhys96}.}. 
To be more specific, I will consider a square billiard model, of size
$L$, with free motion (i.e.\ just a kinetic energy term) inside the
the billiard and Dirichlet boundary conditions on its border. An
illustration of the shortest (with non-zero area) closed orbit and of
the shortest periodic orbit, for the classical version of the billiard, is
given in figure~\ref{fig:square}.

\begin{figure} 
\begin{center}
\includegraphics[totalheight=8cm]{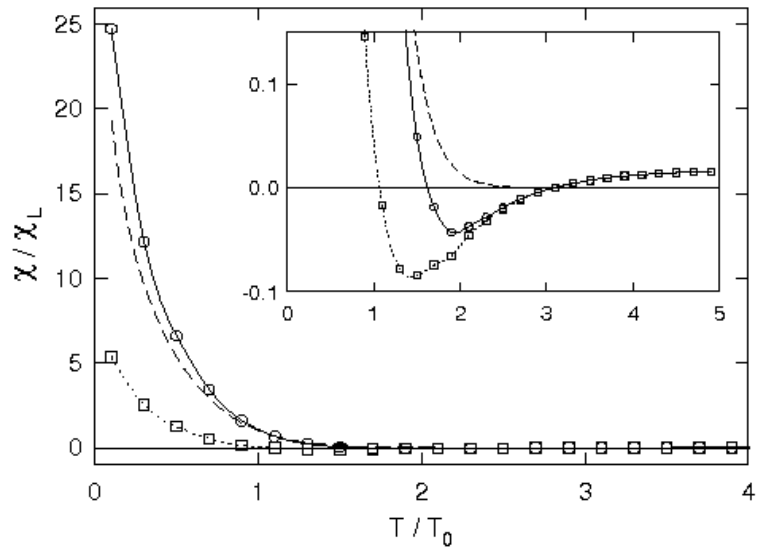}
\end{center}
\caption{ Temperature dependence of the zero-field susceptibility
  (solid line) for an ensemble of squares at $\kf L \smeq 50$. The
  contribution of the non-diagonal channel (dashed, family (11) and
  repetitions) exceeds that of the diagonal Cooper channel (dotted) at
  low temperatures.  Temperatures are expressed in units of {$k_{\rm
      B} T_0 = \hbar \vf/2\pi L$}. 
  Inset: same, but with a different scale for
  the vertical axis.  This figure is taken from
  \citeasnoun{Ullmo98}.}
\label{fig:susceptibility}
\end{figure}

Let us discuss first the diagonal contribution $\Omega^{(D)}$ 
(\ref{eq:cooperon_series}) to the thermodynamic potential.  Because
of the simplicity of the geometry, the explicit expression of
$\Sigma_j(\bfr,\bfr';\omega)$ for any orbit $j$ joining $\bfr'$ to
$\bfr$ can be obtained quite straightforwardly (though the resulting
expressions might be a bit cumbersome and will thus not be given).  At
a given temperature $T$, the operator $\Sigma^{(D)}(\bfr,
\bfr';\omega)$ is then constructed by summing all such contributions
for orbits of lengths shorter than the thermal length $L_T$ (see
(\ref{eq:tT})).  A numerical computation of $\Omega^{(D)}$ (and
therefore, after derivation with respect to the magnetic field, of the
magnetic susceptibility) can then be obtain by representing
$\Sigma^{(D)}(\bfr, \bfr';\omega)$ on a grid, then going to a diagonal
representation, and in this diagonal representation taking the $\log$
and performing the trace.  The temperature dependence of the resulting
contribution to the susceptibility, for a number of particles
corresponding to $\kf L = 50$, is shown as a dotted line in
figure~\ref{fig:susceptibility}.

\begin{figure} 
\begin{center}
\includegraphics[totalheight=8cm]{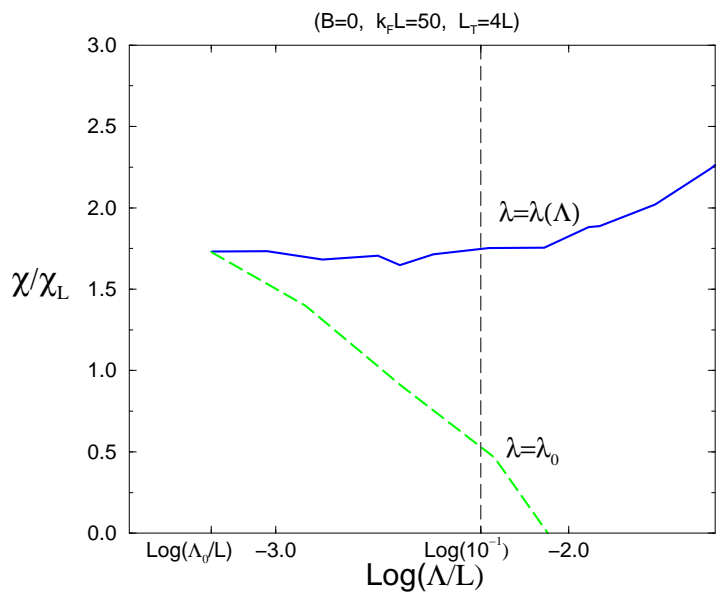}
\end{center}
\caption{
Diagonal part of the interacting contribution to the magnetic
susceptibility as a function of the cutoff $\Lambda$ used to define the
particle-particle propagator $\Sigma$. The data correspond to zero
magnetic field, $\kf L = 50$, and a temperature $T$ such that $L_T  =
4L$. Solid line: computation for which the coupling
constant $\lambda$ varies as a function of $\Lambda$ following 
(\protect\ref{eq:lambda(Lambda)}).  Dash line: $\lambda$ is kept fixed
($=\lambda_0$). The vertical (dash) line corresponds to the limit of
the range for which the condition $\Lambda \ll L$ is fulfilled}.
\label{fig:rg}
\end{figure}

In this way, one can furthermore check that the renormalization
process described above actually works for the square geometry. For
instance figure~\ref{fig:rg} shows the value of the magnetic
susceptibility at zero field, and for a temperature such that $L_T =
4L$, as a function of the cutoff $\Lambda$.  As long as $\Lambda \ll
L$, the diagonal part of the susceptibility is clearly independent of
the cutoff if the renormalised interaction $\lambda(\Lambda)$ given by
(\ref{eq:lambda(Lambda)}) is used.

Let us finally compare the diagonal and non-diagonal contributions.
This latter is built from pairs of orbits with the same action (i.e.\
for a billiard, same length) but different geometry, overlapping at a
given point.  Those may occur quite generically for integrable systems
for which families of periodic orbits exist and two members of the
same family may intersect in configuration space.  An illustration of
such an intersection is given in the right panel of
figure~\ref{fig:square} for the family (labelled $(1,1)$) of shortest
periodic orbits (with non-zero enclosed area) of this system.  In the
square billiard, for a given closed orbit $j$ of length $L_j$ and
enclosing an area $\area_j$, the action in the absence of magnetic
field is given by $S_j^0/\hbar = \kf L_j$, its derivative with respect
to the magnetic field by $dS_j / dB = 2 \pi \area_j /\phi_0$, and the
stability determinant (cf.\ (\ref{eq:Dj})) by $D_j = m/\sqrt{\hbar \kf
  L_j}$.  The contribution $\chi_{(1,1)}$ of the pairs of orbits from
the family $(1,1)$ can for instance be derived quite straightforwardly
from equations (\ref{eq:HFgen3a})-(\ref{eq:HFgen3a}) noting that that
$L_{11} = 2\sqrt{2} L$ and $\area_{11} = 2x_0(L-x_0)$ ($x_0$ is the
abscissa of the intersection of the orbit with the bottom border of
the billiard, and can be use to label an individual orbit within the
family $(1,1)$).  One obtains in this way \cite{Ullmo98}
\begin{equation}
   \label{nondiag} 
   \frac{\langle \chi^{\rm non-diag}_{(1,1)} \rangle }{
     \left|\chi_{\scriptscriptstyle L} \right|} = 
   - \ \frac{3 \kf L }{ 2 (\sqrt{2}\pi)^3} \ 
   \ \frac{ d^2 {\cal C}^2(\varphi) }{ d \varphi^2} 
   \ R^2\left(\frac{L_{11}}{L_T}\right) \; .
\end{equation} 
The temperature dependence is governed by the function $R(x) \smeq {x /
\sinh(x)}$ and the field dependence by $ {\cal
C}(\varphi) \smeq ({2 \varphi})^{-1/2} \left[ \cos(\pi
\varphi) {\rm   C}(\sqrt{\pi   \varphi})  +  \sin(\pi  \varphi)   {\rm
S}(\sqrt{\pi \varphi}) \right]$, with $\varphi \smeq B L^2  / \phi_0 $, and
${\rm C}$  and ${\rm S}$ Fresnel functions.

As seen in figure~\ref{fig:susceptibility}, because the non-diagonal
contribution is not renormalised by higher-order terms, it dominates
the magnetic response as soon as the temperature is low enough not to
suppress it.

\subsection{Discussion}
\label{subsec:exp}

After this overview of some of the results that can be derived for the
magnetic response of mesoscopic rings or dots within a Fermi-liquid, at
equilibrium, description, it is worthwhile to come back to how well
existing experiments can be understood within this framework.  Since
the typical response for a single system
\cite{Mailly93,Chandrasekhar91} may in some case be dominated by the
non-interacting contribution, which is not addressed here, I will
however limit this discussion to experiments made on ensembles of
micro-structures.  

Ballistic and diffusive systems differ with respect to the comparison
between experimental measurements and theoretical predictions, but for
both cases there is clearly not a complete adequacy.

In the case of ballistic systems \cite{Levy93}, the amplitude of the
magnetic response in the low temperature range, as well as scale of
the field dependence, are well in line with the predictions
\cite{VonOppen94,Ullmo95,RichterPhysRep96}.  However, the temperature
dependence, and in particular the fact that a magnetic response is
observed even when the thermal length $L_T$ is smaller than the size
of the system seems very difficult to interpret within the Fermi liquid
framework used here \cite{Ullmo95,Ullmo98}.  There is, up to now, no
real suggestion to explain how a significant magnetic response, with a
field scale unambiguously associated with interference effects, could
survive at such high temperatures.

For diffusive rings \cite{Levy90}, the temperature dependence seems to
be less of an issue than in the ballistic case.  However the magnitude
of the average persistent current appears to be quite a bit larger (a
factor three to five) than what is expected theoretically
\cite{Ambegaokar90epl}.  It has been suggested that this large
magnetic response might be related to non-equilibrium effects, such as
the coupling to a phonon or photon bath
\cite{Mohanty99,Kravtsov00,Entin03}.  In later experiments
\cite{Jariwala01} (see also \citeasnoun{Reulet95} and
\citeasnoun{Deblock02} in this context), a further, and presumably
more dramatic, discrepancy is that the sign of the magnetic response
-- which has not been determined in \citeasnoun{Levy90} -- was furthermore
not the one expected for a repulsive interaction.  This has motivated
consideration of whether a BCS interaction, weak enough to be
compatible with the absence of observed superconducting state for the
considered materials, could be responsible for both the change of sign
(which would be an immediate consequence of having an attractive
interaction) but also the increase in the magnitude of the observed
signal \cite{Schechter03,EckernComment04,SchechterReply04}.  It is
not clear however that such a range of attractive interaction actually
exists in practice.

In the end, it is remarkable that on the one hand a good part
of the experimental findings has a simple and natural interpretation
within the ``confined electron gas with weak interaction'' picture
developed here, when on the other hand no consensus has emerged yet
concerning the interpretation other significant experimental findings.
One of the limiting factors in this respect is presumably the
imbalance between the large number of theoretical works and the much
sparser character of experimental studies on these questions.  In
particular, only one group has measured the magnetic response of an
ensemble of micro-structures in the ballistic regime \cite{Levy93}.  In
that case the low temperature data agree very well with the theoretical
description when there is little hope to interpret what is observed by
the high temperature regime without introducing some new physical
mechanisms.  However, the lack of experimental indications on what
these other mechanisms could be is making difficult any theoretical
progress.  In the same way, for diffusive systems, a complete
understanding would be made easier if one knew how much the sign of
the magnetisation observed in \citeasnoun{Jariwala01} is material
dependant.

Obviously, it would be very non-trivial to improve on the existing
experiments, in particular because it is necessary to deal properly
with very weak magnetic fields.  However, there are enough indications
that some deep physics might be involved to motivate further
experimental work in this field.

\setcounter{footnote}{1}

\section{Mesoscopic Kondo effect}
\label{sec:Kondo}

Turning now to another context where the interrelation
between interferences and interactions plays an important role, I shall 
wander briefly away from Fermi liquids, and consider some (limited)
aspects of the problem of a Kondo impurity within a mesoscopic conductor.

\subsection{A quick background}

The term ``Kondo effect'' refers to the physics of an impurity with
some internal degree of freedom, interacting with a gas of otherwise
non interacting electrons. It represents one of the simplest models in
condensed matter physics for which {\em correlations} play a central
role \cite{HewsonBook}.

In its simplest version, the $s$-$d$ model, the impurity is
just treated as a spin one half interacting locally with the electron
gas. The corresponding Hamiltonian then reads
\begin{equation}
\label{eq:KondoHam1}
H_{\rm K} = \sum_{\alpha \sigma} \epsilon_\alpha 
\hat c^{\dagger}_{\alpha \sigma} \hat c_{\alpha\sigma}
+ H_{\rm int} \; ,
\end{equation}
where $\hat c^{\dagger}_{\alpha \sigma}$ creates a particle with energy
$\epsilon_\alpha$, spin $\sigma$ and wave-function $\varphi_{\alpha}({\bf
  r})$, and the interaction with the impurity is expressed as 
\begin{equation}
\label{eq:KondoHam2}
H_{\rm int} = \frac {J_0}{\hbar^{2}} \; \bi{ S} \cdot
\bi{s} (0) 
\end{equation}
with $J_0$ the coupling strength, $ \mathbf{S} = (S_x,S_y,S_z)$ a spin
operator ($\hbar^{-1} S_i$ is half of the Pauli matrix $\sigma_i$),
$\mathbf{s}(0) = \frac{\hbar}{2} \hat \Psi^{\dagger}_{\sigma}(0)
\mathbf{\sigma}_{\sigma\sigma^\prime} \hat \Psi_{\sigma}(0)$ the spin
density of the electron gas at the impurity position ${\bfr} \equiv
0$, and $\hat \Psi^{\dagger}_{\sigma}(0)= \sum_{\alpha}
\varphi_{\alpha}(0) \hat c^\dagger_{\alpha}$.

Originally, physical realizations of the Kondo Hamiltonian corresponded
to actual impurities (e.g. $Fe$) in a bulk metal (e.g.\ $Cu$). The
wave-functions $\varphi_{\alpha}$ could then be taken as plane waves, and
one could assume a constant spacing $\Delta$ between the
$\epsilon_\alpha$, so that the electron gas could be characterized by
only two quantities: the local density of states $\nu_0 = (\Volume
\Delta)^{-1}$ ($\Volume$ is the volume of the sample), and the
bandwidth $\mathrm{D}_0$ of the spectrum.

What gave (and still gives) to the Kondo problem its particular place
in condensed matter physics is that it is the simplest problem for
which the physics is dominated by renormalization effects.  Indeed,
assuming the dimensionless constant $ J_0 \nu_0 \ll 1$, it can be shown
using one-loop renormalization group analysis \cite{Fowler71}, or
equivalent earlier approaches such as Anderson poor's man scaling
\cite{Anderson70} or Abrikosov parquet diagrams re-summation
\cite{Abrikosov65}, that the low energy physics remains unchanged by the
simultaneous modification of both $J_0$ and $\mathrm{D}_0$ into new values
$J_{\rm eff}$ and $\mathrm{D}_{\rm eff}$ provided they are related by
\begin{equation} \label{eq:RGbulk}
{J}_{\rm eff}(\mathrm{D}_{\rm eff}) = \frac{J_0}{1 - {J_0\nu_0}\ln(\mathrm{D}_0/\mathrm{D}_{\rm eff}) } \; , 
\end{equation}
The renormalization procedure should naturally be stopped when $\mathrm{D}_{\rm
  eff}$ becomes of the order of the temperature $T$ of the system.
Equation (\ref{eq:RGbulk}) defines an energy scale, the Kondo temperature 
\begin{equation} \label{eq:TKbulk}
T_K  = \mathrm{D}_0\exp(-1/(J_0\nu_0)) \; ,
\end{equation}
which specifies the crossover between the weakly and strongly
interacting regime.  For $T \gg T_K$, the impurity is effectively
weakly coupled to the electron gas, and the properties of the system
can be computed within a perturbative approach provided the
renormalised interaction ${J}_{\rm eff}(T)$ is used.  The regime $T
\ll T_K$ is characterized by an effectively very strong interaction
(in spite of the bare coupling value $J_0\nu_0$ being small), in such a
way that the spin of the impurity is almost completely screened by the
electron gas.  Perturbative renormalization analysis (and thus
(\ref{eq:RGbulk}) itself) can obviously not be applied in this
regime, but (for bulk system) a rather complete description has been
obtained by a variety of approach, including numerical renormalization
group \cite{WilsonRMP75}, Bethe anzatz techniques
\cite{Andrei80,Wiegmann80}, and in the very low temperature regime
Nozières Fermi liquid description \cite{Noziere74}.

One important consequence of the scaling law (\ref{eq:RGbulk}) is
that physical quantities can be described by universal functions,
which can be understood by a simple counting of the number of
parameters defining the $s$-$d$ model in the bulk.  I will illustrate
this discussion with a particular physical quantity namely the local
susceptibility
\begin{equation}
   \chi_{\rm   loc} = \int_0^\beta d\tau \langle S_z(\tau) S_z(0) 
   \rangle \; ,
\end{equation}
which is the variation of the impurity spin magnetisation to a field
applied only to the impurity.  The electron gas is characterized by
its local density of states $\nu_0$ and its bandwidth $\mathrm{D}_0$, and the
impurity by the coupling constant $J_0$.  Therefore, for a given
temperature $T$, $\chi_{\rm loc}$, as any physical quantity, can
depend only on these four parameters.  Furthermore, only two
dimensionless parameters can be constructed from them, the ratio $T/\mathrm{D}_0$
and the product $J_0 \nu_0$.  However, because of the scaling law
(\ref{eq:RGbulk}), we see that these two parameters turn out to be
eventually redundant.  A dimensionless quantity can therefore be
expressed as a function of a single parameter, which can be chosen to
be $T/T_K$ so that we have for instance
\begin{equation} \label{eq:fchi}
  T \chi_{\rm loc} = f_\chi(T/T_K) \; ,
\end{equation}
with $f_\chi(x)$ a universal function which has been computed by
 \citeasnoun{WilsonRMP75} using his numerical renormalization group
approach.  Note finally that within the one-loop approximation
(\ref{eq:RGbulk})-(\ref{eq:TKbulk}), 
\begin{equation} \label{eq:Jrho}
T/T_K = \exp(J_{\rm   eff}(T) \nu_0)
\end{equation}
so that another way to express the universal character of physical
quantities is to say that they depend only on the dimensionless
quantity $J_{\rm eff}(T) \nu_0 $.

\subsection{Mesoscopic fluctuations}

We see that the universal character of the Kondo physics is a direct
consequence of the fact that the local density of states is flat and
featureless, and can therefore be characterized by two parameters,
$\nu_0$ and $\mathrm{D}_0$.  There are many situations however for which the
variation in energy (and position actually) of the local density of
states $\nu_{\rm loc}(\bfr;\energy)$ might be significantly more complex, and
it is natural to ask in which way this would modify the description
given above in the bulk (flat band) case.

The origin of fluctuations in the density of states can be of
different nature, but they are always in the end associated to
interference or finite size effects.  One possibility is the presence
of disorder in a bulk material
\cite{Dobrosavljevic92,Kettemann04,Kettemann06,Kettemann07,Zhuravlev08},
in either the metallic or localised regime.  Another is the proximity
of some boundary in the host material \cite{Ujsaghy01}, for instance
in the case of a narrow point contact \cite{Zarand96} or for thin film
\cite{Crepieux00}.  Finally the class of systems where a Kondo
impurity is placed within a fully coherent, finite size electron sea,
as has been realized for instance in the context of ``quantum
corrals'' \cite{Fiete01}, has been also considered
\cite{Thimm99,Affleck01,Cornaglia02a,Cornaglia02b,Simon02,Franzese03,Cornaglia03,KaulEPL05,Lewenkopf05,KaulPRL06,Simon06}.

One important recent development which has made very natural the idea
that a magnetic impurity could be connected to finite size electron
gas is the realization that Kondo physics was actually relevant to
transport properties of quantum dots \cite{GlazmanHouches05}.  Indeed,
as was pointed out by \citeasnoun{Glazman88} and
simultaneously \citeasnoun{Ng88}, a quantum dot in the deep
Coulomb blockade regime (so that particle number fluctuations are
suppressed) containing an odd number of electrons, and sufficiently
small that the temperature can be made much smaller than the mean
level spacing between orbitals can be described by a Anderson impurity
model, which, up to a Schrieffer-Wolff transformation, is essentially
equivalent to a Kondo impurity. When the dot is weakly coupled to
leads, these latter play the role of the electron gas.  In the low
temperature regime $T \ll T_k$, a correlated state is formed which
mixes the quantum dots and both drain and source wave-functions,
leading to a large conductance ($\simeq e^2/h$) in spite of the dot
being in the deep Coulomb blockade regime.  These predictions have
been observed by \citeasnoun{Goldhaber98} a
decade after these predictions, some later experimental realization
even reaching  the unitarity limit \cite{VanDerWiel00}.

The great flexibility in the design and control of nanoscopic systems
further motivated the study of many new configurations involving more
exotic Kondo effects. One can cite for instance the theoretical design
\cite{Oreg03} and experimental observation \cite{Potok06} of the
2-channel Kondo, which has proved elusive in bulk system, the possible
occurrence of SU(4) Kondo
\cite{Borda03,LeHur03,LeHur04,Galpin05,LeHur07} and its relevance to
carbon nanotubes \cite{Choi05,Makarovski07a,Makarovski07b}, or double
dot system where Kondo physics might be in competition with RKKY
interactions \cite{Craig04,Vavilov05,Simon05,Martins06}.

The subject of Kondo physics and quantum dots is a very vast, and
still rapidly developing, field.  It is clearly not realistic to
cover it in any reasonable way here, and I will only consider in more
detail the, admittedly rather specific, aspects more closely
related to the subject of this review. Indeed most of the more exotic
designs imply at some point that a small quantum dot playing the role
of a quantum impurity is connected to larger mesoscopic object, for
which Kondo physics in itself is irrelevant, but such that finite size
effects may become important. In other words, the context of Kondo
physics and quantum dots  makes it almost unavoidable to consider situations
where the ``quantum impurity'' is connected to an  electron gas for
which finite size effects are important.

As a consequence, for each such mesoscopic electron reservoir
connected to the quantum impurity, two new energy scales enter into
the description of the Kondo problem: the corresponding mean level
spacing $\Delta_R$ and Thouless energy $\Eth$.  The existence of
a finite mean level spacing of the electron reservoir will, for
instance, clearly modify the Kondo physics drastically for low
temperature $T \ll \Delta_R$.  This will affect the conductance
\cite{Thimm99,Simon02,Cornaglia03} as well as thermodynamic properties
\cite{Cornaglia02a,Franzese03,KaulEPL05}, and considerable insight can be
gained by considering the properties of the ground states and first
few excited states of the system \cite{KaulPRL06,Kaul08}.

The range of energy between $\Delta_R$ and $\Eth$ is furthermore
characterized by the presence of mesoscopic fluctuations (at all scales
in this range) in the local density of states of the reservoir's
electron, and thus by the fact that the density of states is not flat
and featureless.  In particular, one may wonder whether a (eventually
fluctuating) Kondo temperature can be defined, and if physical
quantities remain universal function of the ratio $T/T_K$.  

To fix the ideas, let us consider the $s$-$d$ Hamiltonian
(\ref{eq:KondoHam1}) with a local density of states at the
impurity site $ \nu(\bfr \smeq 0;\energy) \df \sum_i |\varphi_i(0)|^2
\delta(\epsilon - \epsilon_i) $.  In the semiclassical regime,
$\nu(\bfr \smeq 0;\energy)$ can,  be
written as the sum
\begin{equation}
 \nu(\energy) = \nu_0 + \nu_{\rm fl}(\energy)
\end{equation}
where $\nu_0$ is the bulk-like contribution 
(\ref{eq:nu-0}) [one should of course include here either  a realistic
band dispersion relation or a cutoff at $\mathrm{D}_0$ to account for the finite
bandwidth], and the fluctuating term $\nu_{\rm fl}(\energy)$ is a
quantum correction associated with the  interfering  closed orbit
contributions (\ref{eq:nu-osc}).  Because $\nu_{\rm
  fl}(\energy)$ is the sum of rapidly oscillating terms, it will
fluctuate not only with respect to the energy $\energy$, but also
with respect to the position $\bfr$ of the quantum impurity or with
respect to the variation of any external parameter that may affect the
classical actions $S_j$ (\ref{eq:Sj}) on the scale $\hbar$.  As a
consequence, one may think of $\nu_{\rm   fl}(\energy)$ as a
statistical quantity with different realizations corresponding to
various location of $\bfr$ or obtained by varying a external parameter
in such a way that the classical dynamics remains unmodified, but that
the phases $\exp(i S_j /\hbar)$ are randomised.

Let us denote $T_K^0$ the Kondo temperature of the associated bulk system
for which $\nu(\energy)$ is replaced by $\nu_0$.  For $T \gg T_K^0$
it is possible to use a perturbative renormalization group approach in
the same way as in the bulk, but including the mesoscopic fluctuations
of the density of states.  Following 
\citeasnoun{Zarand96}, this gives in the one-loop approximation
\begin{equation} \label{eq:RGmeso}
J_{\rm eff}(\mathrm{D}_{\rm eff}) = \frac{J_0}{1-J_0
  \int_{\mathrm{D}_{\rm eff}}^{\mathrm{D}_0} 
  ({d\omega}/{\omega}) \, \nubeta(\omega) } \; ,  
\end{equation} 
with
\begin{equation} \label{eq:nubeta}
  \nubeta(\omega) = \frac{\omega}{\pi} \int_{-\infty}^{\infty} d\energy 
  \frac{\nu_{\rm loc}(\energy)}{\omega^2 + \energy^{2}}, 
\end{equation}
the temperature smoothed density of states (note the renormalization
up to order two loops  was given by \citeasnoun**{Zarand96}).

There are two different ways to use the above renormalization group
equation.  For some physical realizations, the fluctuations of the
local density of states may be very significant, yielding even larger
variation of the Kondo properties because of the 
exponential  dependence in (\ref{eq:TKbulk}).  
In that case, one
is mainly interested in the fluctuations of the Kondo temperature
(which is now a functional of the local density of states) defined as
the {\em energy scale} separating the weak and strong coupling
regimes.  One can then use the same approach as in the bulk
and define $T_K [\nubeta]$ as the temperature at
which the one-loop effective interaction diverges, giving the implicit
equation
\begin{equation} \label{eq:TKmeso}
J_0 \int^{\mathrm{D}_0}_{T^*_K[\nubeta]} \frac{d\omega}{\omega}\nubeta(\omega)
    = 1  \; .
\end{equation}
Examples of systems for which the fluctuations of the density are
large enough that one is mainly interested in the fluctuation of the
scale $T_K^*$ defined by (\ref{eq:TKmeso}) include, for instance,
the case of ``real'' (chemical) impurities in a geometry such than one
dimension is not much larger than the Fermi wavelength. This may be
either a quantum point contact in a two dimensional electron gas
\cite{Zarand96} or a thin three dimensional film \cite{Crepieux00}. In
that case there is a large probability that the impurity is located at
a place at which the Friedel oscillations (i.e. the term in the
oscillating part of the density of states associated to the trajectory
bouncing back on the boundary and returning directly to its starting
point) are large.  In that case, the impurities located at a distance
from the boundary such that the interference of the returning orbit
are constructive have a significantly larger Kondo temperature.
Disordered metals near or beyond the localisation transition provide
another kind of systems with large fluctuations of the local density of
states, and can lead in this particular case to a finite density of
``free moments'' for which the $T^*_K[\nubeta]$ corresponding to the
space location is actually zero \cite{Kettemann07,Zhuravlev08}.

In a typical ballistic or disordered-metallic mesoscopic system with
all dimension much larger than the Fermi wavelength, the fluctuating
part of the local density of states $\nu_{\rm fl}$ is a quantum
correction to the secular term $\nu_0$ and is therefore
parametrically smaller.  As a consequence the fluctuations of $T_K
[\nubeta(\omega)]$ defined by (\ref{eq:TKmeso}) are not
very large compared to $T_K^0$, which if it is taken only as an energy
scale is somewhat meaningless.

In the bulk case, $T_k$ has however, beyond being an energy scale,
another meaning which is to be the parameter entering into universal
functions such as $f_\chi$ in (\ref{eq:fchi}).  Within this
framework, $T_k$ is directly (and quantitatively) related to physical
observables, and its fluctuations need not be large to be relevant.

In the mesoscopic case, however, there is a priori no reason for any
physical observable to be a universal function, since the number of
``parameters'' defining the problem can be considered as infinite (in
the sense that one needs an infinite number of parameters to define the
function $\nu(\energy)$). For a given temperature $T > T_K^0$
however, it is possible to perform the perturbative renormalization
group analysis leading to the effective interaction strength $J_{\rm
  eff}(T)$, (\ref{eq:RGmeso}), which amounts to integrating out all
degrees of freedom corresponding to energies larger than $T$.  One can assume
moreover that the features of the density of states at energies smaller
than $T$ will not affect the observables at temperature $T$.  As a
consequence, after renormalization, physical quantities can depend only
on the parameters $T$, $J_{\rm eff}(T)$, and $\nubeta(T)$.  
Dimensional analysis then implies that dimensionless quantities such
as $T \chi_{\rm loc}(T)$ can depend only on one single parameter, the product
${\cal J}(T) = \nubeta(T) J_{\rm eff} (T)$.  In particular
 $T\chi_{\rm loc}(T)$ should be equal for an arbitrary mesoscopic 
realization and for some bulk system with, at temperature $T$, the same
value of the parameter ${\cal J}(T)$.  This implies that if one defines
the {\em  realization and  temperature dependent} Kondo temperature
$T_K[\nubeta](T)$ by the generalisation of (\ref{eq:Jrho}) to
the mesoscopic case, namely
\begin{equation} \label{eq:Jrho-meso}
T_K[\nubeta](T) = \frac{T}{\exp(J_{\rm   eff}(T) \nubeta(T))} \; ,
\end{equation}
one obtains, within the one-loop approximation, a quantitative
prediction for physical quantities in the mesoscopic case.  For
instance one has for the local susceptibility
\begin{equation} \label{eq:chi_th}
T\chi_{\rm loc}(T) = f_\chi(T/T_K[\nubeta](T)) =
f_\chi(1/\exp(J_{\rm   eff}(T) \nubeta(T)).
\end{equation}
which relates the universal function $f_\chi$, {\em computed for the
  bulk} to the impurity susceptibility in the mesoscopic case.

From a practical point of view however, one may first note that with
the definition (\ref{eq:Jrho-meso}), $T_K[\nubeta] (T_K^0) =
T^*_K[\nubeta]$ is the same as the one obtained from
(\ref{eq:TKmeso}).  Furthermore, one finds that
there is in general little difference between $f_\chi(T/T_K[\nubeta](T))$ and
$f_\chi(T/T^*_K[\nubeta])$ in a large range of temperature above
$T_K^0$.  Therefore, although only the form (\ref{eq:chi_th}) can
be formally justified, I will continue the discussion assuming that in
the high temperature regime $T \ge T_K^0$, the susceptibility can be
described by the universal form $f_\chi(T/T^*_K)$ with a realization
(but not temperature) dependent $T^*_K$ defined by
(\ref{eq:TKmeso}).

It turns out that this approach gives a very precise prediction,
provided $T_K^0$ is computed from a fit or the two-loop
approximation, and one uses $T^*_K = T_K^0 + \delta T_K$ with $\delta
T_K$ defined by
\begin{equation} \label{eq:deltaTK}
J_0 \int_{T_K^0 + {\delta T_K}}^{\mathrm{D}_0} \frac{d\omega}{\omega} (\nu_0 +
{\delta \nubeta}(\omega))   
=
J_0 \int_{T_K^0}^{\mathrm{D}_0}  \frac{d\omega}{\omega} \nu_0
   \; .
\end{equation}
\citeasnoun{KaulEPL05} and \citeasnoun{Yoo05}  showed, by comparison with
exact numerical quantum Monte Carlo calculations, that, even if one
neglects the temperature variation of $T_K[\nubeta]$, the
predictions from (\ref{eq:chi_th}) are quantitatively extremely
precise, even up to temperatures somewhat below $T_K^0$.  This approach
therefore provides a simple and quantitative way to discuss the
mesoscopic fluctuations of the Kondo properties in the temperature
regime $ T \geq T_K^0$.

The nice thing here is that in this form, all the fluctuations of
physical quantities in this regime are expressed in terms of the local
density of states that we know how to relate to classical closed
orbits.  Indeed, (\ref{eq:deltaTK}) can be rewritten for $\delta
T_K \ll T_K^0$ as 
\begin{equation} \label{elabet}
\frac{\delta T_K }{ T_K^0} = 
\nu_0^{-1} \int_{T_K^0} \frac{d \omega}{\omega} \nubeta(\omega)\; .
\end{equation}
Furthermore using that $\nubeta(\omega) = \nu_{\rm
  loc}(\energy_F + i \omega)$ and the analytical continuation of
(\ref{eq:1/2classGreen}) for complex energies in the same way as
in (\ref{eq:Green_ie}), one can immediately relate the realization
dependant Kondo temperature $T_K^*[\nubeta]$ to a sum over orbits
starting and ending at the impurity site $\bfr_0 = 0$
\begin{equation}
\fl
\frac{\delta T_K}{T_K^0} \simeq \frac{1}{\sqrt{2\pi^3\hbar^3} \, \nu_0}
\sum_{j:\bfr_0 \to \bfr_0} D_j \sin \left[ \frac{1}{\hbar} S_j(\energy) -  \maslov_j
 \frac{\pi}{2} + \frac{\pi}{4} \right] 
\int_{T_K^0}^\infty \frac{d\omega}{\omega} \exp(-t_j \omega/\hbar) \; .
\end{equation}
The last integral can be approximated by $\log\left( \frac{T_K^0
    t_j}{\hbar} \right) \exp(-t_j T_K^0/\hbar)$, thus providing a
cutoff for trajectories with a time of return $t_j$ larger than
$\hbar/T_K^0$.  From this result, the variance of the Kondo
temperature can then be obtained.  Pairing a trajectory only with
itself and, if the electron gas  is time-reversal invariant, with its
time-reversal  symmetric, and using
the sum rule (\ref{eq:Mformula}), one has
\begin{equation}
\frac{\langle {\delta T_{K}}^2  \rangle}{{{T}_K^0}^2} \simeq 
\frac{\betaRMT}{\pi \hbar \nu_0} \int_{T_K^0}^{D_{\rm cut}} \frac{d
  \omega'_1}{\omega'_1}  
\frac{d \omega'_2}{\omega'_2} \int_0^\infty dt P^\energy_{\rm cl}(t) 
e^{- ( \omega'_1 + \omega'_2) t/\hbar}  \; , 
\end{equation}
with $P^\energy_{\rm cl}(t)$ the classical probability of return and $\betaRMT =
1$ for a time-reversal non-invariant and $2$ for a time-reversal invariant
electron gas.

For chaotic systems, using the model
(\ref{eq:PclChaosa})-(\ref{eq:PclChaosb}) for the classical
probability of return then immediately gives
\begin{equation} \label{eq:varTkchaotic}
\frac{\langle {\delta T_{K}}^2  \rangle}{{{T}_K^0}^2} \simeq 
\frac{( 2 \betaRMT \ln 2)}{\pi} \frac{\Delta}{{T}_K^0} \; .
\end{equation}
Interestingly, the same result can be derived from a random-matrix
description 
\cite{Kettemann04,KaulEPL05}.  Indeed, using (\ref{elabet}), the
variance of $\delta T_K$ can be expressed in terms of the correlator
\begin{eqnarray} 
\fl R_2(\omega_1,\omega_2) & \df & \frac{1}{\nu_0^{2}} \int_{\omega_1} 
\int_{\omega_2} d \omega'_1 d \omega'_2  
\frac{ \langle \nubeta(\omega'_1)\nubeta(\omega'_2) \rangle }
{\omega'_1 \omega'_2} \\
\fl & = & \frac{1}{\pi^2 \nu_0^{2}} \int_{\omega_1} 
\int_{\omega_2} d \omega'_1 d \omega'_2  
\sum_{\alpha_1 \alpha_2} \frac{ \langle \left(|\varphi_{\alpha_1}|^2 -
    1/\Volume \right)  \left(|\varphi_{\alpha_2}|^2 -
    1/\Volume \right)\rangle }
{\left[(\mu-\energy_{\alpha_1})^2 + {\omega'_1}^2 \right]
\left[(\mu-\energy_{\alpha_2})^2 + {\omega'_2}^2 \right]}
\; .
\end{eqnarray} 
Within the random-matrix model, the wave-functions have a
Porter-Thomas distribution ((\ref{eq:Porter_Thomas_GOE}) or
(\ref{eq:Porter_Thomas_GUE})), and, neglecting the correlations between
eigenstates, which introduce only $1/g$ corrections, one has $\langle
\left(|\varphi_{\alpha_1}|^2 - 1/\Volume \right)
\left(|\varphi_{\alpha_2}|^2 - 1/\Volume \right)\rangle =
\delta_{\alpha_1\alpha_2} (\betaRMT/\Volume^2) $.  Neglecting the
fluctuations of the eigenergies $\energy_\alpha$ and performing the
integral, one then gets
\[
R_2(\omega_1,\omega_2) = \frac{\betaRMT \Delta}{\pi} \left(
\frac{1}{\omega_2} \ln\frac{\omega_1+\omega_2}{\omega_1} +
\frac{1}{\omega_1} \ln\frac{\omega_1+\omega_2}{\omega_2} \right) \; ,
\]
and thus $R_2({T}_K^0,{T}_K^0) = 2 \ln 2 (\betaRMT\Delta /
\pi{T}_K^0 )$, from which (\ref{eq:varTkchaotic}) derives
immediately \cite{KaulEPL05}. Within the random-matrix description,
the full distribution of $T_K$ has been obtained by 
\citeasnoun{Kettemann04}.

To conclude this section, we see that the mesoscopic fluctuations of
Kondo properties provide an example of physical systems where a
relatively non trivial information is eventually encoded in the
classical trajectories.  Moreover through the sum rule
(\ref{eq:Mformula}) and the model
(\ref{eq:PclChaosa})-(\ref{eq:PclChaosb}) for the classical
probability of return $P^\energy_{\rm cl}(t)$,  a connection with random-matrix
theory can be made.

It should be stressed however that the most interesting aspect of this
mesoscopic Kondo problem, namely the fluctuations of physical
properties in the deep Kondo regime ($\Delta <$) $T \ll T_K^0$, is
still an open problem.  This regime should be characterized by much
larger fluctuations, which should be therefore easier to observe
experimentally, as well, as has already been seen in exact numerical
Monte Carlo calculations \cite{KaulEPL05}, as a lack of universality.
The perturbative renormalization group point of view used in the high
temperature regime will clearly not be applicable at those
temperatures, but a Fermi-liquid descriptions based on mean-field
slave-boson techniques should nevertheless be able to shed some light
on the physics dominating the mesoscopic fluctuations in this regime
\cite{Burdin08}.

\setcounter{footnote}{1}

\section{Coulomb blockade peak spacing and ground-state spin of
  ballistic quantum dots}
\label{sec:CB}

I will turn now to a third (and last) illustration of physical systems
for which the interplay between interference and interactions plays a
dominant role, namely Coulomb blockade in ballistic quantum dots.
Coulomb Blockade in itself is an essentially classical effect. It can
for instance take place  when a small metallic grain is weakly
connected to electrical contacts and maintained at a temperature low
compared  to  its charging energy $E_c = e^2/C$ ($C$ is the total capacitance
of the grain).  In that case, in the lowest order in the grain-leads
coupling, the conductance through the grain can be seen as a
succession of transitions between states with different number of
electrons $N$ within the grain.  Because of the Coulomb interaction
between the electrons, these various states will have an energy
difference of order $E_c$, and in general the conservation of the
total energy cannot be fulfilled. As a consequence, the transport
through the grain is blocked. If, however, the grain is capacitively
coupled to an external gate, this latter will also affect in a
different way the energy of the states with a different number of
electrons.  The potential of the gate, $V_g$, can therefore be tuned
so as to adjust the energies of the various states.  Thus, as a
function of $V_g$, the conductance through the grain will display an
alternation of valleys and peaks (for  reviews see
\cite{Grabert&DevoretBook,Kouwenhoven97Article}).

The Coulomb-blockade process described above does not imply a quantum
mechanical effect, and can in particular be observed if the mean-level
spacing, or even the Thouless energy, is much smaller than the
temperature.  Using very small systems, such as ballistic quantum dots
built in GaAs/AlGaAs, one can nevertheless increase sufficiently the
one-particle mean-level spacing $\Delta$ so that $\Delta > T$, and the
transport in the dot takes place through a single (or a few) levels.
In this regime, one observe fluctuations in both spacings and 
 heights of the conductance peaks,  fluctuations which encode non-trivial
information about the many-body ground-states (with various particle
numbers), and possibly a few excited states.

\subsection{Constant-interaction model and experimental distributions}

\label{subsec:CIM}

As far as the peak-height fluctuations are concerned, predictions
\cite{Jalabert92} based on the so called ``constant-interaction
model'' appeared to fit the distributions measured in the earlier set
of experiments \cite{Chang96,Folk96}.  In this model, beyond a
classical charging term $(Ne)^2/2C$, interactions among the electrons
are completely neglected, and Porter-Thomas (i.e.\ RMT-like)
fluctuations are assumed for the one-particle wave functions. Even at
this early stage, however, the presence of correlations between
successive peak heights appeared incompatible with a simple RMT
description, and pointed to the role of short periodic orbits
\cite{Narimanov99,Kaplan00,Narimanov01}.  Furthermore, more recent
experiments \cite**{Patel98b} showed decreased probability of having
either very large or very small heights.  Several suggestions have
been put forward to interpret these deviations, among which the effect
of inelastic scattering \cite{Rupp02}, of spin orbit \cite{Held03}, or
of interactions \cite{Usaj03}.  It remains that for peak-height
distributions, the constant-interaction model seems to capture a good
part of the relevant physics.

The status of the peak-spacing fluctuations, however, is very different.
For this quantity, the constant-interaction model gives very striking
predictions, since, because of the spin degeneracy, the alternation
between singly and doubly occupied orbitals, as the number of electrons
$N$ in the dot increases, is associated with a strongly bimodal
distribution  \cite{Sivan96}.  Indeed, in the zero temperature limit, the
position $V_g^*$ of a conduction peak is determined by the energy conservation
condition
\begin{equation}
  {\cal E}^N (V_g^*) + \mu = {\cal E}^{N+1} (V_g^*)
\end{equation}
($\mu$ is the chemical potential in the leads). Writing ${\cal E}^N
(V_g)$,  the ground-state 
energy of the dot with $N$ electrons, 
as the sum of some ``intrinsic'' part $E^N_0$ plus a term $- (C_g/C) eN V_g $ due
to the capacitive coupling with the control gate ($C_g$ is the
capacitance of the gate to the dot), one  obtains that the
spacing in $V_g$ between two successive peaks is proportional to the
second (discrete) energy difference  
\begin{equation} \label{eq:D2E}
  (V_g^*)_{N \to N+1} - (V_g^*)_{N-1 \to N } \propto \delta^{2} E^N_0 \df
  E^{N+1}_0 + E^{N-1}_0 - 2 E^{N}_0 \; . 
\end{equation}

In the constant-interaction model, the ground-state energy of the dot
with $N$ electrons is written as
\begin{equation} \label{eq:CIM}
  { E}^N_0 = \frac {(eN)^2}{2C} + \sum_{{\rm occupied} \; i\sigma}
  \energy_i \; ,
\end{equation}
where the last term is simply the one-particle energy of a system of
non-interacting fermions ($i$ and $\sigma$ are respectively the
orbital and spin index and $\energy_i$ the corresponding one-particle
energy).  One obvious prediction of this model is that the ground
state spin of the dot can be only $0$ (for $N$ even) or $1/2$ (for $N$
odd).  Furthermore, because of the spin degeneracy, one gets
\begin{eqnarray} 
  \delta^{2} E^N_0 = e^2/C && \mbox{for odd $N$} \; , \label{eq:CIpspa}\\
  \delta^{2} E^N_0 = e^2/C + (\energy_{N/2+1} - \energy_{N/2}) &&
  \mbox{for even $N$} \; ,\label{eq:CIpspb}
\end{eqnarray}
and the peak spacing is the superposition of a Dirac delta function
(corresponding to odd $N$, referred below as ``odd spacings'') and of
the nearest neighbor distribution $P_{\rm nns}(s)$ of the $\energy_i$
(corresponding to even $N$, referred below as ``even spacings''). For a
chaotic system, this is described by random-matrix theory, giving
(\ref{eq:Wigner_surmise}).

The distributions observed experimentally share very little
resemblance with this prediction.  The first set of experimental
results \cite{Sivan96,Simmel97,Simmel99} not only did not show any
trace of bimodality, but the width of the distributions seemed on a
scale of the charging energy rather than on the one of the mean-level
spacing (as $P_{\rm nns}(s)$ was expected to be).  It was soon
realized that these very large distributions were dominated by
switching events, that is, by the displacement of trapped charges located
between the quantum dot and the control gate.  Further experiments
\cite**{Patel98a,Luscher01,OngBHPM01} where then performed which took
care of maintaining  the level of the noise associated to
switching events well below the mean-level spacing.

\begin{figure} 
\begin{center}
\includegraphics[totalheight=8cm]{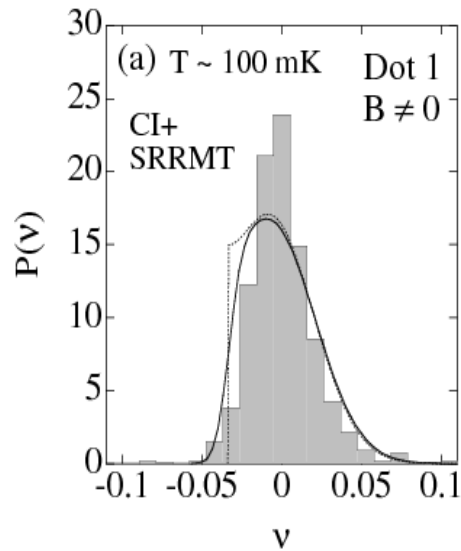}
\end{center}
\caption{Histogram of the peak-spacing data for the dot 1 of
  \citeasnoun**{Patel98a}.  The noise level and temperature correspond
  respectively to $8\%$ and $20\%$ of the mean-level spacing $\Delta$.
  The electronic density $n_s = 2.\,10^{11} \, cm^{-2}$ correspond to a
  gas parameter $r_s \simeq 1.24$.
 This figure is taken from \citeasnoun**{Patel98a}.  The dotted
 and solid lines corresponds to predictions from the constant
 interaction  plus spin resolved RMT model (CI+SRRMT) presented in this
reference.}
\label{fig:patel}
\end{figure}

In the work of \citeasnoun**{Patel98a}, seven dots of
different sizes (and two different densities) have been investigated
for finite as well as zero magnetic field. Measures were taken at an
electron temperature of $100\,mK$, which corresponded to a ratio
$T/\Delta$ ranging from $20\%$ for the smallest dot (largest $\Delta$)
to $60\%$ for the largest dot (smallest $\Delta$).  The noise level,
estimated by comparing the data when reversing the sign of the
magnetic field, ranged from $8\%$ to $94\%$ of the mean-level spacing.
The peak-spacing distribution obtained for the smallest dot, which
turns out to be the one showing the less noise ($T/\Delta =20\%$,
noise $=8\%$ of $\Delta$), is reproduced in figure~\ref{fig:patel}.  As seen
there, this distribution remains essentially uni-modal and Gaussian,
and therefore clearly incompatible with the constant-interaction model
predictions.

\begin{figure} 
\begin{center}
\includegraphics[totalheight=8cm]{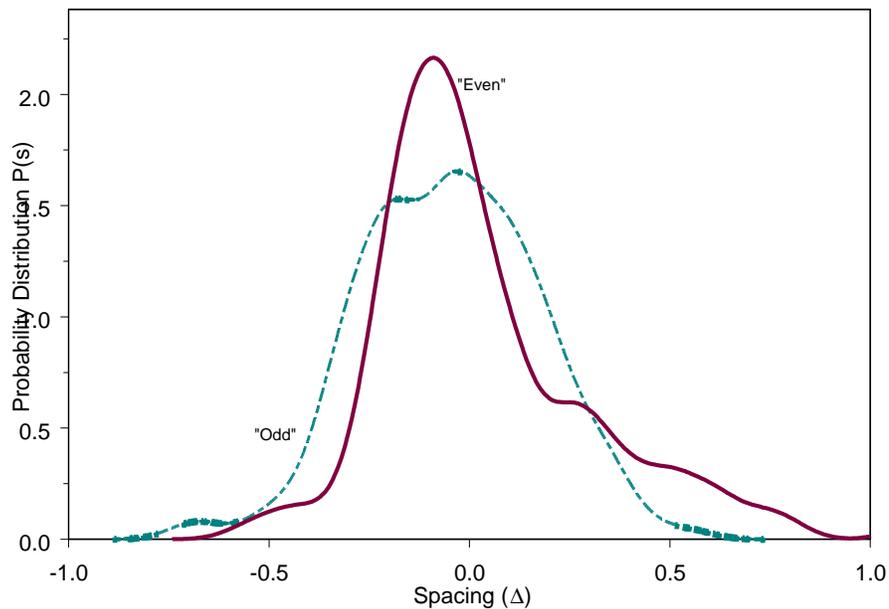}
\end{center}
\caption{Probability distributions of even (solid line) and odd (dash
  line) peak spacings.  The  data correspond to the  dot 1 of
  \citeasnoun**{Patel98a},  but for a different gate configuration which is
  such that, in spite of the slightly lower temperature $T = 90mk$,
  the ratio $T/\Delta$ is about $30\%$.  Some visible difference is
  seen between the odd and even distributions, but these latter clearly do not
  correspond to the constant-interaction model predictions
  (\ref{eq:CIpspa})-(\ref{eq:CIpspb}). 
 This figure is taken from \citeasnoun{OngBHPM01}.}
\label{fig:ong}
\end{figure}

Data from the same series of dots have been re-analysed by
\citeasnoun{OngBHPM01}.  In this latter work, special emphasis has
been put on the study of the difference between odd and even spacings
predicted by the constant-interaction model.  Some odd-even effect
could actually been observed for the dots exhibiting the least amount
of noise and the smallest $T/\Delta$ ratio. This is illustrated in
figure~\ref{fig:ong} which shows separately the distributions of odd
and even peak-spacings for the same dot as in figure~\ref{fig:patel}, but with
a slightly different gate configuration so that the effective area and
thus the ratio $T/\Delta$ are somewhat larger ($T \simeq 30\%
\Delta)$.  In that case, some differences are seen between the odd and
even distributions.  They remain however clearly incompatible with the
constant-interaction model predictions
(\ref{eq:CIpspa})-(\ref{eq:CIpspb}) . In particular, the total
peak-spacing distribution that would be obtained by summing the two
curves would, as the one shown in figure~\ref{fig:patel}, be uni-modal
and essentially Gaussian.

\begin{figure} 
\begin{center}
\includegraphics[totalheight=8cm]{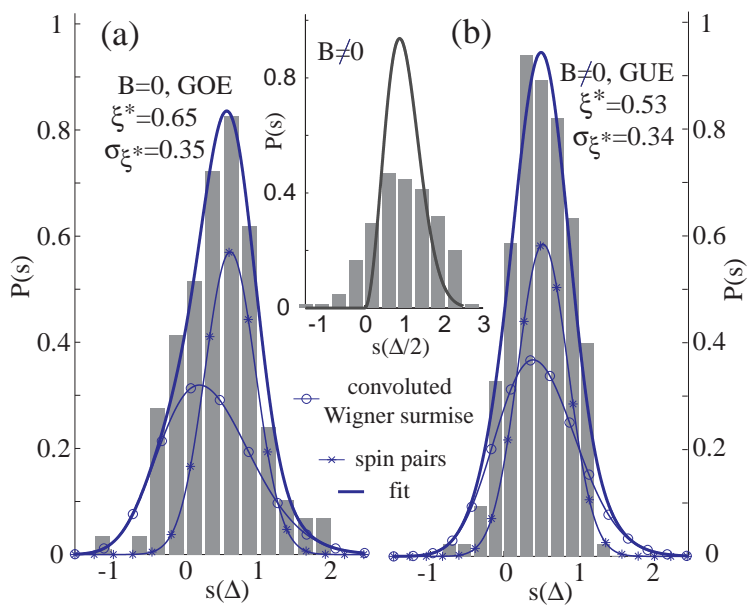}
\end{center}
\caption{Histogram of the peak-spacing distribution for $B=0$ (a) and
  $B \neq 0$ (b) for the quantum dot studied by
  \citeasnoun**{Luscher01}.  The various solid curves are fit to
  theoretical predictions discussed in this reference. Inset: the
  histogram is the same as (b), but with a different scale.  This
  figure is taken from \citeasnoun{Luscher01}.}
\label{fig:luscher}
\end{figure}

The same kind of distributions were obtained by
\citeasnoun**{Luscher01} for a dot defined by local oxidation with an
atomic-force microscope (rather than electrostatic gating as in
\citeasnoun**{Patel98a}).  The steepest confining potential obtained in this
way made it possible: i) to limit the shape variation of the dot as
the lateral gate voltage $V_g$ is changed; ii) to use a back gate to
increase the electronic density ($n_s \simeq 5.9 \, 10^{11} \,
cm^{-2}$) and thus reduce the gas parameter to a value $r_s \simeq
0.72$; and furthermore to define a smaller quantum dot so that the
electronic temperature $T = 120\,mK$ used amounts to a significantly
smaller ratio $T/\Delta \simeq 5\%$.  As illustrated in
figure~\ref{fig:luscher} the peak-spacing distributions nevertheless
remained uni-modal and essentially Gaussian.

Other experimental evidences that the constant-interaction model does
not properly account for the ground-state properties of quantum dots
can be obtained by considering their total spin.  Indeed, as already
mentioned, this model implies a ``naive'' occupation of the orbitals,
and therefore, for the ground state, a total spin zero for even or one
half for odd number of particles respectively.  Already in
\citeasnoun{Luscher01}, the parametric variation of the
Coulomb-blockade peak position as a function of a weak perpendicular
magnetic field, and in particular the existence of kinks
\cite{Baranger00}, pointed to the existence of ground-states with spin
one.  Such spin-one ground-state were further used to observe Kondo
effect in small dots with even number of particles \cite{Kogan03}.
More systematic studies of ground-state spins were later performed
using stronger in-plane magnetic field coupled to the spin of the
electrons through Zeeman effect \cite{Rokhinson01,Folk01,Lindemann02}.
Not enough data have been collected in this way to obtain an
experimental distribution of ground-state spins.
\citeasnoun**{Folk01} have nevertheless obtained clear evidence of the
existence of non-trivial (i.e.\ different from zero or one half)
ground-state spins for semiconductor quantum dots similar to the ones
used for figures~\ref{fig:patel} and \ref{fig:ong}.

\subsection{The universal Hamiltonian}

\label{subsec:univHam}

These discrepancies between the constant-interaction model predictions
and what was observed pointed clearly to the role played by the
residual interactions (i.e.\ beyond simple charging energy) in these
systems, and motivated a large number of works ranging from exact
numerical calculations on small system \cite{Sivan96,Berkovits98prl}
to various kind of self-consistent approximations \footnote{As for
  instance Hartree Fock \cite{Walker99prl,Walker99prb,Cohen99,Ahn99}
  or density functional theory
  \cite{Stopa93,Stopa96,Koskinen97,Lee98,Hirose99,Hirose02,Jiang03prl,Jiang03prb}.}.
In the end however, it seems that, even if not all aspects of the
experimental data can yet be explained, the main features are
compatible with a relatively simple Fermi-liquid description provided
the role of the spins are properly taken into account
\cite{Blanter97,Brouwer99,Baranger00,Ullmo01prbb,Aleiner02PhysRep}.
Once the consequences of a finite temperature are included
\cite{Usaj01,Usaj02}, the predictions for the distributions of
Coulomb-blockade peak-spacing are actually in very decent agreement
with at least some of the experimental data.

What is meant by ``Fermi-liquid'' description here
is the picture given in section~\ref{sec:screening}: an effective
``mean field'' one-particle Hamiltonian $H_{\rm MF}$ which, for a
specific system could be in principle obtained by minimising the
Thomas-Fermi functional (\ref{eq:FTF}) (see for instance the
discussions given in \citeasnoun{Ullmo01prba} and
\citeasnoun{Ullmo04prb}), and a weak residual interaction well
approximated by the RPA-screened Coulomb interaction, or even its long
wave length zero frequency limit (\ref{eq:Vq2d})-(\ref{eq:Vq3d}).
As pointed out in section~\ref{sec:screening} there is up to now, for
a generic system, no real derivation of this picture from basic
principles.  The point here is thus not to define an approximation
scheme to predict quantitatively the energies of a few specific
realizations, but rather to have a model which, once supplemented by some
modelling for the fluctuations of the one-particle eigenstates $\varphi_i$
and energies $\energy_i$, typically RMT for chaotic systems, gives
correct predictions for statistical properties such as the peak
spacing distribution, etc..

\subsubsection{Time-reversal non-invariant systems}
\label{subsec:TRNI}

As for orbital magnetism (see section~\ref{subsec:correlations}), time
reversal invariance introduces a slight complication in the discussion
because of the Cooper series for which, at $r_s \simeq 1$, high-order
terms need to be included and eventually renormalise the effects of
the interactions.  I shall therefore start the discussion 
assuming  the presence of a magnetic field strong enough to break time
reversal invariance, but nevertheless small enough not to change
qualitatively the classical dynamics within the system (and in
particular so that it remains in the chaotic regime).  In that case,
the screened interaction can be treated at the first order of the
perturbation.

In the limit where the residual interactions are neglected, the
many-body eigenstates are Slater determinants characterized by the
occupation numbers $n_{i\sigma} = 0,1$ of the mean-field Hamiltonian's
wave-functions $\varphi_{i}$ .  For the states such that all singly
occupied orbital have the same spin polarisation (which as we shall
see are the only ones that can be the ground-state of the system), the
usual non-degenerate perturbation theory can be used: in first order
the eigenstates are unmodified and have an energy \cite{Ullmo01prbb}
\begin{eqnarray}
E\{n_{i\sigma}\} & = & E_{\rm sm}(N) + \sum_{i\sigma} n_{i\sigma} \energy_i +
E^{\rm RI}\{n_{i\sigma}\} \label{eq:1rstorder} \\
E^{\rm RI}\{n_{i\sigma}\}   & = & 
\frac{1}{2} \sum_{i\sigma,j\sigma'}  n_{i\sigma} \mathrm{M}_{ij} n_{j\sigma'}
- \frac{1}{2} \sum_{i,j,\sigma}  n_{i\sigma} \mathrm{N}_{ij} n_{j\sigma} \;
. \label{eq:RI} 
\end{eqnarray}
Here $E_{\rm sm}(N)$ is a smooth contribution containing essentially
the electrostatic energy $(Ne)^2/2C$, so that the first two terms of
(\ref{eq:1rstorder})
correspond to  the constant-interaction model.  The residual
interaction term $E^{\rm RI}\{n_{i\sigma}\}$ is then expressed in terms of
\numparts
\begin{equation} \label{eq:Mij}
\mathrm{M}_{ij} = \int d\bfr d\bfr' |\varphi_i(\bfr)|^2 \Vsc(\bfr-\bfr')
|\varphi_j(\bfr')|^2 \; ,
\end{equation}
\begin{equation} \label{eq:Nij}
\mathrm{N}_{ij} = \int d\bfr d\bfr' \varphi_i(\bfr) \varphi_j^*(\bfr) \Vsc(\bfr-\bfr')
\varphi_j(\bfr') \varphi_i^*(\bfr') \; .
\end{equation}
\endnumparts

We shall see below (see section~\ref{subsec:beyond_univ_Ham}) that the
fluctuations of the $\mathrm{M}_{ij}$ and $\mathrm{N}_{ij}$ are
parametrically smaller than their mean values $ \langle
\mathrm{M}_{ij} \rangle $ and $ \langle \mathrm{N}_{ij} \rangle $.  We
shall therefore for a moment neglect the former.  The quantities
$\langle \mathrm{M}_{ij} \rangle $ and $ \langle \mathrm{N}_{ij}
\rangle $ can be derived by different methods, in particular from the
knowledge of the two-point correlation function
(\ref{eq:correlationsa})  and a Gaussian
hypothesis for the higher-order correlation functions
\cite{Srednicki96,Hortikar98prl}.  This can also be done directly from
the random-plane-wave model.  For instance, for a billiard system with
only a kinetic energy term in a region of space of volume (or area)
$\Volume$, the insertion of (\ref{eq:RPW}) in (\ref{eq:Mij}) gives
\begin{equation}
\fl \langle \mathrm{M}_{ij}\rangle   =   \int d  \bfr d \bfr'
\sum_{\mu,\mu',\eta,\eta'} \langle a_{i\mu} a^*_{i\mu'} a_{j\eta}
a^*_{i\eta'} \rangle  \exp \left(\frac{i}{\hbar} [\bfr(\bp_{i\mu} -
  \bp_{i\mu'}) + \bfr'(\bp_{j\mu} - \bp_{j\mu'})] \right) \; .
\end{equation}
The Gaussian character of the $a_{i\mu}$ and (\ref{eq:aimu})
imply $\langle a_{i\mu} a^*_{i\mu} a_{j\eta} a^*_{j\eta'} \rangle =
\langle a_{i\mu} a^*_{i\mu'}\rangle \langle a_{j\eta} a^*_{j\eta'}
\rangle +\delta_{ij} \langle a_{i\mu} a^*_{i\eta'} \rangle \langle
a_{i\eta} a^*_{i\mu'}\rangle = ({\Volume}M)^{-1} (\delta_{\mu\mu'}
\delta_{\eta\eta'} + \delta_{ij} \delta_{\mu\eta'}
\delta_{\eta\mu'})$. One obtains in this way 
\begin{eqnarray}
\langle \mathrm{M}_{ij} \rangle & = & 
\langle \mathrm{M}_{i \neq j} \rangle + \delta_{ij} \, \delta \langle \mathrm{M}_{ii}
\rangle \; , \nonumber \\
\langle \mathrm{M}_{i \neq j}\rangle  & = &   \frac{1}{\Volume} \hV(0) \\
\delta  \langle  \mathrm{M}_{ii} \rangle 
 & = &   \frac{1}{\Volume} \langle \hV \rangle_{\rm f.s.}
\end{eqnarray}
where $\langle \hV \rangle_{\rm f.s.}$ is the average of $\hV(\bq)$ for
$\bq = \bp-\bp'$, $\bp$ and $\bp'$ being uniformly distributed on the
Fermi surface.  For two and three dimensions:
\begin{eqnarray}
\langle \hV \rangle_{\rm f.s.} & = & \frac{1}{2\pi} \int d\theta 
\hV \left( \kf \sqrt{2(1+\cos \theta)} \right) \qquad \mbox{($d$=2)}
\; , \\
 & = & \frac{1}{4\pi} \int \sin(\theta) d\theta d\varphi
\hV\left( \kf \sqrt{2(1+\cos\theta)} \right) \qquad \mbox{($d$=3)} \; .
\end{eqnarray}
 A direct calculation gives similarly
\begin{equation}
\langle \mathrm{N}_{i \neq j} \rangle =   \delta  \langle
\mathrm{M}_{ii}\rangle =  
\frac{1}{\Volume} \langle \hV \rangle_{\rm f.s.}  \; .
\end{equation}
The equality between $\langle \mathrm{N}_{i \neq j} \rangle$ and
$\delta \langle \mathrm{M}_{ii}\rangle $, which may seem somewhat
surprising at first sight, can be understood as taking root in the
invariance of the residual-interaction energy under a change in the
spin quantization axis.  Finally $\mathrm{N}_{ii} \! = \! \mathrm{M}_{ii}$
and thus compensates the corresponding $\sigma=\sigma'$ contribution.
 
\begin{figure} 
\begin{center}
\includegraphics[totalheight=8cm]{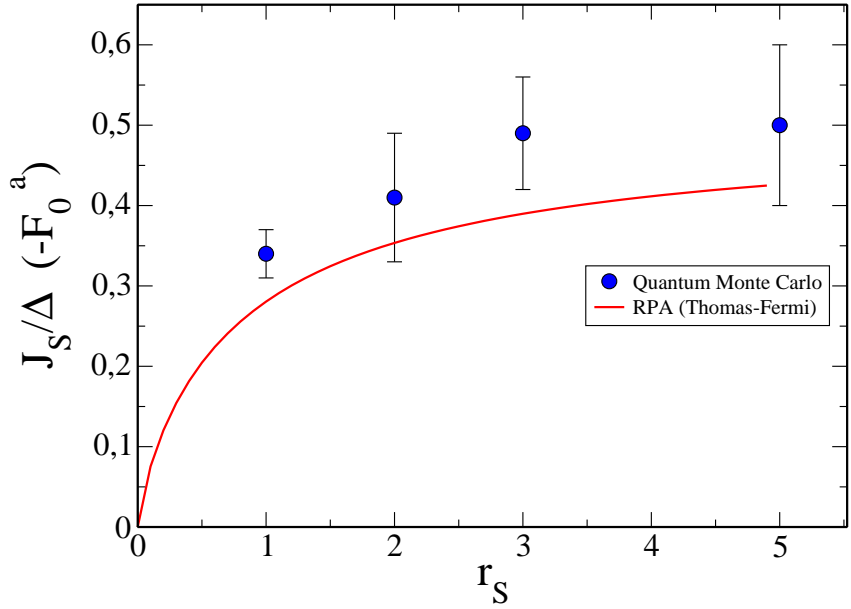}
\end{center}
\caption{Comparison between the value $J_{\rm RPA}$ obtained from the
  RPA (actually Thomas-Fermi) expression (\ref{eq:Vq2d}) and the
  numerical Monte-Carlo evaluation of $f^0_a$ from
  \citeasnoun*{Tanatar89}. (Courtesy of Gonzalo Usaj.)}
\label{fig:foa}
\end{figure}

Using the approximation (\ref{eq:Vq2d})-(\ref{eq:Vq3d}) for the
screened interaction one gets
\begin{eqnarray}
  \frac{1}{\Volume} \hV(0) & = & \frac{\Delta}{\mathrm{g_s}} \; , \\
  \frac{1}{\Volume} \langle \hV \rangle_{\rm f.s.} & = & J_{\rm RPA}
  \Delta \; ,
\end{eqnarray} 
where $\Delta$ is the mean-level spacing between the $\energy_i$'s and
$\mathrm{g_s}=2$ the spin degeneracy.  The last equation actually
defines the dimensionless constant $J_{\rm RPA}$, which, according to
(\ref{eq:Vq2d})-(\ref{eq:Vq3d}), is seen to be of order $r_s$ in
the high density ($r_s \to 0$) limit but of order one (although always
smaller than 0.5) as $r_s$ becomes of order one.  More properly
however, $\langle \hV \rangle_{\rm f.s.} / \Delta$ should be
interpreted as the parameter $f^0_a$ of the Fermi-liquid theory
\cite{PinesNozieresVol1}.  This ratio has been computed, as a function
of $r_s$, in a variety of ways.  Figure~\ref{fig:foa} shows for
instance, for two dimensional systems, a comparison between the value
$J_{\rm RPA}$ obtained from the RPA (actually Thomas-Fermi) expression
(\ref{eq:Vq2d}) and the numerical Monte-Carlo evaluation from
\citeasnoun*{Tanatar89}. For $r_s=1$ for instance, we see that the
Monte-Carlo result $\simeq 0.34$ is slightly larger than $\langle \hV
\rangle_{\rm f.s.}/ {\Volume} = 0.28 $ obtained from RPA.

A few remarks are in order here.  First, we note that, as expected, the
mean value of the interaction contribution to the total energy is much
smaller than the mean value of the non-interacting contributions
(whether  the electrostatic term $E_{\rm sm}(N) \simeq (eN)^2/2C$
or the smooth part of the one-particle energy $\sum n_{i\sigma}
\energy_i$).  In the same way, we shall see below that the
fluctuations of the interaction contributions are parametrically smaller
than that of the non-interacting term.  However, {\em the mean
  values} of the $\mathrm{M}_{ij}$'s and $\mathrm{N}_{ij}$'s are on the same scale
($\Delta$) as {\em the fluctuations} of the one-particle energies
$\energy_i$.  In the end, it is the interplay between these
fluctuations of a large term, and the mean value of a smaller one
which makes possible a qualitative change from the
interactions although they can be treated perturbatively.

The second point that should be stressed is that, since the variation
with the orbital indices of the $\mathrm{N}_{ij}$'s and $\mathrm{M}_{ij}$'s is
neglected, the residual-interaction term (\ref{eq:RI}) depends only
on the number of same-spin and different-spin pairs of particles,
which themselves can be expressed in terms of the total number of
particle $N$ and of the difference $(N_+ \! - \! N_-)$ between the
numbers of majority and minority spins.  We consider here only
states such that all singly occupied orbitals have the same spin
polarisation, hence $(N_+ \! - \! N_-)$ is just twice the total spin $S$ of
the system.  Keeping in mind that $\delta \langle  \mathrm{M}_{ii}
\rangle = \langle 
\mathrm{N}_{ij} \rangle$, we get from simple algebra 
\begin{equation}
 \fl E^{\rm RI}(N,S) = \frac{N(N-1)}{2}  \langle \mathrm{M}_{i\neq j} \rangle  - 
  \frac{N(N-4)}{4} 
  \langle  \mathrm{N}_{ij} \rangle/2  - S(S+1)\langle  \mathrm{N}_{ij}
  \rangle \; . 
\end{equation}
The $N$-dependent part can be safely aggregated with the large smooth term 
$E_{\rm sm}(N)$, and it turns our that the residual-interaction term
is in this case proportional to the eigenvalue $S(S+1)$ of the total spin
square $\mathbf{\hat S}_{\rm tot}^2$ of the dot.

As was pointed out by Aleiner and coworkers
\cite{Kurland00,Aleiner02PhysRep} this result is to be expected, and
is just the consequence of the general symmetries of our problem.
Indeed, the description we have used for the eigenstates of the
non-interacting system is, within an energy band centered at the Fermi
energy and of width the Thouless energy, equivalent to a random-matrix
model.  This means that as long as one considers an energy scale
smaller than the Thouless energy (which is assumed here), there is no
preferred direction in the Hilbert space, and any physical quantity
should, after averaging over the ensemble, yield a result which is
invariant under rotation of the Hilbert space.  As a consequence, if
fluctuations are neglected, the residual-interaction Hamiltonian
\begin{equation}
\hat  H_{\rm RI} = \frac{1}{2} \sum_{\fraczero{i,j,k,l }{\sigma
    \sigma'}} V_{ijkl}  
   c^\dagger_{i\sigma} c^\dagger_{j\sigma'} c_{l\sigma'}c_{k\sigma}
\end{equation}
with 
\begin{equation} \label{eq:Vijkl}
   V_{ijkl} \df \int d\bfr d \bfr' \varphi^*_{i}(\bfr) \varphi^*_{j}(\bfr') 
\varphi_{l}(\bfr') \varphi_{k}(\bfr) 
\end{equation}
should have a ``universal form'', i.e.\
should be expressed only in terms of the invariants, which, without
time-reversal symmetry, are the number operator $\hat N$ and the total
spin square $\mathbf{\hat S}_{\rm tot}^2$.

Using the random-plane-wave model we can get an explicit expression of
this universal-Hamiltonian form, including the values of
the various parameters.  Indeed, inserting (\ref{eq:RPW}) in 
(\ref{eq:Vijkl}) one has
\begin{equation} \label{eq:Vijkl_bis}
\fl V_{ijkl} = \sum_{\mu\mu'\nu\nu'}
a^*_{i\mu} a^*_{j\nu} a_{l\nu'} a_{k\mu'} 
\int d\bfr d \bfr' \exp \left(\frac{i}{\hbar} [\bfr\cdot(\bp_{\mu'} - \bp_\mu)
+ \bfr'\cdot(\bp_{\nu'} - \bp_\nu) ] \right) \Vsc(\bfr-\bfr') \; .
\end{equation}
With $\langle a^*_{i\mu} a^*_{j\nu} a_{l\nu'} a_{k\mu'} \rangle = 
\langle a^*_{i\mu}  a_{k\mu'} \rangle \langle a^*_{j\nu} a_{l\nu'}  \rangle
+\langle a^*_{i\mu} a_{l\nu'} \rangle \langle  a^*_{j\nu}  a_{k\mu'} \rangle
= \frac{1}{M \Volume} [\delta_{ik}\delta_{\mu\mu'} \cdot
\delta_{jl}\delta_{\nu\nu'}  + \delta_{il}\delta_{\mu\nu'} \cdot
\delta_{jk}\delta_{\nu\mu'}]$, one gets
\begin{equation} \label{eq:HRImean}
\fl \hat H_{\rm RI} = \frac{1}{2} \sum_{\fraczero{i,j }{\sigma \sigma'}}
\left[ \frac{\hV(0)}{\Volume} c^\dagger_{i\sigma} c^\dagger_{j\sigma'}
  c_{j\sigma'}c_{i\sigma}  
  + 
  \frac{\langle \hV \rangle_{\rm f.s.}}{\Volume} 
c^\dagger_{i\sigma} c^\dagger_{j\sigma'} 
  c_{i\sigma'}c_{j\sigma} \right] \quad + \quad 
\mbox{fluctuating terms} \; .
\end{equation}
We can then use the equalities
\begin{eqnarray*}
 \sum_{\fraczero{i,j}{\sigma \sigma'}} c^\dagger_{i\sigma}
c^\dagger_{j\sigma'} c_{j\sigma'}c_{i\sigma} & = & \hat N(\hat N -1)
\\
\sum_{\fraczero{i,j }{\sigma \sigma'}}c^\dagger_{i\sigma}
c^\dagger_{j\sigma'} c_{i\sigma'} c_{j\sigma} & = & -\hat N^2/2 + 2\hat N
- 2 \mathbf{\hat S}_{\rm tot}^2 
\end{eqnarray*}
to write the residual-interaction part of the Hamiltonian as the sum
of two terms $\hat H_{\rm RI} = \hat H^{(N)}_{\rm RI} + \hat
H^{(S)}_{\rm RI}$.  The  $\hat N$-dependent part $\hat H^{(N)}_{\rm
  RI} = ({\hV(0)}/{2\Volume}) \hat N(\hat N -1) - ({\langle \hV
  \rangle_{\rm f.s.}}/{\Volume}) [(\hat N/4)(\hat N -4)] $ can be
aggregated with the smooth term $E_{\rm sm}(N)$ in
(\ref{eq:1rstorder}), while the spin part can be expressed as
\begin{equation} \label{eq:univH}
\hat H^{(S)}_{\rm RI} = -  \frac{\langle \hV
  \rangle_{\rm f.s.}}{\Volume} \mathbf{\hat S}_{\rm tot}^2= - \Delta \,
f_0^a \mathbf{\hat S}_{\rm tot}^2 \; . 
\end{equation} 
We obtain in this relatively pedestrian way that, as expected, when
fluctuations are neglected, from general symmetry considerations,
the residual interaction is, up to a term which depend smoothly on the
total number of particles, proportional to $\mathbf{\hat S}_{\rm
  tot}^2$. We moreover obtain that the proportionality constant is $-
\langle \hV \rangle_{\rm f.s.}/ \Volume$ which should, as before, be
interpreted as $- \Delta \, f_0^a$.

\begin{figure} 
\begin{center}
\includegraphics[width=8cm]{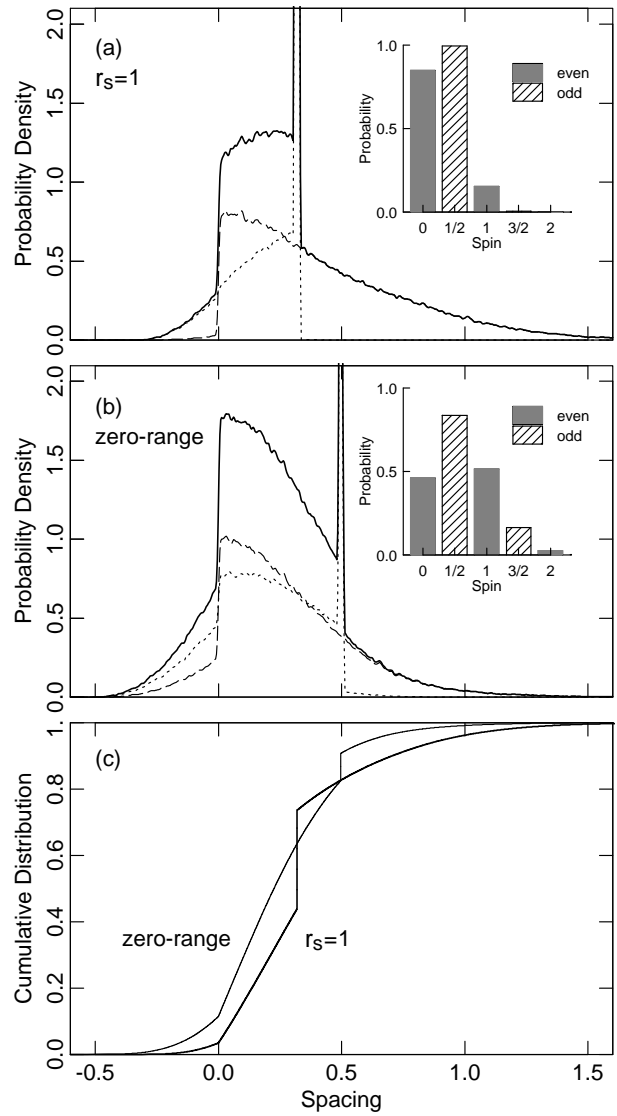}
\end{center}
\caption{ The probability density of Coulomb-blockade peak spacings
  obtained from the universal-Hamiltonian model at zero temperature. The
  total distribution (solid) as well as that for $N$ even (dashed) and
  odd (dotted) is given for two strengths of the interactions: (a)
  $r_s \smeq 1$ and (b) the zero-range interaction limit $\Vsc(\bfr -
  \bfr') = \mathrm{g_s}^{-1} \nu_0 \, \delta(\bfr - \bfr')$.  The presence of a
  $\delta$-function in the distribution is particularly clear in the
  cumulative distribution functions (the integral of the probability
  density) shown in (c). Insets show the probability of occurrence of
  ground-state spins in the two cases.  The spacing is in units of the
  mean-level separation $\Delta$, and the origin corresponds to the
  classical spacing $e^2/C$.  This figure is taken from
  \citeasnoun{Ullmo01prbb}. }
\label{fig:PRB2001}
\end{figure}

We now understand that the basic ingredient missing in the
constant-interaction model was the existence of the term $\hat
H^{(S)}_{\rm RI}$.  This term tends to polarise the spin of the
quantum dot, and is of the same order (the mean-level spacing
$\Delta$) as the cost in one-particle energy associated with moving a
particle in higher orbitals.  Because the numerical value of $f^0_a$
is smaller than one half, the one-particle energy cost is {\em on
  average}, always numerically larger than the spin polarisation term.
However the fluctuations of the one-particle energies $\energy_i$ taking place
also on the scale $\Delta$, imply that, depending on the actual value
of the $\energy_i$'s, the ground state may
actually have a non-minimal spin.  This in return will modify the
quantities such as the peak-spacing distributions.
Figure~\ref{fig:PRB2001} shows the distributions obtained within this
model for  $r_s = 1$.

\subsubsection{Time-reversal invariant systems}
\label{subsec:TRI}

Physical effects not contained in the universal Hamiltonian are going
to further modify the peak-spacing and spin distributions.  Before
considering them, let us return to the question of time
reversal invariance, assuming again chaotic dynamics.  When this
symmetry is present the random matrix is no more invariant under
all unitary transformations, but instead under the smaller group of orthogonal
transformations.  In addition to $\hat N$ and $\hat S$, such
transformations leave also $\hat T = \sum_i c_{i\uparrow}
c_{i\downarrow}$ invariant.  Taking into account that the
residual-interaction Hamiltonian should be invariant under a global
phase change of the unperturbed one-particle eigenstates (and thus of
the $c_{i\sigma}$, implying that only the product $\hat T^\dagger
\hat T$ should be involved), and that $\Hint$ contains only
four-fermion products, we see that an extra term
\begin{equation} \label{eq:univHTR}
\hat H^{(T)}_{\rm RI} = - J_T \hat T^\dagger \hat T
\end{equation} 
should be added to the general form of the universal Hamiltonian
\cite{Kurland00,Aleiner02PhysRep}.

Using the form of the plane-wave model valid for time-reversal
invariant systems [see discussion below (\ref{eq:aimu})], we can
as before express the interaction matrix element $V_{ijkl}$ through
(\ref{eq:Vijkl_bis}).   Now, however, $\langle a^*_{i\mu}
a^*_{j\nu} a_{l\nu'} a_{k\mu'} \rangle  = \frac{1}{M
  \Volume} [\delta_{ik}\delta_{\mu\mu'} \cdot
\delta_{jl}\delta_{\nu\nu'} + \delta_{il}\delta_{\mu\nu'} \cdot
\delta_{jk}\delta_{\nu\mu'} + \delta_{ij}\delta_{\mu, -\nu} \cdot
\delta_{kl}\delta_{\nu',-\mu'} ] $, where the last contribution is
specifically due to time-reversal invariance.  One  obtains
\begin{equation}
  \langle V_{iijj} \rangle_{\rm TRI}  = 
 \langle V_{iiii}\rangle_{\rm TRNI} \delta_{ii} +
  \frac{1}{\Volume} \langle 
  \hV \rangle_{\rm f.s.} \; ,
\end{equation}
where $\langle V_{iiii}\rangle_{\rm TRNI}$ is the mean value of the
completely diagonal matrix element in the time-reversal non-invariant
case.

We obtain in this way that $J_T$ equals $J_S$.  What makes somewhat
more difficult the discussion of systems with time-reversal invariance
is the need to consider higher-order perturbation terms.  Indeed,
while $\hat N$ and $\hat S^2$ were commuting with the unperturbed
Hamiltonian, the product $\hat T^\dagger \hat T$ is not.  In
particular, the evaluation of second-order corrections in the residual
interactions (see (\ref{eq:E(2)GS}) in the next subsection) shows that
the matrix elements associated with the promotion to an empty orbital
of two electrons (with opposite spins) occupying the same orbital are
now of the  order of the mean-level spacing $\Delta$.  This correction is
thus by no means small.

We recognise however in $\hat H^{(T)}_{\rm RI}$ the usual pairing
Hamiltonian used in the study of superconductivity (except that the
interaction here is repulsive).  It is well known \cite{AGD}  that
the important higher-order terms are, as in section
\ref{subsec:correlations}, the Cooper series shown in
figure~\ref{fig:cooper}.
In the same way as for the
magnetic response, the main role of these higher-order terms is therefore
to renormalise the interaction {\em in the Cooper channel} according to
(\ref{eq:renormalized_int}) (for $d=2$). 
This argument is often used to neglect $\hat H^{(T)}_{\rm RI}$ since
the renormalised coupling constant goes to zero in the limit $g \to
\infty$.  However, for typical numbers of particles in the studied
experimental dots (e.g.\ from 340 to 1000 for
\citeasnoun**{Patel98a} or about 200 for
\citeasnoun{Luscher01}) the ``large'' logarithm
$\mathrm{g_s}^{-1} \ln{(\kF L)} \simeq \mathrm{g_s}^{-1} \ln{(4\pi N
  /\mathrm{g_s})}$ remains in the range $[3.5,4.4]$.  The higher-order
terms therefore reduce the effects of the interactions in the Cooper
channel (i.e.\ a smaller effective  parameter $J_{\rm T}$ should be
used) but do not eliminate them completely.

The role of the renormalization in the Cooper channel is presumably
the reason why density functional calculations in the local spin
density approximation, while they agree very well with the kind of Fermi-liquid
description given here for time-reversal non-invariant systems
\cite{Ullmo05prb(R)},  significantly overestimate the role of the
residual interactions in the time-reversal invariant case.  The
higher-order terms appear not to be treated correctly in this approach
\cite{Jiang03prl,Jiang03prb,Ullmo04prb}.

\subsection{Beyond the universal Hamiltonian}
\label{subsec:beyond_univ_Ham}

In figure~\ref{fig:PRB2001} one sees that predictions derived at zero
temperature from the universal Hamiltonian are noticeably different
from the ones ((\ref{eq:CIpspa})-(\ref{eq:CIpspb})) obtained within
the constant-interaction model.  In particular, as shown in the
inset, the exchange term (\ref{eq:univH}) may give rise to a non-zero
proportion of non-naive (i.e.\ different from zero or one half)
ground-state spins.  However, even if one accounts for some
experimental noise that would smooth some of the sharpest features,
the peak-spacing distributions derived in this way do not have the
uni-modal / Gaussian-like shape observed experimentally.  It is clear
that the universal Hamiltonian cannot be the full story. Nevertheless
it seems not unreasonable to assume that it provides a decent starting
point, which needs to be supplemented by a few other physical effects
before a complete description is reached.  Among the possible
candidates, come naturally to mind: i) scrambling
\cite{Blanter97,Ullmo01prbb,Usaj02,Jiang05prb} and gate effects
\cite{Vallejos98,Ullmo01prbb,Usaj02,Jiang05prb}; ii) fluctuations of
the residual-interaction terms \cite{Blanter97,Ullmo01prbb,Usaj02},
and contributions beyond order one \cite{Jacquod00,Jacquod01,Usaj02} ;
and finally iii) temperature effects \cite{Usaj01,Usaj02}.  I shall
consider each of them separately, and end with a discussion about the
importance of the more or less chaotic character of the actual
dynamics within the dots \cite{Ullmo03prl}.

\subsubsection{Scrambling and gate effects}

\label{subsec:scrambling}

In addition to residual-interaction terms (\ref{eq:RI}), there
is one noticeable difference between the Fermi-liquid approach that we
follow here and the constant-interaction model.  Indeed, we have in
principle a well defined procedure to specify the self-consistent
confining potential $\Umf(\bfr)$ seen by the electrons: for a given
experimental configuration, it can be obtained by solving the
Thomas-Fermi equations (\ref{eq:TFSC1})-(\ref{eq:TFSC2}).  [Note it is important
not to use here for $\Umf(\bfr)$ the potential obtained from a
self-consistent Hartree Fock, or density functional, calculation, as
this would mix the scrambling and the gate effects with (some
treatment of) the residual interactions]. As one extra electron is
added into the quantum dot, or as the voltage of the control gate is
changed (without changing the number of electrons in the dot) from
$V^*_g(N-1 \to N)$ to $V^*_g(N \to N+1)$, the potential $\Umf(\bfr)$
will be modified to a new value $\Umf(\bfr) + \delta \Umf(\bfr)$.  As a
consequence, the one-particle energies $\energy_i$, and therefore the
corresponding contribution to the ground-state energy, will be
modified.  The fluctuating part of the this energy change is referred
to as scrambling (for the variation of $N$) and gate effect
\cite{Vallejos98} (for the variation of $V_g$).

An evaluation of the magnitude is possible analytically for the simple
geometry of a circular billiard, with in addition (for the gate effect)
the assumption that one is using an ``universal gate'', i.e. one which
is featureless and of the same size (or larger) than the quantum
dot.  Under these assumptions, both effects appear to have a variance
scaling as $1/g$, to be of the same
size, and to be comparable in magnitude with the fluctuations of the
residual interactions that I shall discuss below
\cite{Ullmo01prbb,Usaj02,Aleiner02PhysRep}. 

A more careful analysis reveals that a realistic modelling of the
geometry of the quantum dots, which implies the numerical solution of
the Thomas-Fermi self-consistent equations, gives a drastically
different picture \cite{Jiang05prb}.  Indeed because in practice
lateral plunger gates are used rather than the kind of ``universal''
ones assumed above, the quantum dots are affected asymmetrically by
the gates, and gate effects turns out to be stronger than scrambling.
On the other hand, the very smooth confinement associated with a
realistic potential $\Umf(\bfr)$ is much easier screened than the hard-wall
boundaries of a billiard.  This turns out to diminish noticeably the
magnitude of both scrambling and gate effects:  In the
configuration studied in \citeasnoun{Jiang05prb}, very small
fluctuations (less than a percent of the mean-level spacing for the
root mean square (r.m.s.)) is obtained for the scrambling, so that it
can presumably be neglected in actual experiments.  Because of the use
of a lateral plunger gate, gate effects are -- again in the
configuration studied in \citeasnoun{Jiang05prb} -- not as small
as scrambling, giving rise to fluctuations of the order of $7\%$ of a
mean-level spacing.  This figure remains however smaller than the
noise level for all the dots studied in \citeasnoun**{Patel98a}
(though marginally smaller for the quieter one, used for
figure~\ref{fig:patel}).

Since some particular care has been exercised by L\"uscher et al.\ to
minimise gate effects (the lateral gate is not so narrow, and the
higher electronic density makes screening more effective) it can be
assumed that the quantum dot studied in \citeasnoun{Luscher01} shows
even smaller gate effects than their  evaluation in
\citeasnoun{Jiang05prb}.

\subsubsection{Fluctuations of the residual-interaction terms and
  higher-order correction}
\label{subsec:higherorder}

Screening and gate effects correspond to changes of the one-particle
energies when the smooth effective confining potential is modified for
one or another reason. They lead to  variations of the
parameters of the universal Hamiltonian, but remain in some sense
contained within this description.

One may, however, consider terms, associated with the fluctuations of
the residual interactions, which are corrections to the universal
Hamiltonian.  For instance, the off-diagonal (i.e.\ $(i,j) \neq
(k,l)$) coefficients $V_{ijkl}$ (\ref{eq:Vijkl}) are zero on
average, but have a non-zero variance which can be evaluated easily
within the random-plane-wave model.  For instance if $i \neq k,l$ and
$j \neq k,l$, considering for simplicity the zero range approximation
$\Vsc(\bfr-\bfr') = \nu_0^{-1} \delta(\bfr-\bfr')$, and using
(\ref{eq:RPW}) and (\ref{eq:aimu}) together with the Gaussian
character of the $a_{i\mu}$'s, one can write
\begin{equation}
\langle |V_{ijkl}|^2 \rangle = \frac{1}{ \nu_0^2} \frac{1}{(M {\Volume})^4}
\sum_{\mu \nu \tau \eta} \int d\bfr d\bfr' 
\exp \left[\frac{i}{\hbar}{(\bp_\mu + \bp_\nu - \bp_\tau -
    \bp_\eta)(\bfr-\bfr')} \right]
\; .
\end{equation}
Once the momenta are constrained to be on the Fermi
circle (or sphere, this seems to be irrelevant here), the condition
$\bp_\mu + \bp_\nu = \bp_\tau + \bp_\eta$ implied by the integration
over space imposes $\bp_\mu = \bp_\tau + \Or(\hbar/L)$ and $\bp_\nu =
\bp_\eta + \Or(\hbar/L)$ or the
converse (the $ \Or(\hbar/L)$  originates from the width given to 
 the Fermi surface in the RPW model).  As a consequence, using that
the number of plane waves $M$ scales as the dimensionless conductance
$g$, we get
\begin{equation}
\frac{\langle |V_{ijkl}|^2 \rangle}{\Delta^2} \propto \frac{1}{g^2}
\; .
\end{equation}
We see that the smallness of the typical size of the off-diagonal
elements is not so much due to the weakness of the interactions (in
some sense, $\nu_0^{-1} \delta(\bfr-\bfr')$ is of order one once put in
the proper units), but rather originates from  the self averaging associated with
the integration over space of fluctuating quantities.  

As the matrix elements are small in the semiclassical limit $g \to
\infty$, one can expect for instance that second-order correction to
the ground-state energy
\begin{equation} \label{eq:E(2)GS}
  E^{(2)}_{0} = \sum_{j \neq 0} \frac{|\langle \Psi_j^N | H_{RI} |
  \Psi_0^N \rangle|^2}{E^{(0)}_{0} - E^{(0)}_j} 
\end{equation}
also scales as some power of $1/g$.  The analysis is made somewhat more
complicated by the fact that the number of one-particle levels within
the Thouless energy grows with $g$, and thus also the number of
one-particle--one-hole and two-particle--two-hole excitations involved
in the summation.  A careful analysis performed by
\citeasnoun*{Usaj02} shows nevertheless that the typical size of the
second-order correction to the addition energy scales as $\Delta/g$,
is numerically already extremely small (less than $2\%$ of $\Delta$)
for $N=500$, and can be safely neglected.  This furthermore gives us
extra confidence that the kind of Fermi-liquid perturbative approach
that we are following is indeed valid.

The largest terms neglected by the universal-Hamiltonian description
is therefore associated with the fluctuations of the diagonal residual
interaction terms $\mathrm{M}_{ij} = V_{ijij}$ and $\mathrm{N}_{ij} =
V_{ijji}$ that we 
are going to evaluate more carefully now in the case of a two
dimensional quantum dot.  

As before, we shall base our calculation on the random-plane-wave
model.  It turns out that the variances of the $\mathrm{M}_{ij}$'s and
$\mathrm{N}_{ij}$'s (expressed in units of the mean-level spacing $\Delta$)
keep a dependence on the size of the system.  To reproduce correctly
this  size dependence one needs the second version of the
random-plane-wave model, in which the condition (\ref{eq:quantcond})
is imposed on the states used in the expansion (\ref{eq:RPW}), giving
a width $\sim \hbar /L $ to the Fermi surface.   In this way, 
one can write the $\mathrm{M}_{ij}$'s as 
\begin{eqnarray}
	\mathrm{M}_{ij} & = & \sum_{\mu_1,\mu_2,\mu_3,\mu_4}
	\!\!\! a_{i\mu_1} a^*_{i\mu_2} a_{j\mu_3} a^*_{j\mu_4}
	\delta_{\bp_{\mu_1}-\bp_{\mu_2},-\bp_{\mu_3}+\bp_{\mu_4} }
          \hV(\bp_{\mu_1}-\bp_{\mu_2}) 
            \nonumber \\
	 & = & \sum_{\bq} \hV(\bq) W_{i \bq} W^*_{j \bq}
\end{eqnarray}
where one introduces the definition
\begin{equation}
	W_{i \bq} \equiv \sum_{\bp_{\mu_1}-\bp_{\mu_2}=\bq} 
        v_{i\mu_1} v^*_{i\mu_2} \; .
\end{equation}

The random-matrix model implies
\begin{eqnarray}
	\langle a_{i \bk_1} a^*_{j\bk_2} \rangle & = & (\kf L)^{-1} \delta_{ij}
	\delta_{\bp_1 \bp_2} 	\qquad \mbox{ if }  \delta k < \pi/L \\
	& = & 0 \; 	\qquad\qquad \qquad \qquad \mbox{ if }  
	\delta k > \pi/L
        \nonumber
\end{eqnarray}
with $\delta k = \left| |\bk_1| - k_i \right|$ and 
$k_i = \sqrt{2 m \energy_i}/\hbar$.  From this, one deduces
\begin{equation}
	{\rm var} ( \mathrm{M}_{i \neq j} ) \simeq \frac{1}{A^2} \sum_{\bq\neq 0}
	\hV^2(\bq) \langle |W_{i \bq}|^2 \rangle  
	\langle |W_{j \bq}|^2 \rangle \; .
\end{equation}
$|W_{j \bq}|^2$ can be interpreted as $(2\pi k_i)^{-2}$ times the
area of the intersection of two rings of diameter $k_i$ and width $2
\pi /L$, centered at a distance $|\bq| = q$.  Simple geometry
 gives, for $2\pi /L \leq |\bq| \leq 2k_i - 2\pi/L$ 
\begin{equation}
	\langle |W_{i \bq}|^2 \rangle \simeq \frac{4}{(q L)
		\sqrt{(2 k_i L)^2 - (q L)^2}} \;.
\end{equation}
One obtains for $i \simeq j$
\begin{equation}
     {\rm var} ( \mathrm{M}_{i \neq j} )  \simeq 
        \frac{8}{\pi A^2} \int_{\pi/L}^{2k-\pi/L} \frac{dq }{q}
           \frac{\Vsc(q)^2}{(2kL)^2 - (qL)^2} \; .
\end{equation}

The variance of $\mathrm{M}_{ii}$ and $\mathrm{N}_{ij}$ and the
covariance between $\mathrm{M}_{ij}$ and $\mathrm{N}_{ij}$ can be
computed along the same lines, and one gets
\begin{eqnarray}
\fl     {\rm var} ( \mathrm{N}_{i \neq j} ) & \simeq &
        \frac{2}{\pi A^2} \int_{\pi/L}^{2k-\pi/L} \frac{dq}{q}
           \frac{[\hV(q) + \hV(2k)]^2 }{(2kL)^2 - (qL)^2} \; ,
                 \label{eq:N_var} \\
\fl	\langle \mathrm{M}_{i \neq j} \mathrm{N}_{i \neq j} \rangle - 
	\langle \mathrm{M}_{i \neq j}\rangle \langle \mathrm{N}_{i
          \neq j} \rangle  
         & \simeq & 
        \frac{4}{\pi A^2} \int_{\pi/L}^{2k-\pi/L} \frac{dq}{q}	
           \frac{\hV(q)[\hV(q) + \hV(2k)] }{(2kL)^2 - (qL)^2} \; .
\end{eqnarray}
The diagonal part of the direct residual interaction
has an extra contribution because of the additional fashion in which the
wave-functions can be paired:
   \begin{equation} \label{eq:Mii_var} \fl
      {\rm var}( \mathrm{M}_{ii} ) \simeq
          2 {\rm var}( \mathrm{M}_{i \neq j}^2 )
         + \frac{8}{\pi A^2} \int_{\pi}^{2k-\pi/L} \frac{dq}{q}
           \frac{\hV(q) [\hV(0)+\hV(\sqrt{(2k)^2 - q^2})]}{(2kL)^2 -(qL)^2}
        \; . 
   \end{equation}
In the zero-range interaction limit, the expressions for the variance of the
$M$'s and $N$'s simplify considerably and one finds in this case
\begin{equation} \label{eq:sr_var}
      {\rm var} ( \mathrm{M}_{ij} ) = {\rm var} ( \mathrm{N}_{i \neq j} )=
     \frac{3 \Delta^2}{4\pi} \frac{\ln(kL)}{(kL)^2} (1+3\delta_{ij})
        \; .
\end{equation}
Note that the decay of the wave-function correlations appearing in the
variance produces a factor of $1/\kf L$ in the root mean square
compared to the mean.  The $\ln(kL)$ factor is special to two
dimensions; it comes from the $1/kL$ decay of the wave-function
correlator in this case. 

The variance of  $\mathrm{M}_{ij}$'s and $\mathrm{N}_{ij}$'s
 scale as $1/{g}^2$.  When adding an extra electron of spin $\tilde \sigma$
 in the orbital $\tilde \jmath$, the variation of the residual interactions 
[cf.\ (\ref{eq:RI})] 
\begin{equation} \label{eq:dRI}
\delta E^{\rm RI}   = 
 \sum_{i\sigma}  n_{i\sigma} \mathrm{M}_{i \tilde \jmath}
- \sum_{i}  n_{i \tilde \sigma} \mathrm{N}_{i \tilde \jmath} \; ,
\end{equation}
involves in principle a summation over all occupied orbitals.  The
energy range within which fluctuations take place is however given by
the Thouless energy $\Eth$. Thus when evaluating the fluctuations of
$\delta E^{\rm RI}$, only a number scaling as $g = \Eth/\Delta$ of
$M_{i\tilde \jmath}$ and $N_{i\tilde \jmath}$ should be considered as
independent.  As a result the variance of $\delta E^{\rm RI}$ scales as
$g \times 1/g^2 = 1/g$ (and the r.m.s.\ as $1/\sqrt{g}$).

The contributions of the residual interactions to the second
difference (\ref{eq:D2E}) then depend on whether or not the $N$'th and
$N+1$'th particles occupy the same orbital $\tilde \jmath$.  In the
former case (assuming for simplicity that all orbitals are either
doubly occupied or empty before the $N$'th electron is added) only
$M_{\tilde \jmath,\tilde \jmath}$, survive the difference $ \delta^{2} E^{\rm
  RI} \equiv \delta E^{\rm RI}[N \! \to \! (N\!+\!1)] - \delta E^{\rm
  RI}[(N\!-\!1) \! \to \! N]$.  The variance of $ \delta^{2} E^{\rm RI}$ then
scales as $1/g^2$. This implies in particular that the Dirac delta peak
visible in figure~\ref{fig:PRB2001} is only marginally affected by the
fluctuations of the residual interaction. If however the last electron is
added into a different orbital, there is no
cancellation between $\delta E^{\rm RI}[N \!\to\! (N\!+\!1)]$ and $\delta
E^{\rm RI}[(N\!-\!1) \! \to \! N]$.  
Typical values of $\delta^{2} E^{\rm RI}$ then scales as $\delta E^{\rm
  RI}$, i.e.\ as $1/\sqrt{g}$. 
For relatively small dots ($ N \simeq  100 $) this can lead to fluctuations
of the order of $10\%$ of a mean-level spacing.  The fluctuations of
the residual interactions therefore give rise to a contribution to the
fluctuations of the second energy difference $\delta^2 E_N$ which, for
$N$ even (and more precisely when the $N$'th and $N+1$'th electrons
are added to different orbitals),  is parametrically larger (r.m.s.\
$\sim 1/\sqrt{g}$) than the fluctuations due to higher-order terms,
and numerically larger than the scrambling and gate effects discussed
in section~\ref{subsec:scrambling}.

\subsubsection{Finite temperature}
\label{subsec:finiteT}

Up to this point, we have considered scrambling and gate effects,
which can be considered as due to variations of the one-particle part
of the universal Hamiltonian, and fluctuations of the residual
interactions terms which are genuine corrections to this Hamiltonian.  They
give rise to fluctuations with a r.m.s.\ scaling\footnote{Note that
  for scrambling and gate effects, this scaling has been properly
  derived only in the case of billiards systems, and not for the more
  realistic case of smooth confinement.} as $1/\sqrt{g}$ and
prefactors which, for experimentally realistic parameters, are $ < 1\%
\Delta$ for scrambling, about $7\% \Delta$ for gate effects, and of
the order of $10\% \Delta$ for the residual-interaction fluctuations.

These figures have  to be compared with the r.m.s.\ of
the experimental noise  which, for the dots
studied in \citeasnoun**{Patel98a} is equal
to $8\% \Delta$ for dot 1 (whose data have been used to construct
figure~\ref{fig:patel} and \ref{fig:ong}) $17\% \Delta$ for dot 5,
$22\% \Delta$ for dot 4, and $40\% \Delta$ or above for the four
remaining ones.

For dots with a reasonably low level of experimental noise, these
extra fluctuating terms (namely scrambling, gate effects, residual
interaction fluctuations, and noise itself) can blur the sharp
features of distributions such as the one depicted in
figure~\ref{fig:PRB2001}.  Nevertheless, they cannot  prevent
the distribution of the second ground-state energy difference
$\delta^2 E_N$ to remain bimodal, and therefore incompatible with the
experimental peak-spacing distributions.

Usaj and Baranger pointed out  that an analysis done at
zero temperature, and thus concerning only ground-state properties,
was however inadequate to interpret the present experiments.  For
instance, the dots studied in \citeasnoun**{Patel98a} were at a
temperature ranging from $20\%$ to $60\%$ of a mean-level spacing.  If one
had in mind that the mean-level spacing gives the scale of the first
many-body excitation,  such a difference between $T$ and $\Delta$ seemed
enough to justify a zero temperature approach, at least for the
smallest of the dots (largest $\Delta$).  This is especially true
for chaotic system, since level repulsion ensure that the proportion
of small spacings is small ($\simeq 8. \, 10^{-3}$ for spacings smaller than
$0.2 \, \Delta$ in the Gaussian unitary ensemble).  Therefore, within the
constant-interaction model, the probability for an excited state to be
populated because of thermal fluctuations could be considered as
negligible for the smallest dots, and reasonably small for the
larger ones.

The presence of the exchange term (\ref{eq:univH}) makes however a
drastic difference in this respect. Consider for instance $\delta
E_{S=1,S=0}$, the difference of energy between the $S=1$ and the $S=0$
lowest-energy states (assuming $N$ even).  It equals on average
$\Delta - 2J_S$, that is, for $r_s=1$ ($J_s \simeq 0.34$), about
thirty percent of the mean-level spacing.  Moreover level repulsion
does not help anymore since it affects only the small
spacings.  If one assumes GUE fluctuations for the one-particle
energies, the probability that $|\delta E_{S=1,S=0}| \leq 0.2\Delta$
is $\simeq 0.29$.  In other words, even for the smallest of the dot
studied in \citeasnoun**{Patel98a}, there is almost one chance out of
three that both the $S=0$ and the $S=1$ states are both significantly
occupied.

In addition, even when most of the conduction is provided by the
ground-state, the fact that different spin states have different
degeneracies, and thus different entropies ${\cal S} = \ln(2S+1)$, will
modify the conduction peak positions when $T \neq 0$.

The finite temperature linear conductance near a $N-1
\to N$ transition can be obtained in the rate equation approximation
as \cite{Beenakker91,Meir92,Usaj01,Usaj02}
\begin{equation}
\label{eq:G}
G(V_g) = \frac{e^2}{\hbar \kt}\, P_{\rm eq}^{N}\,
                \sum_{\alpha}{\frac{\Gamma_{\alpha}^L\Gamma_{\alpha}^R}
                 {\Gamma_{\alpha}^L\smpl\Gamma_{\alpha}^R}\,w_{\alpha}}
\end{equation}
where $P_{\rm eq}^{N}$ is the {\em equilibrium} probability that the
quantum dot contains $N$ electrons, $\Gamma^{L(R)}_{\alpha}$ is the
partial width of the 
single-particle level $\alpha$ due to tunneling to the left (right)
lead, and $w_{\alpha}$ is a weight factor given by
\begin{equation} \label{eq:weight}
w_{\alpha} = \sum_{i,j,\sigma} F_{\rm eq}(j|N)
\left|\langle\Psi_{j}^{N} |c_{\alpha,\sigma}^{\dagger}
     |\Psi_{i}^{N\smmi1}\rangle\right|^2 \,
              \frac{1}{1 + \exp[-\beta(E_j - E_i)]} \; . 
\end{equation}
In (\ref{eq:weight}), $\Ha_{\rm QD}|\Psi_{j}^N\rangle \smeq
E_{j}|\Psi_{j}^N\rangle$, so that ``$j$'' labels the many-body states
of the quantum dot, and $F_{\rm eq}(j|N)$ is the conditional probability that
the eigenstate $j$ is occupied given that the quantum dot contains $N$
electrons.

From this, one can derive the temperature dependence of
the peak positions (maxima of $G(V_g)$) when  the sole ground states
have a notable occupation probability  as \cite{Usaj01,Usaj02}
\begin{equation} \label{eq:UB2002}
  \mu = E_0^N -  E_0^{N-1} - k_{\rm B}T 
  \ln\left[\frac{2S_0^N +1}{2S_0^{N-1}+1}\right] +
  2 (C_g E_c/e) \, \delta V_g
\end{equation}
which, up to a factor $1/2$, amounts to replacing the ground-state
energies $E_0^N$ by the free energies $F^N = E_0^N + k_{\rm B}T
{\cal S}$.

\begin{figure}
\begin{center}
\includegraphics[width=3.3in]{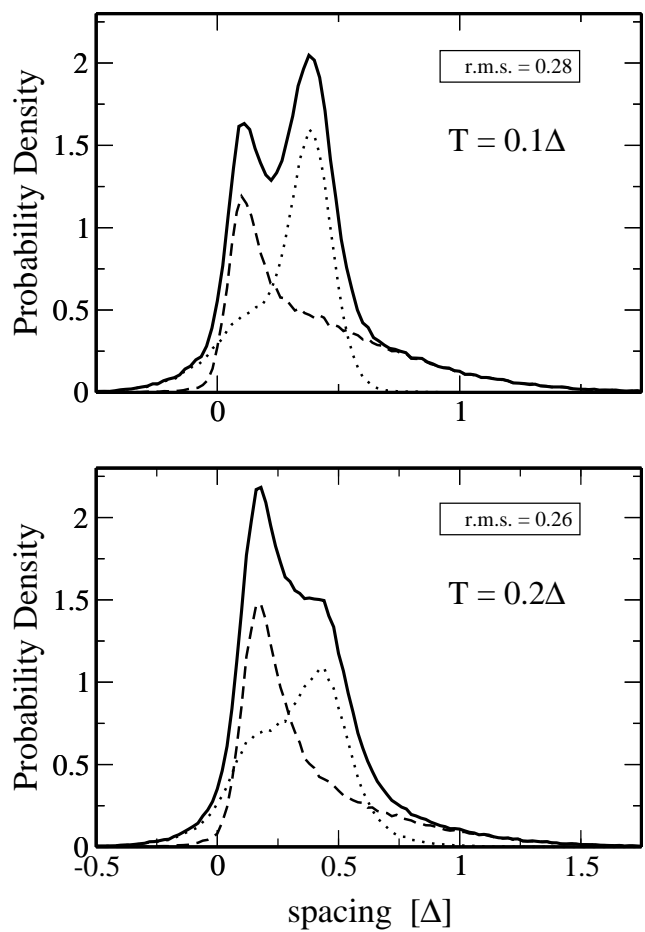}
\end{center}
\caption{\label{fig:usaj_finitT} Finite temperature Coulomb-blockade peak-spacing
  distribution obtained once, in addition to the universal
  Hamiltonian, scrambling, gate effects, fluctuations of the residual
  interactions, and, more importantly, finite temperature effects, are
  taken into account.  The dotted line correspond to $N$ odd, the
  dashed line to $N$ even, and the solid one to the total (odd plus
  even) distribution. $N\smeq500$ ($g \!\approx\! 6$), $J_S \smeq 0.32
  \Delta$ and $\kt \smeq 0.1 \Delta$ $(0.2 \Delta)$ in the top
  (bottom) plot.  This figure is taken from \citeasnoun{Usaj01}. }
\end{figure}

The general case (conduction through more than one state) leads to
more complicated expressions, but can be obtained
\cite{Usaj01,Usaj02}.  Both effects -- conduction through excited
states and displacement of the peaks due to the entropy -- affect the
spacing distribution: at $k_{\rm B}T \simeq 0.2 \Delta$ they already
dominate the fluctuation of the residual interactions.  This is
illustrated in figure~\ref{fig:usaj_finitT} -- taken from
\citeasnoun{Usaj01} -- where it is seen that, once scrambling, gate
effects, fluctuation of the residual interactions, are already taken
into account, raising the temperature from $10\%$ to $20\%$ of the
mean-level spacing suppresses the bi-modality of the peak-spacing
distribution.  Already at $T = 20\% \Delta$, this
distribution is in reasonable agreement with the
experimental result shown in figure~\ref{fig:patel}.  As soon as the
temperature is further increased (e.g. $T=30\%\Delta$ for the data in
figure~\ref{fig:ong}) or the level of experimental noise becomes
significant, one should expect little (or none) even/odd effect or
asymmetry in the peak-spacing distributions.

\subsubsection{Non-chaotic dots}
\label{subsec:NCdots}
  
As we just have seen, starting from a universal-Hamiltonian
description, the peak-spacing distributions measured by Patel et al.\
can be understood within an approach where finite temperature effects
and (experimental) noise -- except for the quietest dot -- are the most
important ingredients.  Conversely, the fact that noise and
temperature play an important role implies that other features of the
modelling are less critical, and in that sense are not really probed by these
experiments.

The size of the dot studied by \citeasnoun**{Luscher01} is
however significantly smaller, leading to a ratio $T/\Delta \simeq
5\%$ such that temperature cannot be responsible for the absence of
bi-modality in the peak-spacing distribution.  The noise level in
\citeasnoun{Luscher01} is not quoted, but is presumably small enough
not to affect significantly the spacing distributions either 
\cite{EnsslinPrivate}.  The approach based on the universal
Hamiltonian plus corrections does not therefore appear to be able to explain the
distributions shown in figure~\ref{fig:luscher}.  Other mechanisms have
to be introduced to understand the statistical properties
of peak-spacings.

Interestingly, one issue, to which usually little attention is paid,
is to know whether using a chaotic model for the dynamics of the
electrons in the dot is adequate.  

There are in fact good reasons to focus on chaotic dynamics, the main
one being that the behaviour of chaotic systems is universal (in the
sense that this behaviour does not depend on the details of the system
as long as this latter is in the chaotic regime).  Working in the
chaotic regime, both experimentally and theoretically, one (happily)
avoids messy system-specific considerations.

It remains nevertheless that, even when some care has been exercised,
as in \citeasnoun**{Patel98a}, to bring the quantum dots in the
chaotic regime, it is extremely difficult to design a system with
smooth confining potential that is without question in the chaotic
regime: No known smooth Hamiltonian system is mathematically proved to
be in the chaotic regime;  Furthermore, as any numerical simulations
of the classical dynamics of low dimensional systems readily
demonstrates, it is extremely difficult, especially in the presence of
magnetic field, to choose a confining potential without any obvious
regions of regular motion.

Clearly the point is not to decide whether any experimental system in
the chaotic regime in the strict mathematical sense, but, for a given
setup, how close the system is from fully developed chaos, and,
from a general perspective, how much the predictions based on the
assumption that the dynamics is chaotic are robust with respect to
that hypothesis.  In other words, the question which has to be
answered is how much chaotic dynamics are representative, for what
concerns the residual interactions, of the larger class of mixed (i.e.\
partly chaotic partly regular) dynamics that one is going to find in
practice in experiments.

One consideration may indicate that residual interactions, and as a
consequence the ground-state spin and peak-spacing distributions, are
sensitive to the degree of regularity of the system under
investigation.  Indeed, using for simplicity the short range
approximation $\Vsr (\bfr-\bfr') = \nu_0^{-1} \delta(\bfr-\bfr')$, one
can express the
completely diagonal matrix element as
$V_{iiii} = \mathrm{M}_{ii} = \nu_0^{-1} \int d\bfr
|\varphi_i(\bfr)|^4$. Up to the factor $\nu_0^{-1}$, $\mathrm{M}_{ii}$
is therefore the inverse participation ratio of the state $\varphi_i$,
a measure of how much this latter is localised.

Of all possible dynamics, chaotic systems are however the one showing
the least localisation, as their wave-functions are spread
out uniformly across the whole energetically accessible domain.
Integrable and mixed systems on the other hand display
various forms of phase space localisation (in the sense that the
Wigner or Husimi transforms of the wave-functions are concentrated in
some portion of the phase space).  The mechanisms underlying this kind
of localisation (not to be confused  with Anderson
localisation) range from the relatively trivial quantization on
invariant torus to more subtle effects of partial barrier
\footnote{For instance cantori
  \cite{MacKay84prl,MacKay84physicaD,MacKay87physicaD} or stable and
  unstable manifolds \cite{Bohigas90prlb,Bohigas93}).}, but are known
in any case to be quite pervasive in mixed systems, even when the
phase-space proportion of genuinely regular motion  is not very
large.

What this considerations imply is that although chaotic systems
represent most presumably a universality class as far as residual
interaction effects are concerned, they might be the class for which
{\em residual interactions are the least effective} when considered
from the more general point of view of the range of possible dynamics.
To go beyond the kind of qualitative arguments used above is made
complicated by the lack of simple models (such as the
random-plane-wave model) for the wave function fluctuations of mixed
systems.  What is possible however is to consider some specific
example of smooth Hamiltonian systems showing, as a function of some
external parameter, various kinds of dynamics.  One then can see in
these particular cases whether the actual wave-functions fluctuations
used as an entry for the computation of the residual interactions
induce strong modifications with respect to the chaotic predictions
for the ground-state spin or addition-energy distributions.

A convenient choice, introduced by \citeasnoun{Ullmo03prl}, is the
coupled quartic oscillator system ($\bfr = (x,y)$, $r = |\bfr|)$)
\begin{equation} \label{eq:MixedHam}
   \hat H = \frac{\left(\bp - \kappa \sqrt{a(\lambda)} x^2 \frac{\bfr}{
       r} \right)^2}{2}
   +
   a(\lambda) \left( \frac{x^4}{b} + b y^4 + 2\lambda x^2 y ^2 \right)
   \; .
\end{equation}
Here, $b=\pi/4$, $a(\lambda)$ is a convenient scaling factor chosen so
that the mean number of states with energy smaller than $\energy$ is given
by $N(\energy) = \energy^{3/2}$ and $\lambda$ is the coupling between the
oscillators.  The parameter $\kappa$ is chosen such that
time-reversal invariance is completely broken. 

\begin{figure}[hr]
\begin{center}
\includegraphics[width=3.3in]{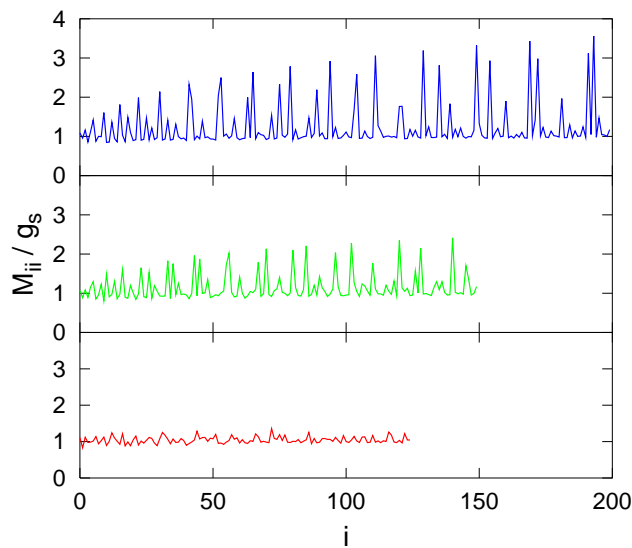}
\end{center}
\caption{Inverse participation ratio as a function of the orbital index
for the symmetry class $(+,+)$.   From top down: $(\lambda,\kappa) =
(+0.20,1.00)$ 
[nearly integrable], $(-0.20,1.00)$ [mixed], and $(-0.80,1.00)$ [mostly
chaotic]. This figure is taken from \citeasnoun{Ullmo03prl}.
\label{fig:miiScaled}}
\end{figure}
\begin{table}[h]
\caption{\label{table:spinpol} Probabilities $P(S=2)$ , $P(S=5/2)$ to
  find a spin two (even $N$) or five halves (odd $N$) ground-state, and
  average value $\langle  \delta S  \rangle$ of the ground-state spin
  augmentation ($\delta S =  S$ or $(S-1/2)$ for even or odd number of
  particles, respectively), for the  various dynamical regimes (values of
  $\lambda$) with $\kappa=1.0$ and $J_s=0.4$.  The last column is the
  RMT/RPW prediction~\cite{Ullmo01prbb} valid for the hard-chaos
  regime.  This table is taken from 
  \citeasnoun{Ullmo03prl}.} 
\begin{center}
\begin{tabular}{|c|cccc|}
\hline
$\lambda$ & \ ~~+0.20~~ \ & \ ~~-0.20~~ \ & \  ~~-0.80~~ \ & ~RMT/RPW~
\\ \hline
$P(S=2)  $  & 0.13  &  0.16 & 0.07  & 0.01   \\
$~P(S=5/2)~$  & 0.08  &  0.10 & 0.02  & 0.00    \\
$\langle \delta S \rangle$ & 0.51  &  0.54 & 0.38   & 0.23  \\
\hline
\end{tabular}
\end{center}
\end{table}

In figure~\ref{fig:miiScaled}, values of sets of diagonal terms
$\mathrm{M}_{ii}$ are represented for one symmetry class in various
dynamical regimes, showing as expected that their behaviour is very
sensitive to the nature of the classical dynamics.  Table
\ref{table:spinpol} further shows that this also affects drastically
the spin distribution.  Indeed, for the model (\ref{eq:MixedHam})
there is, in the mixed regime, a significant proportion of ground
state spin 2 or 5/2, whereas, as predicted by the universal-Hamiltonian
approach, such ``large'' spins are essentially absent in the chaotic
case. As discussed in \citeasnoun{Ullmo03prl}, the presence of a
large proportion of non-trivial spin is associated with distributions
of spacings $\delta^2 E_N$ which differ significantly from the ones expected
in the hard-chaos regime. This example illustrates that the question
of non-chaoticity, although it has received less attention than other
issues such as scrambling or fluctuations of residual interactions, may
play a significant role in explaining the difference between
universal-Hamiltonian predictions and experimental observations, and
possibly be as important as the question of temperature.

\setcounter{footnote}{1}

\section{Conclusion}
\label{sec:Conclusion}

In this review, I have introduced some of the tools from the field of
quantum chaos which may be applied to the understanding of various
many-body effects in mesoscopic physics.  Among these tools, a first
group is related to the semiclassical approximations of the Green's
function.  They provide a link between the quantum properties of fully
coherent systems and the classical propagation of trajectories within
their classical counterparts.  This link gives a very intuitive
picture for many quantal properties of interest.  In particular it
makes it possible to introduce naturally the classical probability
$P^\energy_{\rm cl}(\bfr,\bfr',t)$ (cf.\ the M-formula
(\ref{eq:Mformula})) which is central in the understanding of
diffusive or chaotic quantum dots.  Another set of tools is related to
random-matrix theory, and the closely related random-plane-wave
models, allowing for a statistical description of individual
eigenstates of classically chaotic systems.

The main body of the review, however, has been devoted to the
application of theses tools to a selected set of examples of physical
interest.  In particular, I have discussed in detail how the
``semiclassical-Green's-function-based'' approximations can be used to
compute the interaction contribution to the orbital magnetic response
of mesoscopic systems.  For diffusive dynamics, I have shown that it
was possible to recover in this way, in a simple and transparent
manner, most of the results derived previously by diagrammatic
perturbation techniques.  This approach made it possible furthermore
to address the ballistic regime, and in particular non-chaotic
geometries which are beyond the scope of applicability of the more
traditional methods. Another illustration was provided by the mesoscopic
fluctuations of Kondo properties for a magnetic impurity in a bounded,
fully coherent electron gas, in the temperature regime such that a
perturbative renormalization group treatment can be applied.  Here
again, semiclassical techniques made it possible to relate some
relatively non-trivial properties of the quantum system to the
propagation of classical trajectories.

In the last section, turning to an energy scale $\sim \Delta$ for
which the semiclassical approximations of the Green's function are
generally not applicable, I have discussed how the statistical
properties of individual wave-functions derived from random ensembles
could be applied to the study of peak-spacings and ground-state--spin
statistics in Coulomb-blockade experiments.  In the chaotic regime, I
have presented in particular the ``universal-Hamiltonian'' description
derived by Aleiner and coworkers, in a leading semiclassical (i.e.
here $1/g$) approximation, from a generic random-matrix argument.  I
have furthermore shown how the parameters of this universal
Hamiltonian, as well as the evaluation of the leading corrections, can
be obtained from the random-plane-wave models.  Finally I have
discussed some important modifications expected for non-chaotic dots,
and their possible experimental relevance.

From these various examples emerges a general picture of what the
``quantum chaos'' based approach advocated in this review brings to
the understanding of the many-body physics of mesoscopic systems.  I
would like, for the remaining of this concluding section, to discuss
it on more general grounds.

In this respect, one important class of systems is such that
either the dynamics is genuinely diffusive (i.e.\ disordered systems),
or it is assumed that there is no interesting difference
between the actual dynamics and (some limit of) a diffusive one.
Ballistic chaotic systems under some circumstances fall into
the latter category.  Furthermore, extension of the nonlinear
super-symmetric $\sigma$-model to billiard systems with diffusive
boundaries permits addressing many important statistical properties of
chaotic systems \cite{Blanter98,Blanter01} (as long as the former
represents a good model for the latter).  In these cases, compared to
traditional diagrammatic
\cite{AltshulerAronov85,AkkermansMontambauxBook} or more modern
\cite{EfetovBook,Blanter98,Blanter01} techniques, the approach
proposed in this review provides mainly an alternative route to the
results. Its main advantage is then that this alternative route is
closer to the actual physics considered.  Indeed, most of the relevant
parameters (Thouless energy, transport mean free path), and more generally
the physics (in the sense of extracting scales and qualitative
behaviours) of these systems are usually discussed in terms of 
classical quantities (classical probability of return, time to diffuse
to the boundary, etc..).  The semiclassical approach developed here
makes it therefore possible to perform the calculations with the same
``language'' as the physical discussion, and is therefore usually more
transparent and intuitive.

For disordered systems this advantage has however to be balanced
with the respective strengths of the other approaches.  The
super-symmetric nonlinear $\sigma$-model techniques \cite{EfetovBook}
for instance can address stronger disorders, for which Anderson
localisation begins to develop, whereas semiclassical methods, as well as as
diagrammatic perturbations, are limited to the ``good conductor,''
i.e. diffusive, regime.  In the same way, when the properties of
individual wave functions are considered, and more generally when the
physics of interest takes place on the scale of the mean level spacing
$\Delta$, super-symmetric nonlinear $\sigma$-models provide more
systematic (and controlled) ways of performing calculations in the
diffusive regime.  The manipulation of the random-plane-wave
models may on the other hand sometime involve  a little bit of
artwork, but is more flexible in nature and can be adapted to other
dynamical regimes \cite{Baeker2002a,Baecker2002b}.

\begin{figure} 
\begin{center}
\includegraphics[totalheight=5cm]{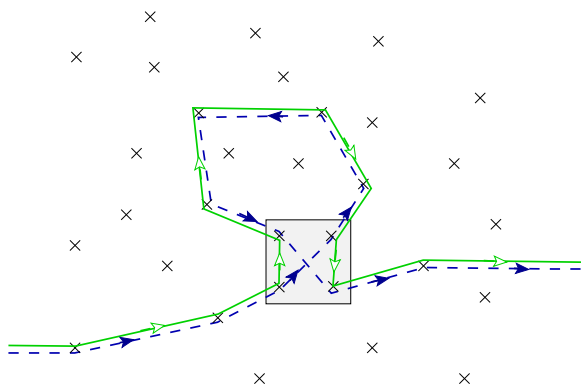}
\end{center}
\caption{Illustration of an ``Hikami box'' (shaded region in this
  figure).  In this particular example the two trajectories switch
  from following the same path to following time reversal path.  Such
  configuration would typically be found in the calculation of weak
  localization.}
\label{fig:hikami}
\end{figure}

Another issue which has to be kept in mind when addressing disordered
systems is the one of Hikami boxes \cite{Hikami81}.  In a
semiclassical language (see section C3.1 of
\citeasnoun{AkkermansMontambauxBook} for a more detailed
discussion), a Hikami Box can be described as a restricted region of
space where two pairs of trajectories following closely each other
change partner.  An illustration is given in figure~\ref{fig:hikami}.

Because the length scale associated to Hikami boxes (namely the mean
free path) is well separated from the classical scales associated to
diffusion, their computation are not particularly difficult in the
diagrammatic approach.  Hikami boxes turn out to be more delicate to
handle in a purely semiclassical description, in particular because
questions of current conservation have to be treated with care.
Recent works \footnote{See for instance
\cite{Aleiner96prb,Richter02prl,Adagideli03,Muller04,Heusler04,Muller05,Whitney05,Rahav05prl,Heusler06,Jacquod06,Whitney06prl,Brouwer06,Muller07}.}
show that these difficulties can actually be overcome, but the level
of technicality involved becomes then at least as high as with the
diagrammatic approach.  The usefulness of the semiclassical approaches
when processes related to Hikami boxes are involved is then not
anymore their simplicity but the mere fact that they can be
applied to a wider class of problems, being not by construction
limited to disordered problems.

For questions related to many-body effects in mesoscopic systems, the
quantum chaos based approach described in this review serves therefore
two purposes.  One is to provide a way to perform needed calculations
which remains very closely related to the qualitative physics
discussed, and is therefore very transparent.  In this respect it is
presumably the best first step into this field for either
theoreticians or experimentalists.

It is, furthermore, the only way to address any experimental system
which cannot be properly described by diffusive or fully chaotic
dynamics.  From this perspective, it is interesting to come back to
the interpretation of the experimental results of
\citeasnoun**{Patel98a} and of \citeasnoun**{Luscher01}.  In the first
case \cite{Patel98a}, some care has been exercised to bring the dots
under study in the chaotic regime (although it is presumably not
genuinely there).  Furthermore, the value of $r_s$ (i.e. the strength
of the interactions), and the relatively high ($ \geq 20\% \Delta$)
temperature make it such that experimental data are compatible with a
chaotic modelling of the dynamics \cite{Usaj01,Usaj02}.  On the other
hand, the dots studied by L\"uscher et al.\ have an essentially square
shape, and it is much less likely that the associated dynamics are in
the fully chaotic regime.  Furthermore a smaller value of $r_s$, and
more importantly of the temperature ($\sim 0.05 \Delta$) make their
data incompatible with a chaotic modelling even if one includes the
effects of the gate, of scrambling, of residual interaction
fluctuations, and of temperature.

As seen in the last subsection of Sec.~\ref{sec:CB}, the qualitative
behaviour of non-chaotic systems can, in some circumstances, be
drastically different.  The experimental results of \citeasnoun{Luscher01}
provides at least one unambiguous example where assuming  diffusive or
chaotic behaviour is not a reasonable starting point. 
It therefore demonstrates the interest of having an approach not
making too much of an assumption concerning  the nature of the classical dynamics
within the system, so that at least the question of whether anything
new, or interesting, could happen in other dynamical regimes could, at
least, be asked.  My hope is that this review provides a step in this
direction.

\ack

This review has benefited from many discussions on various of the
aspects it covers, and in particular with Harold Baranger, Oriol
Bohigas, Rodolfo Jalabert, Patricio Leboeuf, Gilles Montambaux,
Christophe Texier, and Gonzalo Usaj, who has furthermore provided some
of the figures included in the Coulomb-blockade section.  Harold
Baranger, Rodolfo Jalabert, Gilles Montambaux, Christophe Texier,
Steve Tomsovic, and Marcel Veneroni, have in addition read preliminary
versions of the manuscript.  Their comments and suggestions have
greatly improved the readability of its content.

\setcounter{footnote}{1}

\appendix

\section{ Screening of the coulomb interaction in a
  generic quantum dot}
\label{sec:appendix}

The purpose of this appendix is to address the question of screening
of the Coulomb interaction in a generic quantum dot.  This issue has
been studied in the case of diffusive systems
\cite{Blanter97,Aleiner02PhysRep}, and I will essentially follow here the
approach proposed in section~2.3.2 of
\citeasnoun**{Aleiner02PhysRep}. The main motivation will be to
illustrate how the semiclassical approach presented in
section~\ref{sec:basic} can be used to generalise some results derived
in the more traditional framework of diffusive systems.  Here however
the exercise will turn out to be somewhat academic since I will need
to perform at one point an ``uncontrolled'' (read incorrect here)
approximation, leading in the end to an unphysical result (the
equations (\ref{eq:WTFSC1})-(\ref{eq:WTFSC2}) in place of the expected
ones (\ref{eq:TFSC1})-(\ref{eq:TFSC2})).  Nevertheless, the
calculation is in itself instructive enough to be detailed.
Identifying clearly the uncontrolled step of the derivation may
furthermore helping clarifying the condition of applicability of the
original result derived for the diffusive regime, which might turn out
to be useful.

In the bulk, the RPA screened interaction, is obtained by considering the Dyson
equation for the dressed interaction (see the discussion in
section~9 of \citeasnoun*{Fetter&Walecka}) 
\begin{equation} \label{eq:RPA} \fl
V_{\rm dressed}(\bfr_1,\bfr_2,\omega) = V_{\rm coul}(\bfr_1-\bfr_2) +
\int d\bfr \int  d\bfr' V_{\rm coul}(\bfr_1-\bfr) \Pi(\bfr,\bfr',\omega) 
V_{\rm dressed}(\bfr',\bfr_2,\omega) \; ,
\end{equation}
which is exact if all the one-particle irreducible diagrams are
included for the polarisation operator $\Pi$  but gives the RPA
approximation if only the (lowest order) bubble diagram
\begin{equation}
\Pi^0(\bfr,\bfr',\omega) = \mathrm{g_s} \int_{-\infty}^{+\infty}
\frac{d\omega'}{2i\pi} 
G(\bfr,\bfr',\omega+\omega')G(\bfr',\bfr,\omega')
\end{equation}
is kept.  $G(\bfr,\bfr',\omega) = \Theta(\omega) G^R(\bfr,\bfr',\omega) +
\Theta(-\omega) G^A(\bfr,\bfr',\omega)$ is the unperturbed time ordered
Green's function, with $\Theta(x)$ the Heaviside function, and
$\mathrm{g_s} = 2$ is the spin degeneracy factor.  In the zero
frequency low momentum limit one gets (in the bulk)
$\Pi^0(\bfr,\bfr',\omega=0) \simeq - \mathrm{g_s} \nu_0
\delta(\bfr-\bfr')$, with $\nu_0$ the local density of states per spin
(\ref{eq:rho-0}).  Inserting this expression for $\Pi^0$ in
(\ref{eq:RPA}) gives (\ref{eq:Vq2d})-(\ref{eq:Vq3d}).

Let us consider now a mesoscopic systems, and assume that its typical
dimensions are much larger than the screening length.  One then expects
that the residual screened Coulomb interaction should be very similar
to the one in the bulk, and it is therefore natural to approach the
question from the same viewpoint.  In that case however the Green's
function are not known exactly, so one needs to resort to semiclassical
approximations of $G^{R,A}$ in the expression of $\Pi^0$.  The
difficulty encountered then is that semiclassical approximations are
valid for high energies (high $\omega$), and in particular one cannot
expect the semiclassical expressions for $G(\bfr,\bfr',\omega)$ to
be accurate if $\omega$ is not much larger than the mean level spacing
$\Delta$ of the system.

Following \citeasnoun**{Aleiner02PhysRep}, the idea is then, in the spirit of
the renormalization group approach, to integrate out only the ``fast
variable'' (high-energy part) for which a semiclassical approximations
can be used, and to deal with the low energy physics by some other
methods (based for instance on a random-matrix description
\cite{Murthy02,Murthy03}).  Using the exact expression for the
polarisation bubble
\begin{equation}
\Pi^0(\bfr,\bfr',\omega) = {\mathrm{g_s}}
\sum_{nn'} \Theta(- \epsilon_n \epsilon_{n'}) 
\frac{ \varphi^*_n(\bfr') \varphi_n(\bfr)  \varphi^*_{n'}(\bfr)
  \varphi_{n'}(\bfr')}
{\omega 
+ \epsilon_{n'} - \epsilon_{n}} (-{\rm sgn}(\epsilon_{n}))
\end{equation}
with $(\epsilon_{n},\varphi_n(\bfr))$ the one-particle energies and
eigenstates, we see that this can be achieved by restricting the sum
in the above expression to pair $(n,n')$ such that at least one energy
is outside a band centered at the Fermi energy $\energy_F$ (taken as
the origin of energies) and of width $\energy^*$ chosen such that
$\Delta \ll \energy^* \ll E_{\rm Th}$, and which precise value (once
in this range) is expected to be irrelevant.  Up to an unimportant
boundary term, this is equivalent to restricting the sum to
particle-hole energies $\epsilon_{n'} - \epsilon_{n}$ larger (in
absolute value) than $\energy^*$. Introducing
$\Pi^{R,A}(\bfr,\bfr',\omega) \df \lim_{\eta \to 0+}
\Pi^0(\bfr,\bfr',\omega \pm i \eta)$ the retarded
and advance polarisation bubbles, one can therefore write the polarisation
operator in which only the fast modes are integrated out as
\begin{equation} \label{eq:hatPi} 
\hat \Pi_{\energy^*} (\bfr',\bfr,\tilde
  \omega \smeq 0) = \frac{1}{2i\pi} \int \frac{d\omega}{\omega} \left[
    \Pi^R(\bfr,\bfr',\omega) - \Pi^A(\bfr,\bfr',\omega) \right]
  \Theta(|\omega| -\energy^*) \; .
\end{equation}
The insertion of $\hat \Pi_{\energy^*}$ in (\ref{eq:RPA}) will then give
the effective interaction describing the low energy ($\leq \energy^*$)
physics of the quantum dot.

\subsection{Calculation of the polarisation loop}

Let us first consider positive energies $\omega>0$.  Noting that phase
cancellation is possible only for the product $G^AG^R$, but not for
$G^RG^R$ or $G^AG^A$, one has
\begin{equation}
\Pi^R(\bfr,\bfr',\omega) = \mathrm{g_s} \int_{-\omega}^{0} \frac{d\omega'}{2i\pi}
G^R(\bfr,\bfr',\omega' + \omega + i \eta) G^A(\bfr',\bfr,\omega')
\end{equation}
and 
\begin{equation} \label{eq:PiA}
\Pi^A(\bfr,\bfr',\omega) = \Pi^R(\bfr',\bfr,\omega)^* \; .
\end{equation}
Using the semiclassical expressions (\ref{eq:1/2classGreen}) and
keeping only the diagonal approximation in which a trajectory $j$ is
paired with itself to cancel the oscillating phases, one gets
\begin{equation} \fl
[G^R(\bfr,\bfr',\omega+\omega')G^A(\bfr',\bfr,\omega')]_{\rm diag}
= \sum_{j: \bfr \to \bfr'} \frac{4\pi^2}{(2\pi\hbar)^{d+1}}
|D_j|^2 \exp \left[i (S_j(\omega +\omega') - S_j(\omega'))  /\hbar \right] \; .
\end{equation}
In this equation, one would like then to perform a Taylor expansion of
the action
\begin{equation} \label{eq:becarefullwiththataxeeugen}
(S_j(\omega +\omega') - S_j(\omega')) = (\partial S_j/\partial
\energy) \omega = t_j \omega \; 
\end{equation}
where the last equality comes from (\ref{eq:dS/dE}).  Inserting
the unity $\int_0^\infty \delta(t-t_j)$ we obtain
\begin{eqnarray} \fl
[G^R(\bfr,\bfr',\omega+\omega')G^A(\bfr',\bfr,\omega')]_{\rm diag}
& = & \frac{4\pi^2}{(2\pi\hbar)^{d+1}}
\int_0^\infty  dt  \sum_{j: \bfr \to \bfr'}  |D_j|^2   \delta(t-t_j)
\exp \left[i t \omega  /\hbar \right] \\ \fl
& = & \frac{2\pi \nu_0(\bfr')}{\hbar} \int_0^\infty dt P^\energy_{\rm cl}(\bfr,\bfr',t) 
\exp \left[i t \omega  /\hbar \right] \label{eq:tohatP} \\ \fl
& = & {2\pi} \nu_0(\bfr) \hat P^\energy_{\rm cl}(\bfr,\bfr',\omega) \; ,
\end{eqnarray}
where we have used the M-formula (\ref{eq:Mformula}) and $\hat
P^\energy_{\rm cl}$ is the Fourier transform of the classical probability
$P^\energy_{\rm cl}$.  Interestingly enough
$[G^R(\omega+\omega')G^A(\omega')]_{\rm diag}$ is independent of
$\omega'$, so that finally 
\begin{equation} \label{eq:PiR+}
\Pi^R(\bfr,\bfr',\omega) = - i\omega  \mathrm{g_s} \nu_0(\bfr') \hat
P^\energy_{\rm cl}(\bfr,\bfr',\omega) \; . 
\end{equation}

Note that the fact that we have computed $\Pi^R$, i.e.\ that $\omega
\equiv \omega+i\eta$, is what is making the Fourier transform in
(\ref{eq:tohatP}) convergent. If we had computed $\Pi^A$ the above approach
would have lead to divergences. $\Pi^A$ should therefore be derived from
$\Pi^R$ using (\ref{eq:PiA}), giving 
\begin{equation} \label{eq:PiA+}
\Pi^A(\bfr,\bfr',\omega) = i \omega  \mathrm{g_s} \nu_0(\bfr) \hat P^\energy_{\rm
  cl}(\bfr',\bfr,\omega) \; . 
\end{equation}
For negative $\omega$, $\Pi^A$ should be calculated first and $\Pi^R$
derived from it with (\ref{eq:PiA}), leading to the same result.

Here, one rather important remark is in order.
The expression (\ref{eq:becarefullwiththataxeeugen}) assumes obviously that
$\omega$ is small.  This is usually not a significant constraint since
the actions $S_i$ are classical quantities, so that the relevant scale
is the Fermi energy (or bandwidth) $\energy_F$.  It is therefore enough that
$\omega \ll \energy_F$ to apply (\ref{eq:becarefullwiththataxeeugen}).
However the integral in the left hand side of (\ref{eq:hatPi}) is
not limited to the neighbourhood of the Fermi surface.  Replacing
$\Pi^{R,A}$ by the approximations (\ref{eq:PiR+})-(\ref{eq:PiA+})
will be incorrect on the edge of the energy band, which will be
associated to short distances $|\bfr - \bfr'| < \lF$.  This will
be the cause of the problems we shall encounter later on.  Let us
ignore this issue for the time being, and come back to this discussion
when it will become obvious that the results obtained in this way are
unphysical.

Then, inserting (\ref{eq:PiR+})-(\ref{eq:PiA+}) into 
(\ref{eq:hatPi}) and writing $\Theta(x) = 1 - \Theta(-x)$ we get
\begin{equation} \label{eq:hatPiSC1} \fl
\hat \Pi_{\energy^*} (\bfr',\bfr,\tilde \omega \smeq 0) =
- \mathrm{g_s}  \int_{-\infty}^{+\infty} \frac{d\omega}{2\pi}
\left[\nu_0(\bfr') \tilde P^\energy_{\rm cl}(\bfr,\bfr',\omega) + 
\nu_0(\bfr) \tilde P^\energy_{\rm cl}(\bfr,\bfr',\omega) \right]
\left(1 - \Theta(\energy^* \! - \! |\omega|) \right) \; .
\end{equation}
The term proportional to one in the integrand of (\ref{eq:hatPiSC1})
gives rise to $\int ({d \omega}/{2\pi})\tilde P^\energy_{\rm
  cl}(\bfr,\bfr',\omega) = P^\energy_{\rm cl}(\bfr,\bfr',t\smeq=0)$.
To evaluate the remaining term, it is useful to discuss the weight
function $\Theta(\energy^* - |\omega|)$.  Its precise form is
irrelevant here, and, rather than the actual Heaviside step function,
I shall assume that $\Theta(\energy^* \! - \! |\omega|)$ is actually a
smooth function $\Theta_{\energy^*}(\omega)$ which is zero for
$|\omega| \gg \energy^*$ and one for $|\omega| \ll \energy^*$.  To fix
the idea one can think for instance of $\Theta_{\energy^*}(\omega) =
\exp(-(1/2)(\omega/ \energy^*)^2$), but this precise form will not
play any particular role.  If one introduces $\tilde
\Theta_{\energy^*}(t)$ the Fourier transform of
$\Theta_{\energy^*}(\omega)$, one has, with $t^* = \hbar / \energy^*$
\begin{eqnarray}
\tilde \Theta_{\energy^*}(t) & \simeq & 1/t^* \qquad \mbox{for $t \ll t^*$} \\
& = & 0 \qquad \quad \mbox{for $t \gg t^*$} \\
\int_0^\infty dt \, \tilde \Theta_{\energy^*}(t)  & = &
\Theta_{\energy^*}(\omega \smeq=0) = 1 \; .
\end{eqnarray}
Assuming furthermore that $\tilde \Theta_{\energy^*}(t)$ is a positive function
(this hypothesis can be easily relaxed), we see that $\tilde
\Theta_{\energy^*}(t)$ is a density probability (since it is positive and
normalised to one) which selects trajectory shorter than $t^*$. We
thus can write
\begin{equation} \label{eq:hatPiSC2} \fl
\hat \Pi_{\energy^*} (\bfr',\bfr,\tilde   \omega \smeq 0) = 
       - \mathrm{g_s} \left[ \nu_0(\bfr') P^\energy_{\rm cl}(\bfr,\bfr', t\smeq 0) 
       - \frac{1}{2} 
       \left( \nu_0(\bfr') \langle P^\energy_{\rm cl}(\bfr,\bfr', t) 
                         \rangle_{t \leq t^*} 
      + \nu_0(\bfr) \langle P^\energy_{\rm cl}(\bfr',\bfr, t) 
                    \rangle_{t \leq t^*} \right) \right] \; ,
\end{equation}
where $\langle f(t)  \rangle_{t \leq t^*} \df \int_0^\infty  dt \, f(t)
\tilde \Theta_{\energy^*}(t) $ is the average over time $t$ lesser than $t^*$
of the function $f(t)$.

Now $P^\energy_{\rm cl}(\bfr,\bfr',t\smeq=0) = \delta(\bfr-\bfr')$.  Furthermore,
the condition $\Delta \ll \energy^* \ll E_{\rm Th}$ is equivalent to $ t_f
\ll t^* \ll t_H$, with $t_H = \hbar/\Delta$ the Heisenberg time and
$t_f$ the time of flight across the system (for ballistic systems) or
time needed to diffuse to the boundary (for diffusive systems).  We
see that the choice of $\energy^*$ is made precisely so that i)
semiclassical approximations are valid, but also ii) that most of the
range $[0,t^*]$ is such that {\em for diffusive or chaotic systems}
(the case of integrable or mixed system should be investigated in this
respect), the motion can be assumed randomised.  Assuming ergodicity
we can therefore write
\begin{equation} \label{eq:<P>}
\langle P^\energy_{\rm cl}(\bfr,\bfr', t) \rangle_{t \leq t^*}
\simeq \frac{\int d\bp \delta(\energy_F - H(\bfr,\bp)) }
{\int d\bfr^{''} d\bp^{''} \delta(\energy_F - H(\bfr^{''},\bp^{''})) } =
\Delta \nu_0(\bfr) \; .
\end{equation}
This eventually leads to 
\begin{equation} \label{eq:hatPiSC3}
\hat \Pi_{\energy^*} (\bfr',\bfr,\tilde \omega \smeq 0) =
- \mathrm{g_s}  \left[\nu_0(\bfr) \delta(\bfr-\bfr') - \nu_0(\bfr) \nu_0(\bfr') \Delta \right]
\; ,
\end{equation}
where one recognise the first term as the zero frequency low momentum
bulk polarisation$\Pi^0_{\rm bulk} (\bfr',\bfr,\tilde \omega \smeq 0) =
-\mathrm{g_s} \nu_0(\bfr) \delta(\bfr-\bfr')$, and I will denote by 
\begin{equation} \label{eq:PiLR}
\Pi_{l.r.} \df \mathrm{g_s} \nu_0(\bfr) \nu_0(\bfr') \Delta
\end{equation}
 the remaining long range part.
For billiard systems for which $\nu_0(\bfr) = ({\Volume}  \Delta)^{-1}
={\rm const.}$,  with ${\Volume}$ the volume of the system,
(\ref{eq:hatPiSC3}) is for instance exactly the equation~(60) of
\citeasnoun**{Aleiner02PhysRep}.

\subsection{Self-consistent equation}

In the bulk, both the Coulomb interaction (\ref{eq:bare}) and the
polarisation operator $\Pi^0_{\rm bulk}$ are translation invariant and
the Dyson equation (\ref{eq:RPA}) can be solved in momentum
representation as
\begin{equation} \label{eq:Vscr-app}
\hat V_{\rm dressed}(\bq) = \frac{\hat V_{\rm coul}(\bq) }{
1 - \hat V_{\rm coul}(\bq) \hat  \Pi^0_{\rm bulk}(\bq) } \; .
\end{equation}
The resulting interaction is then short range, effectively much
weaker than the original Coulomb interaction, and is therefore well
adapted for a perturbative treatment.

The difficulty one encounters in the mesoscopic case is twofold.  First,
lack of translational invariance for $\Pi_{\energy^*}$ makes in principle
(\ref{eq:RPA}) impossible to be solved in closed form for a
generic spatial variations of $\nu_0(\bfr)$.  Second, we know that even at
the level of electrostatics, the effects of the interactions cannot be
small since they will at minima rearrange considerably the static
charges within the system.  Therefore, even if (\ref{eq:RPA})
could be solved, there is little chance that the resulting dressed
interaction could be effectively used in a perturbative approach
starting from the non-interacting electrons Hamiltonian.

For a quantum dot with a fixed number $(N+1)$ of electrons, one way to
solve both of these difficulties is to derive a self-consistent
equation following one of the standard derivation of the Hartree Fock
approximation \cite{ThoulessBook}.  For this purpose, let us note that
any one-body potential $\tilde U(\bfr)$, can be written formally as
the two-body interaction
\begin{equation}
\tilde V(\bfr,\bfr') = \frac{1}{N}(\tilde U(\bfr) + \tilde U(\bfr') )
\end{equation}
since, using for instance a second quantization formalism
\begin{equation}
\frac{1}{2} \int d\bfr  d\bfr' \hat \Psi^\dagger(\bfr) \hat \Psi^\dagger(\bfr')
\tilde V(\bfr,\bfr') \hat \Psi(\bfr') \hat \Psi(\bfr) 
=
\int d\bfr   \hat \Psi^\dagger(\bfr) \tilde U(\bfr)  \hat \Psi(\bfr) \; .
\end{equation}
As a consequence, the total Hamiltonian, as well as the formalism
presented in the first part of this appendix, are unmodified if
the confining potential $U_{\rm ext}(\bfr) $ and the Coulomb potential
$V_{\rm coul}(\bfr,\bfr')$ are respectively replaced by
\begin{eqnarray}
U(\bfr)  & = & U_{\rm ext}(\bfr)   + \tilde U_(\bfr) \\
V(\bfr,\bfr') & = & V_{\rm coul}(\bfr,\bfr') -  \tilde V(\bfr,\bfr') \; 
\end{eqnarray}
One can then now use the freedom in the choice of the function $\tilde
U(\bfr)$ to simplify the Dyson equation.  In particular, if we can impose
that
\begin{equation} \label{eq:constraint}
\int d\bfr \int  d\bfr' V(\bfr_1,\bfr) \Pi_{\rm l.r.}(\bfr,\bfr') 
V_{\rm dressed}(\bfr',\bfr_2) \equiv 0 \; ,
\end{equation}
the Dyson equation (\ref{eq:RPA}) would then have the usual
``bulk-like'' form 
\begin{equation} \label{eq:RPAbulk} \fl
V_{\rm dressed}(\bfr_1,\bfr_2) = V(\bfr_1,\bfr_2) -
\int d\bfr \int  d\bfr' V(\bfr_1,\bfr) \Pi^0_{\rm bulk}(\bfr-\bfr') 
V_{\rm dressed}(\bfr',\bfr_2) \; ,
\end{equation}
which, if $\nu_0(\bfr)$ and $\tilde U(\bfr)$ vary slowly on the scale of
the bulk screening length $\kappa^{-1}$ has the same solution
(\ref{eq:Vq2d})-(\ref{eq:Vq3d}) as in the bulk.

Now, equation (\ref{eq:constraint}) might seem at first sight difficult to
solve since it involves the unknown function $V_{\rm
  dressed}(\bfr',\bfr_2)$.  However, since $\Pi_{\rm l.r.}(\bfr,\bfr')$
does actually not correlate $\bfr$ and $\bfr'$, the two integrals in 
(\ref{eq:constraint}) actually decouple, and a sufficient
condition to solve this equation is that $\int d\bfr  V(\bfr_1-\bfr) \nu(\bfr)  
\equiv 0$, i.e.
\begin{equation} 
\int d\bfr  V_{\rm coul} (\bfr_1-\bfr) \nu(\bfr)
=\frac{1}{N}
\int d\bfr   \nu(\bfr) (\tilde U(\bfr) + \tilde U(\bfr_1)) \; .
\end{equation}
The constant term $\int d\bfr \nu(\bfr) \tilde U(\bfr)/N$ is
irrelevant here as it can be eliminated by a constant shift of
$\tilde U$. One therefore obtain in this way the self-consistent
equation 
\begin{equation} \label{eq:WTFSC1-app}
  \tilde U (\bfr_1) = N\Delta \int d\bfr  V_{\rm coul} 
(\bfr_1-\bfr) \nu(\bfr) \; .
\end{equation}
In other words what we have obtained for the self-consistent potential
are the equations
\begin{eqnarray}  
  \Umf(\bfr) &  = & \Vext(\bfr) + \tilde N \Delta \int d\bfr' \nu_0(\bfr')
  \Vcoul(\bfr,\bfr')  \label{eq:WTFSC1}
  \\ 
  \nu_0(\bfr) & = &\int \frac{d\bp}{(2\pi\hbar)^d}
  \delta(\mu-\Umf(\bfr)-\bp^2/2m) \; . \label{eq:WTFSC2}
\end{eqnarray}

\subsection{Discussion (C\&P)}

The above result has a few nice characteristics, and one, rather
unpleasant, feature.  On the bright side, we see that it allows to
clearly separate the two consequences of the long range Coulomb interaction:
the appearance of a self-consistent potential (obtained from equation
(\ref{eq:WTFSC1-app})) on the one hand, and on the other hand the
screening process leading to the usual ``bulk'' form
((\ref{eq:Vq2d})-(\ref{eq:Vq3d})) of the screened interaction
(provided that $\tilde U(\bfr)$ (and thus $\tilde V(\bfr,\bfr')$) varies
slowly on the scale of the screening length).

What makes (\ref{eq:WTFSC1-app}) less useful however is that it is
obviously incorrect.  Indeed we know that whatever self-consistent
equation we end up writing, it should contain in some approximation
the electrostatic equilibrium of the problem.  This is not the case
here.  If the self-consistent potential $U_{\rm mf}(\bfr) \df U_{\rm
  ext}(\bfr) + \tilde U(\bfr)$ obtained from (\ref{eq:WTFSC1-app})
is well approximated by a constant (giving for instance a billiard
system with weak disorder as was considered in
\citeasnoun**{Blanter97} and \citeasnoun**{Aleiner02PhysRep}), and
assuming $(N+1) \gg 1$, one can do the replacement $N \Delta \nu(\bfr)
\to n(\bfr)$ in (\ref{eq:WTFSC1-app}) and write instead
\begin{equation} \label{eq:TFSC1-app}
  \tilde U (\bfr_1) = \int d\bfr  V_{\rm coul} 
(\bfr_1-\bfr) n(\bfr) \; ,
\end{equation}
i.e.\ (\ref{eq:TFSC1})-(\ref{eq:TFSC2}), which is just the
Thomas Fermi equation, from which plain electrostatic is obtained by
neglecting the kinetic energy term  $\TTF[n]$ in (\ref{eq:FTF}).
However for a generic confining potential $U_{\rm ext}(\bfr)$,
solutions of (\ref{eq:WTFSC1-app}) will not in general be an
approximation of the solution of (\ref{eq:TFSC1-app})

What we see is that, in some sense, equation (\ref{eq:WTFSC1-app}) is
``aware'' of the properties of the system near the Fermi energy (the
density of states $\nu(\bfr)$), but misses the relevant information at
large energies, of the order of the bandwidth.  This is to be expected
since the polarisation operator $\Pi^*_\energy$ (\ref{eq:hatPiSC3})
involves only the local density of states at the Fermi energy
$\nu(\bfr)$.  This can be tracked back to the approximation
(\ref{eq:becarefullwiththataxeeugen}) where the action $S(\omega)$
has been linearised near the Fermi energy, eliminating in this way any
information relevant to the large (i.e.\ $\sim \energy_F$) energies.

\section{Magnetisation and persistent current}
\label{sec:AppB}

In this appendix, I re-derive briefly (for sake of completeness) the 
basic expressions (\ref{eq:Mz}) and (\ref{eq:Iorb}) of the
magnetisation and persistent current which are the starting points of
the discussion in sections~\ref{sec:OrbMag1} and \ref{sec:OrbMag2}.  I
follow here the presentation given in \citeasnoun**{Desbois98}.

Let us therefore consider a two dimensional gas ($d\!=\!2$) of
electrons confined by a potential $U(\bfr)$ ($\bfr = (x,y)$ are the
coordinate inside the plan and $hat \bi{z}$ the unit vector in the
perpendicular direction) and interacting through $V(\bfr-\bfr')$. The
total Hamiltonian of the system is therefore expressed as
$\hat H = \hat H_0 + \hat H_{\rm int}$ with 
\[
\hat H_0 \df \int d\bfr \hat\Psi^\dagger(\bfr) 
\left[ \frac{1}{2m} (-i\hbar {\bnabla} - e {\bi{A}(\bfr)})^2 +
  U(\bfr) \right] \hat\Psi(\bfr)
\] 
the non interacting part (${\bi{A}(\bfr)}$ is the vector potential),
and 
\[
\hat H_{\rm int} \df \frac{1}{2} \int d\bfr d\bfr' \hat\Psi^\dagger(\bfr)
\hat\Psi^\dagger(\bfr') 
 V(\bfr-\bfr') \hat\Psi(\bfr') \hat\Psi(\bfr) \; ,
\] 
the interacting part (which is however not going to play any role
here).  One may furthermore introduce the current density operator
\[
\hat \bi{ \jmath}(\bfr)  ={e} \hat\Psi^\dagger(\bfr) {\bf v} \hat\Psi(\bfr)
\]
with
\[
\bi{v} \df  \frac{1}{m} (-i\hbar {\bnabla} - {e} {\bi{A}(\bfr)})
\]
the velocity.  The variation of the Hamiltonian corresponding to a
variation $\delta {\bi{A}}$ of the vector potential is then expressed as 
\[
\delta \hat H = -\frac{1}{2} \int d\bfr \, [\hat \bi{ \jmath}(\bfr) \delta
\bi{A}(\bfr) + \delta \bi{A}(\bfr) \hat \bi{ \jmath}(\bfr)] \; .
\]

\subsection{Uniform perpendicular magnetic field}

Let us consider first the case where the variation $\delta \bf A$
corresponds to a uniform magnetic field $\bi{B} = \delta B_z \hat \bi{
  z}$.  The equation (\ref{eq:Mz}) basically state that the
magnetisation $M_z \df \langle \hat M_z \rangle$, with
\[
\hat  M_z \df \frac{1}{2} \int d\bfr (\bfr \times \hat \bi{ \jmath}) \cdot
\hat \bi{ z} \; ,
\] 
is the variable conjugated to  $\delta B_z$.
Indeed, choosing for convenience the symmetric gauge 
\[
\delta \bi{A} = \frac{\delta B_z}{2} (\hat \bi{ z} \times \bfr) 
\]
and, noting that $(\hat \bi{ z} \times \bfr)$ and $\bi{v}$ commute, we have
\[
\delta \hat H  = -
\frac{\delta B_z}{2} \int d\bfr 
[\hat \bi{ \jmath}(\bfr)\cdot(\hat \bi{ z} \times \bfr) + (\hat \bi{ z} \times
\bfr) \cdot\hat \bi{ \jmath}(\bfr) ] 
 = 
- \delta B_z \hat M_z \; .
\]
 We  see that $\hat M_z$ is indeed conjugated to $\delta B_z$, and the variation of the grand potential (\ref{eq:Omega})  gives (\ref{eq:Mz}).

\subsection{Flux line}

Let us now consider a half infinite line ${\cal D} = \{ \bfr_0 + \alpha
\hat \bi{ u}; \, (\alpha>0) \}$ originating from $\bfr_0$ 
and directed along the unit vector $\hat \bi{ u} = (\cos \theta_0, \sin
\theta_0)$.  One can define the current operator across ${\cal D}$ 
\[
\hat I^{\rm orb} (\bfr_0,\theta_0) \df 
   \int_{\cal D} d|\bfr-\bfr_0| \frac{(\bfr-\bfr_0)
  \times \hat \bi{ \jmath}(\bfr)}{|\bfr-\bfr_0|} \cdot \hat \bi{ z} \; ,
\]
or, if no current escape to infinity so that $\langle
I^{\rm orb} \rangle $ has no $\theta_0$ dependence 
\[
\hat I^{\rm orb} (\bfr_0) \df \frac{1}{2\pi} \int d\bfr 
   \frac{(\bfr-\bfr_0)
  \times \hat \bi{ \jmath}(\bfr)}{|\bfr-\bfr_0|^2} \cdot \hat \bi{ z} \; .
\]
Introducing the vector potential
\[
\bi{A}_{\bfr_0}(\bfr)  \df \frac{1}{2\pi} \Phi \frac {\hat \bi{ z}
  \times (\bfr-\bfr_0)}{|\bfr-\bfr_0|^2}
\]
we see that the corresponding magnetic field is $\bi{B} = {\bf
  \nabla} \times \bi{A}_{\bfr_0} = \Phi \delta(\bfr-\bfr_0) \hat \bi{
  z}$, and therefore describes a flux lines $\Phi$ at $\bfr_0$.
Following the same steps as above, we find the variation of the
Hamiltonian associated to a variation $\delta \Phi$ of the flux to be 
\[
\delta \hat H = -\delta  \Phi \int d\bfr \hat \Psi^\dagger(\bfr) \frac{1}{2\pi}
\frac {\hat \bi{ z} 
  \times (\bfr-\bfr_0)}{|\bfr-\bfr_0|^2} \cdot \bi{v} \hat \Psi(\bfr)
= -\delta  \Phi \hat I^{\rm orb} (\bfr_0)
\]
The orbital current $\hat I^{\rm orb} (\bfr_0)$ is thus conjugated to
the flux $\Phi$, which, noting $I \df \langle I^{\rm orb} \rangle$, as
before directly imply (\ref{eq:Iorb}).

\section{List of symbols}

\small
\begin{itemize}

\item $a_{i\mu}$: random-plane-wave coefficient (see (\ref{eq:RPW})).

\item $\area_j$: area enclosed by the trajectory $j$.

\item $\bi{A}(\bfr)$: vector potential.

\item $\bi{B}$: magnetic field.

\item $\beta$: $1/\kt$.

\item $\betaRMT$: random-matrix ensemble parameter.

\item $C$: quantum dot total capacitance.

\item $C_g$: gate dot capacitance.

\item $\hat c_{\alpha \sigma}^\dagger$, $\hat c_{\alpha \sigma}$:
  creation and destruction operators.

\item $\chi$: magnetic susceptibility.

\item $\cl$: Landau susceptibility.

\item $\chi_{\rm loc}$: local susceptibility.

\item $D$: diffusion coefficient.

\item $D_j$: determinant describing the stability of the trajectory $j$
  (cf. (\ref{eq:Dj})).

\item $\mathrm{D}_0$: original bandwidth of the Kondo problem.

\item $\mathrm{D}_{\rm eff} $: running  bandwidth of the Kondo problem.

\item $\Delta$: one-particle mean level spacing.

\item $\epsilon_\kappa$: one-body energy.

\item $E_N$: many-body ground-state energy of a N-particle quantum dot.

\item $E\{n_{i\sigma}\}$: energy of the many-body state corresponding
  to the occupation numbers $\{n_{i\sigma}\}$.

\item $E_{\rm sm}(N)$: smooth part of $E\{n_{i\sigma}\}$.

\item $E^{\rm RI}\{n_{i\sigma}\}$: contribution of the residual
  interactions to $E\{n_{i\sigma}\}$.

\item $E_c$: charging energy.

\item $\Eth$: Thouless energy.

\item $\Eext$: Confinement part of the Thomas Fermi functional.

\item $\Ecoul$: Coulomb part of the Thomas Fermi functional.

\item $f_0^a$: Fermi liquid parameter.

\item $f(\energy-\mu)$: Fermi function.

\item $f_\chi(T/T_K)$: universal function describing the
  susceptibility for the [bulk]  Kondo problem.

\item $\FTF$: Thomas Fermi functional.

\item $g$: dimensionless conductance $\Eth/\Delta$.

\item $\mathrm{g_s}$: ($= 2$) spin degeneracy.

\item $G(\bfr,\bfr';\energy)$: (unperturbed) Green's function.

\item $G^A(\bfr,\bfr';\energy)$: (unperturbed) advanced Green's function.

\item $G^R(\bfr,\bfr';\energy)$: (unperturbed) retarded Green's function.

\item $G^R_j(\bfr,\bfr';\energy)$: semiclassical contribution of the
  orbit $j$ to the retarded Green's function (see
  (\ref{eq:1/2classGreen})).

\item ${\cal G}(\bfr,\bfr';\omega)$: (unperturbed) Matsubara Green's
  function. 

\item $\Gamma^{L(R)}_{\alpha}$:  partial width of the single-particle level 
$\alpha$.

\item $\hbar$: Planck constant.

\item $J_0$: bare coupling constant of the Kondo Hamiltonian.

\item $J_{\rm eff} $: renormalised coupling constant of the Kondo
  Hamiltonian.

\item $J_S$: Universal-Hamiltonian coupling constant ($\mathbf{\hat
    S}_{\rm tot}^2$ term).

\item $J_T$: Universal-Hamiltonian coupling constant ($\hat
  T^\dagger \hat T$ term).

\item $J_{\rm RPA}$: RPA approximation of $J_S$.

\item $\kf$: Fermi wave-vector.

\item $k(\bfr)$: $\hbar^{-1} \sqrt{2m(E-U(\bfr))}$.

\item $\kappa$: screening wave-vector.

\item $\ell$: mean free path.

\item $L$: characteristic size of the system.

\item $L_T$: thermal length.

\item $L_\phi$: coherence length.

\item $\lF$: Fermi wavelength.

\item $\lambda_0$: electron-electron bare coupling constant.

\item $\lambda(\Lambda)$: electron-electron running coupling constant.

\item $\Lambda_0$: bare cutoff.

\item $\Lambda$: running cutoff.

\item $m_e$: particle (electron) mass.

\item $\mathrm{M}_{ij}$: screened interaction matrix element (see
  (\ref{eq:Mij})). 

\item $\langle \hat M_z \rangle $: magnetisation.

\item $\mu$: chemical potential.

\item $\pmc$: micro-canonical distribution (\ref{eq:pmc}).

\item $\hat N$: number operator.

\item $\mathrm{N}_{ij}$: screened interaction matrix element (see
  (\ref{eq:Nij})). 

\item $\nu_{\rm loc}(\bfr,\energy)$: local density of states (see
  (\ref{eq:nuloc})). 

\item $\nu_0(\bfr,\energy)$: Weyl (smooth) part of the local density of
  states (see (\ref{eq:nu-0})).

\item $\nu_{\rm osc}(\bfr,\energy)$: oscillating part of the local density of
  states (see (\ref{eq:nu-osc})).

\item $\nu_{j}(\bfr,\energy)$: Contribution of the orbit $j$ to the
  oscillating part of the local density of states.

\item $\nubeta(\bfr,T)$: thermally averaged local density of  states (see
  (\ref{eq:nubeta})).

\item $n(\bfr)$: density of particles.

\item $n(\bfr,\bfr')$: non diagonal element of the particle density
  matrix (see (\ref{eq:nrr})).

\item $\Omega$: grand potential.

\item $\Omega^C$: Cooper series contribution to the grand potential.

\item $P^\energy_{\rm cl}(\bfr,\bfr',t)$: classical probability to go
  from $\bfr'$ to $\bfr$ in a time $t$ at energy $\energy$. 

\item $\tilde P^\energy_{\rm cl}(\bfr,\bfr',\omega )$: Fourier transform
  of  $P^\energy_{\rm cl}(\bfr,\bfr',t)$.

\item $\varphi_\kappa(\bfr)$: one-body eigenstate.

\item $\left[\varphi\right]_{\rm W}(\bfr,\bp)$: Wigner transform of
  the state $\varphi$.

\item $\pf$: Fermi momentum.

\item $P_{\rm nns}$: nearest neighbor distribution.

\item $\phi_0$: quantum of flux.

\item $\phi_j$: magnetic flux enclosed by the trajectory j.

\item $\Psi^N_j$: many-body eigenstate of a quantum dot with $N$-particles.

\item $\hat \Psi^{\dagger}_{\sigma}(\bfr)$, $\hat \Psi_{\sigma}(\bfr)$:
  creation and destruction operators.

\item $r_s$: electron gas parameter.

\item $R(x)$: $x/\sinh(x)$.

\item $\rho(\energy)$: total density of states.

\item $\rho_0(\energy)$: Weyl (smooth) part of the total density of
  states.

\item $\bi{S} = (S_x,S_y,S_z)$: spin operator ($=\frac{1}{2} \hbar
\bsigma$).

\item $\mathbf{\hat S}_{\rm tot}$: quantum dot total spin.

\item ${\bsigma} = (\sigma_x,\sigma_y,\sigma_z)$: Pauli matrices.

\item $\Sigma(\bfr, \bfr'; \omega)$: (free) particle-particle
  propagator (see (\ref{eq:sigma})).

\item $\Sigma^{(D)}(\bfr, \bfr'; \omega)$: diagonal part of the particle-particle
  propagator.

\item $\Sigma_j(\bfr, \bfr'; \omega)$: contribution of the orbit $j$ to
  the diagonal part of the  particle-particle
  propagator.

\item $\orbittime_j$: time of travel of the orbit $j$.

\item $t_{\rm fl}$: time of flight across a ballistic structure.

\item $t_T$: thermal time (see (\ref{eq:tT})).

\item $\hat T^\dagger$: $\sum_i \hat c^\dagger_{i\uparrow} \hat
  c_{i\downarrow}$. 

\item $T_K$: Kondo temperature (bulk case).

\item $T^0_K$: average Kondo temperature (mesoscopic case).

\item $T^*_K[\nubeta]$: realization dependent Kondo temperature
  (mesoscopic case).

\item $T_K[\nubeta](T)$: realization and  temperature dependent Kondo
  temperature   (mesoscopic case).

\item $\TTF$: kinetic energy part of the Thomas Fermi functional.

\item $\Theta(u)$: Heaviside function.

\item $U(\bfr)$: one-body potential.

\item $U_{\rm ext}(\bfr)$: one-body external potential.

\item $U_{\rm mf}(\bfr)$: one-body mean-field potential.

\item $V_g$: gate voltage.

\item $V^*_g$: gate voltage corresponding to a conductance peak.

\item $V(\bfr-\bfr')$: two-body interaction.

\item $\Vcoul(\bfr-\bfr')$: bare Coulomb interaction.

\item $\Vsc(\bfr-\bfr')$: screened Coulomb interaction.

\item $\Vsr (\bfr-\bfr')$: short range approximation ($ = \nu_0^{-1}
  \delta(\bfr-\bfr')$ of the screened interaction.

\item $V_{ijkl}$: screened interaction matrix element (see (\ref{eq:Vijkl})).

\item $\hV(\bq)$: Fourier transform of the screened Coulomb interaction.

\item $\langle \hV \rangle_{\rm f.s.}$: Fermi surface average of $\hV(\bq)$.
\item $\maslov_j$: Maslov index of orbit $j$.

\end{itemize}

\newpage

\bibliography{rmt,nano,general_ref,magnetism,kondo}
\end{document}